\documentclass{article}

\usepackage{setspace} 

\usepackage{booktabs}
\usepackage{graphicx}
\usepackage{amsmath}
\usepackage{amsfonts}
\usepackage{natbib}
\usepackage{commath}
\usepackage{algorithm}
\usepackage{algpseudocode}
\usepackage{amssymb}
\usepackage{mathrsfs}
\usepackage{amsthm}
\usepackage{bbm}
\usepackage[colorlinks, citecolor={blue}]{hyperref}
\usepackage{cleveref}
\usepackage{mathtools}

\newtheorem{theorem}{Theorem}

\newtheorem{proposition}{Proposition}
\newtheorem{assumption}{Assumption}

\theoremstyle{definition}

\DeclarePairedDelimiterX{\infdivx}[2]{(}{)}{%
  #1\;\delimsize\|\;#2%
}
\newcommand{\infdiv}{\infdivx}

\begin{document}

\title{\bf Adaptive Generalized Elliptical Slice Sampling}
\author{Nicholas Marco\thanks{Corresponding author: nicholas.marco@duke.edu}\hspace{.2cm}\\
    Department of Statistical Science, Duke University\\
    and\\
    Surya T. Tokdar\\
    Department of Statistical Science, Duke University}
\maketitle
\bigskip

\begin{abstract}
A central challenge in gradient-free MCMC is designing algorithms that simultaneously bypass manual tuning, scale efficiently with dimension, and adapt to local target geometry. While adaptive strategies can auto-tune generic frameworks like random walk Metropolis, they offer slow, linear-order scaling of mixing times with dimension. Elliptical slice sampling (ESS) offers a promising alternative: it is tuning-free, adjusts to local geometry, and can achieve nearly dimension-free scaling under favorable conditions. However, its efficiency degrades rapidly if there is a mismatch between the target distribution and the distribution used to generate the ellipse-defining auxiliary variables, precluding its use in high-dimensional settings. We demonstrate that a careful synthesis of ESS and diminishing adaptation directly resolves these bottlenecks. The resulting adaptive generalized elliptical slice sampler (AGESS) self-corrects from a slow-mixing to a fast-mixing regime, while preserving ergodicity across a wide variety of target densities satisfying mild regularity conditions. The algorithm’s utility is demonstrated across a broad collection of challenging applications, including generalized regression, deep Gaussian process surrogate modeling, and high-dimensional sparse regression. Together, our theoretical results and the case studies give evidence of the efficiency and robustness of AGESS across target distributions that are non-elliptical, non-differentiable, multi-modal, or high-dimensional.
\end{abstract}

\noindent%
\vfill

\newpage
\onehalfspacing

\section{Introduction}
Barring a few bespoke applications, Markov chain Monte Carlo (MCMC) methods for Bayesian computation broadly rely on three updating strategies: random walk, gradient-based exploration, and slice sampling. Random walk Metropolis proposals are widely applicable, but their performance degrades in high dimensions: even with optimized step size, mixing times grow linearly with dimension \citep{roberts2001optimal}. Self-tuning random walk methods, such as adaptive random walks \citep[ARW,][]{haario2001adaptive}, can automate optimization of the step size, but may require carefully constructed localized proposal distributions for good mixing when the target distribution has an anisotropic geometry with localized features \citep{roberts2009examples, andrieu2008tutorial}. Gradient-based methods, such as the Hamiltonian Monte Carlo \citep[HMC,][]{betancourt2017conceptual, neal2011mcmc, hoffman2014no}, can exploit local gradient information and achieve sub-linear gradient complexity (defined as the mixing time multiplied by the number of gradient evaluations per iteration), albeit under strong structural assumptions and a warm start  \citep{chen2023does}. However, these methods require differentiable log posterior densities, limiting their applicability only to posteriors arising from smooth likelihood functions. HMC also suffers from divergent transitions in the presence of high posterior curvature \citep{piironen2017sparsity} and can mix poorly in multimodal posteriors where modes are separated by low-density regions \citep{dunson2020hastings}.

Slice sampling \citep{neal2003slice} offers a compelling alternative that does not require tuning, can adapt to local shapes without gradient information, and can potentially traverse well separated modes. Elliptical slice sampling \citep[ESS,][]{murray2010elliptical} brought this philosophy to multivariate settings with Gaussian priors, enabling transitions along elliptical trajectories defined by the prior, and generalized elliptical slice sampling \citep[GESS,][]{nishihara2014parallel} subsequently extended this framework to a broad class of continuous target distributions. As demonstrated in Section \ref{sec: Mixing_Times}, an optimally specified elliptical slice sampler scales remarkably well with dimension: in certain settings, its mixing time increases only logarithmically with dimension.
However, the sampling efficiency of ESS can rapidly degrade as the discrepancy increases between the target distribution and the distribution of the auxiliary variable that defines the ellipse, resulting in mixing times that scale poorly with dimension. This highlights a fundamental gap and opportunity shared by ESS and GESS: the auxiliary variable defining the ellipse is drawn without reference to the history of the Markov chain and is instead drawn based on the prior distribution. Since the prior distribution is often a poor approximation to the target distribution---particularly in high-dimensional settings---a typical elliptical slice sampler is likely to operate in the slow-mixing regime with poor dimension-scaling behavior. However, unbeknownst to the user, there may be an optimal choice of distribution that would have produced significantly faster mixing.
Can the optimal choice be discovered by gradually adapting the auxiliary variable distribution to match the shape of the target? In this paper, we show that such an adaptation strategy is practicable and succeeds in helping the sampler move from slow-mixing to fast-mixing regimes through online learning (Figure \ref{fig: Isotropic_gauss}). 
The resulting algorithm, which we call 
adaptive generalized elliptical slice sampling (AGESS), demonstrates that online adaptation is essential: it scales well with dimension and retains the gradient-free, mode-traversing strengths of ESS while outperforming ARW, and even HMC in certain scenarios.

The proof of the pudding, however, is in Section \ref{sec: sim_studies}, where we demonstrate compelling performance gains across three challenging posterior computation problems: (1) {\bf Generalized ReLU regression} in which the posterior is non-differentiable, ruling out gradient-based methods entirely, and becomes progressively less elliptically contoured as the degree of inequality constraint increases. AGESS degrades gracefully across this range but retains superiority over ARW, while ESS and GESS are clearly worse in higher dimensions. (2) {\bf Deep Gaussian process surrogate modeling} in which the posterior is high-dimensional and strongly multimodal with complex inter-parameter dependencies. AGESS is the only method among those considered---including HMC, block ESS, and GESS---to provide reliable inference regardless of initialization, while HMC requires an order of magnitude more computation time and still fails. (3) {\bf Sparse regression under horseshoe prior} in which the 202-dimensional posterior has heavy-tailed geometry that causes HMC to suffer divergent transitions in 30--60\% of iterations. AGESS produces well-mixing chains and outperforms HMC as a general-purpose sampler, despite neither method being able to match a bespoke conjugate sampler that exploits the specific model structure. Taken together, these results establish AGESS as a compelling general-purpose MCMC method for the broad and practically important class of posteriors that are non-differentiable, multimodal, or high-dimensional. In Section \ref{sec: Theory}, we establish ergodicity of the adaptive scheme under mild regularity conditions. We conclude by providing additional comments in Section \ref{sec: discussion}.

\section{Mixing Times of the Elliptical Slice Sampler}
\label{sec: Mixing_Times}

When analyzing the efficiency of MCMC algorithms, a key quantity is the \textit{mixing time}, which describes the rate at which the Markov chain converges to its stationary distribution in total variation distance. The mixing time quantifies how many MCMC iterations are needed for an $n$-step transition to be sufficiently close to the target distribution and therefore serves as an important measure for determining how many MCMC iterations are required in practice \citep{meyn1994computable}. Due to the complicated nature of the transition kernel of the elliptical slice sampler, we are unable to derive meaningful (tight) bounds on the mixing time for arbitrary target distributions. We therefore focus on the special case where the target distribution is a $P$-dimensional Gaussian distribution ($\mu = \mathcal{N}(\mathbf{0}, \boldsymbol{\Sigma})$), and derive bounds on the mixing time of the elliptical slice sampler as the dimension of the target distribution $P$ grows. In this setting, we prove that, under optimal specification of the elliptical slice sampler, the mixing time is upper bounded by $\mathcal{O}(\log(P))$. As a result, an optimally configured elliptical slice sampler can mix faster than an optimally tuned adaptive random walk \citep[$\mathcal{O}(P)$,][]{roberts2001optimal} and can be more scalable than optimal HMC, which has a gradient complexity of $\mathcal{O}(P^{1/4})$ \citep{chen2023does}. In contrast, if the distribution of the auxiliary variable used to define the ellipse---determined by the prior distribution in the elliptical slice sampler---does not coincide with the target distribution, the mixing time grows substantially, illustrating a significant decrease in sampling efficiency.

The elliptical slice sampler \citep{murray2010elliptical} is conventionally used in settings where the target distribution ($\mu$) can be decomposed into the product of a likelihood function ($\mathcal{L}$) and a Gaussian prior ($\pi_0$), that is,  $\mu(\mathbf{x}) \propto \pi_0(\mathbf{x}) \mathcal{L}(\mathbf{x})$. Consider the case in which the target distribution and the Gaussian prior coincide and are both Gaussian distributions centered at the origin (i.e., $\mu = \pi_0 = \mathcal{N}(\mathbf{0},\boldsymbol{\Sigma})$).  In this scenario, we have $\mathcal{L}(\mathbf{x}) = 1$, which means that each proposed move on the elliptical slice is accepted with probability 1; representing an optimal elliptical slice sampler for the given target distribution. In this optimal scenario, we can directly derive an upper bound on the Kullback–Leibler (KL) divergence between the target distribution and the $n$-step transition kernel of the elliptical slice sampler, allowing us to bound the mixing time.
\begin{proposition}
    \label{prop: log_p_convergence}
    Consider the $P$-dimensional target distribution $\mu = \mathcal{N}(\mathbf{0},\boldsymbol{\Sigma})$. Let $\mathbf{x}_0$ be the initial state of the Markov chain, and let $H^n(\mathbf{x}_0, \cdot)$ be the $n$-step transition kernel of the elliptical slice sampler with $\pi_0 = \mathcal{N}(\mathbf{0},\boldsymbol{\Sigma})$. The KL divergence between $\mu(\cdot)$ and $H^n(\mathbf{x}_0, \cdot)$ for $n \ge 3$ can be bounded as follows:
    \begin{align}
        \nonumber D_{\text{KL}}\infdiv{\mu(\cdot)}{H^n(\mathbf{x}_0, \cdot)}  & \le \left(\mathbf{x}_0^\intercal\boldsymbol{\Sigma}^{-1}\mathbf{x}_0 + P\right)\left(2^{-(n+1)} + \pi^{-n/2}\right).
    \end{align}
\end{proposition}
Using Proposition \ref{prop: log_p_convergence}, if one wants to ensure that $D_{\text{KL}}\infdiv{\mu(\cdot)}{H^n(\mathbf{x}_0, \cdot)} < \epsilon$, it is sufficient to ensure that $n > \frac{2}{\log{(2)}}\log\left(\frac{\mathbf{x}_0^\intercal\boldsymbol{\Sigma}^{-1}\mathbf{x}_0 + P}{\epsilon/2}\right)$, illustrating that the dimension of the target distribution ($P$) has a log-scale dependence on the number of iterations needed ($n$). Applying Pinsker's inequality to bound the total variation distance between $\mu(\cdot)$ and $H^n(\mathbf{x}_0,\cdot)$, we obtain that the dimension of the target distribution ($P$) has a log-scale dependence on the mixing time. Although mixing time is a fundamental theoretical construct for characterizing the stopping time of a Markov chain, in practice, measures such as the multivariate effective sample size \citep{vats2019multivariate} are typically used to assess how many MCMC iterations are required. In addition to establishing that the mixing time is $\mathcal{O}(\log(P))$, we can further demonstrate that the resulting MCMC samples are uncorrelated, implying that the multivariate effective sample size is equal to the total number of MCMC iterations; see Section 2 in the Supplementary Materials for a more detailed discussion. 

Although the elliptical slice sampler is extremely efficient under optimal conditions, the sampling efficiency quickly degrades when the prior distribution differs from the target distribution. To illustrate this, consider the elliptical slice sampler where $\mu = \mathcal{N}(\mathbf{0}, \sigma^2\mathbf{I}_P)$ and $\pi_0 = \mathcal{N}(\mathbf{0}, (1 + \alpha)\sigma^2\mathbf{I}_P)$ for some $\alpha > 0$ ($\mathcal{L}(\mathbf{x}) = \exp\{- \norm{x}^2/ (2(1 + \alpha^{-1})\sigma^2)\}$). In this context, increasing $\alpha$ results in a higher proportion of rejected moves on the elliptical slice, which in turn increases the autocorrelation of the Markov chain and the computational cost per iteration. To establish a lower bound on the mixing time of this sub-optimal elliptical slice sampler, we show that, in sufficiently high dimensions, a Markov chain initialized at $\mathbf{x}_0 = \mathbf{0}$, with high probability, will need at least $N \propto \sqrt{ P}/(\log P)^2$ steps to reach a high posterior mass region.
\begin{proposition}
    \label{prop: sub_opt_bound}
    Consider an elliptical slice sampler where the target distribution is $\mu = \mathcal{N}(\mathbf{0}, \sigma^2\mathbf{I}_P$) and the prior is $\pi_0 = \mathcal{N}(\mathbf{0}, (1 + \alpha)\sigma^2\mathbf{I}_P)$ for some $\alpha > 0$. For any $\epsilon \in (0,1)$ and $\alpha > 0$, there exists $P_\alpha \in \mathbb{N}$ such that with $N = \left\lfloor \frac{\sqrt{P(1 + \alpha)}}{4(\log(P))^2} \right\rfloor$,
$$\norm{H^N(\mathbf{0}, \cdot) - \mu(\cdot)}_{TV} \ge 1 - \epsilon \quad \forall P > P_\alpha.$$
\end{proposition}
A direct consequence of Proposition \ref{prop: sub_opt_bound} is that the mixing time is not faster than $\mathcal{O}(\sqrt{P}/ (\log(P))^2)$ in this sub-optimal setting. In addition, in sufficiently high dimensions, an increase in $\alpha$ will require a larger minimum number of iterations needed to achieve the same control over the total variation distance between $H^n(\mathbf{x},\cdot)$ and $\mu(\cdot)$; indicating a decrease in sampling efficiency as $\alpha$ increases.

\begin{figure}
    \centering
    \includegraphics[width=\linewidth]{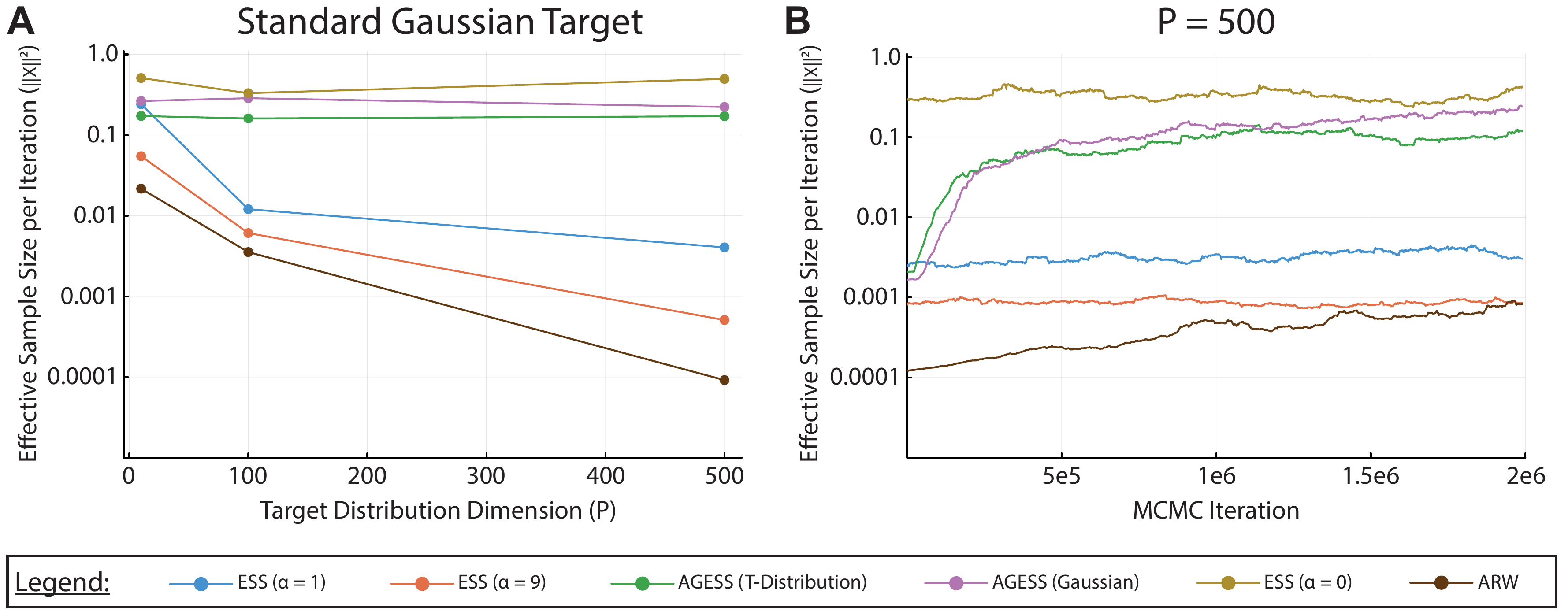}
    \caption{Sampling performance of the various MCMC algorithms when targeting a standard Gaussian distribution. \textbf{Subfigure A} shows the effective sample size per iteration of $\norm{\mathbf{x}}^2$ (computed using the final 40\% of iterations). \textbf{Subfigure B} illustrates how the adaptive schemes are able to adapt to the target distribution and achieve higher effective sample size per iteration as the Markov chain runs.}
    \label{fig: Isotropic_gauss}
\end{figure}

Although the sub-optimal results hold in high-dimensions, decreases in sampling efficiency can be seen in relatively low-dimensional target distributions, as illustrated in Figure \ref{fig: Isotropic_gauss}. Here, we consider the sampling efficiency under a standard Gaussian target distribution ($\mu = \mathcal{N}(\mathbf{0},\mathbf{I}_P)$) using (1) an optimal elliptical slice sampler ($\alpha = 0$), (2) sub-optimal elliptical slice samplers ($\alpha = 1,9$), (3) adaptive random walk \citep[ARW,][]{haario2001adaptive}, and (4) our proposed adaptive generalized elliptical slice sampler (AGESS), where we consider both a Gaussian and t-distribution for the distribution of the auxiliary variable. 
In line with the theoretical results, we observe a significant decrease in sampling efficiency when the prior and target distributions differ, and this decrease becomes more pronounced as the mismatch between the prior and target distributions increases. Alternatively, we can see that an optimal elliptical slice sampler exhibits an effective sample size that is seemingly independent of the dimension of the target distribution. Although AGESS is initialized with a covariance matrix that differs substantially from that of the target distribution ($\boldsymbol{\Sigma}_0 = 10\mathbf{I}_P$), after sufficient adaptation it achieves sampling performance comparable to that of the optimal elliptical slice sampler (Subfigure B). 
While we only provide theoretical results for Gaussian target distributions, \citet{natarovskii2021geometric} found that the elliptical slice sampler exhibited similar dimension-independent effective sample sizes for the \textit{volcano distribution}---an elliptically contoured but not monotonically decreasing distribution---suggesting that, with an optimally specified elliptical slice sampler, fast mixing may be attainable for a much wider class of target distributions than just Gaussian distributions; see Section 2 of the Supplementary Materials for a detailed discussion.

In many Bayesian computation scenarios, the prior distribution is not a good approximation of the target distribution; especially when using diffuse priors or when the likelihood is highly informative. In such situations, the use of a standard elliptical slice sampler \citep{murray2010elliptical} will lead to a slow-mixing Markov chain, particularly when considering moderate- to high-dimensional target distributions. By constructing an adaptive elliptical slice sampler, we can substantially improve sampling efficiency compared to a non-adaptive elliptical slice sampler and, in some cases, be more scalable than Hamiltonian Monte Carlo \citep{betancourt2017conceptual, neal2011mcmc, hoffman2014no} and adaptive random walk methods \citep{haario2001adaptive}.

\section{The Adaptive Generalized Elliptical Slice Sampler}
\label{sec: AGESS}

\begin{figure}
    \centering
    \includegraphics[width=0.8\linewidth]{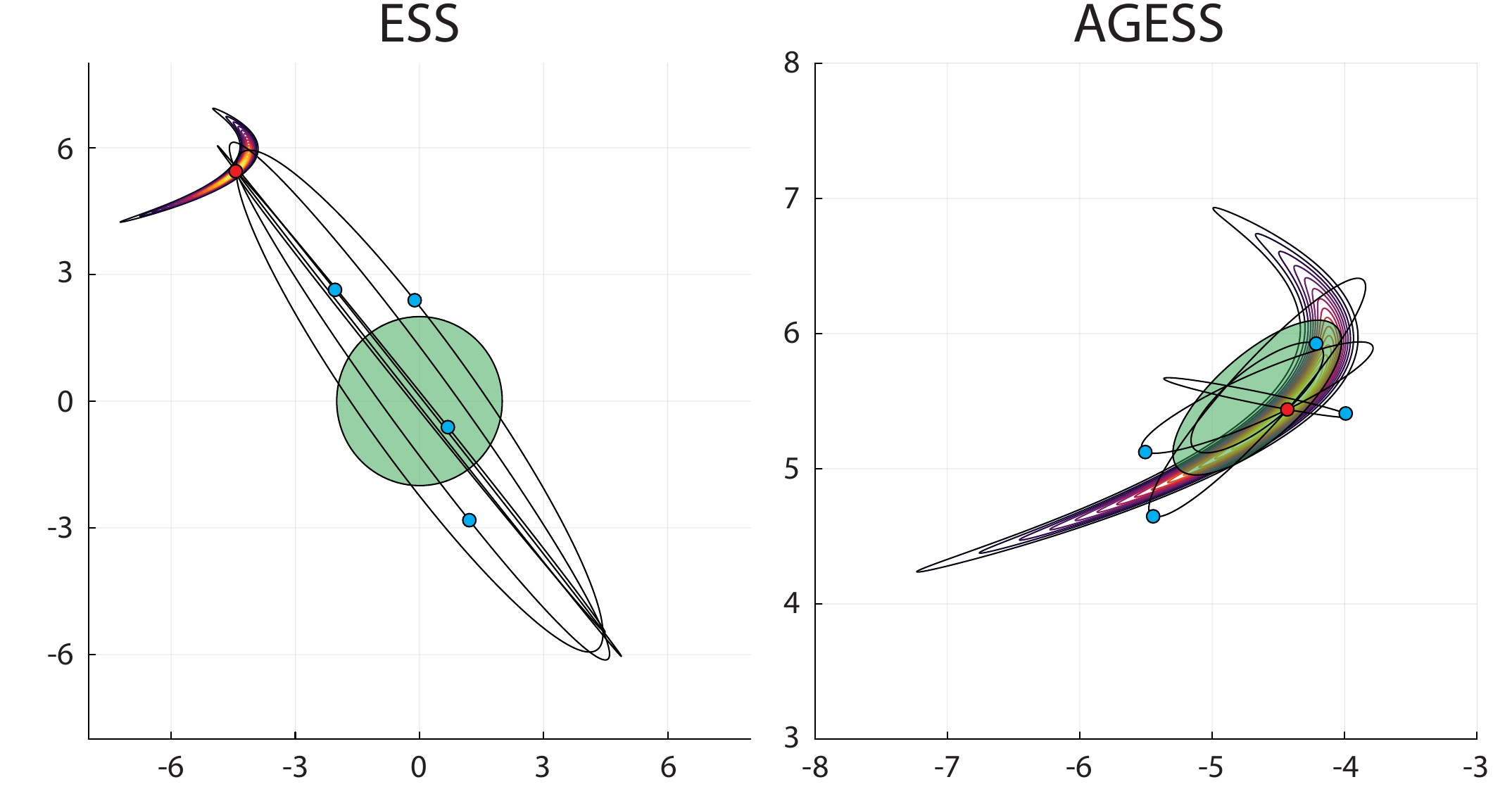}
    \caption{Conceptual illustration of how adaptation can produce a more efficient sampler when the prior distribution greatly differs from the target distribution. Here, the target is a banana distribution centered away from the origin. The red point shows the current Markov chain state, while the blue points represent four draws of the auxiliary variables that define the ellipses, whose covariance is shown by the green ellipses. Under ESS, only a small part of each ellipse lies in high-mass regions, causing slow exploration; adaptation yields more efficient steps.}
    \label{fig: conceptual}
\end{figure}

In MCMC-based Bayesian inference, the primary objective is to generate samples from a target, or posterior, distribution, which we denote by $\mu$. Consider a $P$-dimensional random variable $\mathbf{X}$ of interest with prior distribution $\pi_0$, and let $\mathcal{L}(\mathbf{x})$ represent the likelihood function. In this setting, the target distribution is given by $\mu(\mathbf{x}) \propto \pi_0(\mathbf{x})\mathcal{L}(\mathbf{x})$. Although elliptical slice sampling (ESS) requires $\pi_0$ to be a Gaussian distribution, we relax this assumption and let $\pi_0$ be a relatively arbitrary continuous prior distribution. Following \citet{nishihara2014parallel}, we can express the target distribution as the product of an elliptical distribution and a \textit{transformed likelihood} function, which brings us closer to a setting where ESS can be utilized. Specifically, we can express the target distribution as follows:
\begin{equation}\label{eq: target_distribution}
    \begin{aligned}
    \mu(\mathbf{x}) =& \frac{1}{Z} \mathcal{E}_P(\mathbf{x}; \boldsymbol{\mu}_{\boldsymbol{\gamma}}, \boldsymbol{\Sigma}_{\boldsymbol{\gamma}}, g) \frac{\pi_0(\mathbf{x})}{\mathcal{E}_P(\mathbf{x}; \boldsymbol{\mu}_{\boldsymbol{\gamma}}, \boldsymbol{\Sigma}_{\boldsymbol{\gamma}}, g)}\mathcal{L}(\mathbf{x})\\
     = & \frac{1}{Z} \mathcal{E}_P(\mathbf{x}; \boldsymbol{\mu}_{\boldsymbol{\gamma}}, \boldsymbol{\Sigma}_{\boldsymbol{\gamma}}, g) \mathcal{L}^*(\mathbf{x},\boldsymbol{\mu}_{\boldsymbol{\gamma}}, \boldsymbol{\Sigma}_{\boldsymbol{\gamma}}), 
\end{aligned}
\end{equation}
where $Z$ is the normalizing constant and $\mathcal{E}_P(\cdot; \boldsymbol{\mu}, \boldsymbol{\Sigma}, g)$ is a $P$-dimensional elliptical distribution \citep{frahm2004generalized, fang2018symmetric} with a median vector $\boldsymbol{\mu}$, a positive-definite scale matrix $\boldsymbol{\Sigma}$, and a continuous functional parameter $g(\cdot)$; see Section 1 of the Supplementary Materials for a review of elliptical distributions. 
Although $\mathcal{L}^*$ depends on $g$, we suppress the dependence on $g$ in the notation, since $g$ is considered fixed. As illustrated in Equation \ref{eq: target_distribution}, the Bayesian computation task can be expressed as performing posterior inference on a random variable with an elliptical prior distribution, given a \textit{transformed likelihood} $\mathcal{L}^*$. The general idea of the adaptive scheme is to adapt $\boldsymbol{\mu}_{\boldsymbol{\gamma}}$ and $\boldsymbol{\Sigma}_{\boldsymbol{\gamma}}$ to perform ESS in a more optimal \textit{transformed space}, leading to a more efficient sampling scheme; see Figure \ref{fig: conceptual} for a conceptual illustration.

\begin{algorithm}
\caption{Adaptive Generalized Elliptical Slice Sampling}
\label{alg: AGESS}
\hspace*{\algorithmicindent} \textbf{Input}: initial state $\mathbf{x}_1$, initial mean vector $\boldsymbol{\mu}_0$, initial scale matrix $\boldsymbol{\Sigma}_0$, likelihood function $\mathcal{L}(\cdot)$, $N$, family of elliptical distributions $\mathcal{E}$, $\beta > 0$, and schemes to update the adaptive parameters: \texttt{Update\_Mean} and \texttt{Update\_Scale}\\
\hspace*{\algorithmicindent} \textbf{Output}:  Markov chain $\{\mathbf{x}_t|1 \le t \le N\}$
\begin{algorithmic}
\State $\boldsymbol{\mu}_{\boldsymbol{\gamma}} \gets \boldsymbol{\mu}_0$
\State $\boldsymbol{\Sigma}_{\boldsymbol{\gamma}} \gets \boldsymbol{\Sigma}_0$
\State $i \gets 2$
\While{$i \le N$}
    \State $\mathbf{z} \sim \mathcal{E}_{P}(\boldsymbol{\mu}_{\boldsymbol{\gamma}}, \boldsymbol{\Sigma}_{\boldsymbol{\gamma},\mathbf{x}_{i-1}}, g_{\boldsymbol{\gamma},\mathbf{x}_{i-1}})$ \Comment{Draw $\mathbf{Z}$ conditionally on  $\mathbf{x}_{i-1}$}
    \State $u \sim \mathcal{U}_{[0,1]}$  
    \State $y \gets \log{\mathcal{L}^*(\mathbf{x}_{i-1}, \boldsymbol{\mu}_{\boldsymbol{\gamma}}, \boldsymbol{\Sigma}_{\boldsymbol{\gamma}})} + \log{u}$
    \State $\theta \sim \mathcal{U}_{[0, 2\pi)}$ \Comment{Propose initial angle}
    \State $[\theta_{min}, \theta_{max}] = [\theta - 2\pi, \theta]$
    \State $\mathbf{x}_{i} \gets (\mathbf{x}_{i-1} - \boldsymbol{\mu}_{\boldsymbol{\gamma}})\cos\theta + (\mathbf{z} - \boldsymbol{\mu}_{\boldsymbol{\gamma}})\sin\theta + \boldsymbol{\mu}_{\boldsymbol{\gamma}}$
    \While{$\log{\mathcal{L}^*(\mathbf{x}_{i}, \boldsymbol{\mu}_{\boldsymbol{\gamma}}, \boldsymbol{\Sigma}_{\boldsymbol{\gamma}})} \le y$} \Comment{Shrink possible angles}
        \If{$\theta < 0$}
            \State $\theta_{min} \gets \theta$
        \Else
            \State $\theta_{max} \gets \theta$
        \EndIf
        \State $\theta \sim \mathcal{U}_{(\theta_{min}, \theta_{max})}$ \Comment{Propose new angle}
        \State $\mathbf{x}_{i} \gets (\mathbf{x}_{i-1} - \boldsymbol{\mu}_{\boldsymbol{\gamma}})\cos\theta + (\mathbf{z} - \boldsymbol{\mu}_{\boldsymbol{\gamma}})\sin\theta + \boldsymbol{\mu}_{\boldsymbol{\gamma}}$ \Comment{Propose new state}
    \EndWhile

    \If{$i \in \{N_j\}_{j=1}^{\infty}$ ($N_j := \sum_{i=1}^j \lfloor i^\beta \rfloor$)} \Comment{AirMCMC \citep{chimisov2018air}}
        \State $\boldsymbol{\mu}_{\boldsymbol{\gamma}} \gets \texttt{Update\_Mean}(\boldsymbol{\mu}_0, \mathbf{x}_1, \dots, \mathbf{x}_i)$  \Comment{Update mean}
        \State $\boldsymbol{\Sigma}_{\boldsymbol{\gamma}} \gets \texttt{Update\_Scale}(\boldsymbol{\Sigma}_0, \mathbf{x}_1, \dots, \mathbf{x}_i)$ \Comment{Update scale}
    \EndIf
    \State $i \gets i + 1$
\EndWhile
\end{algorithmic}
\end{algorithm}

The adaptive algorithm (AGESS) is presented in Algorithm \ref{alg: AGESS}. While this approach appears similar to applying ESS in the \textit{transformed space} and updating the parameters of the elliptical prior using past states of the Markov chain, the crucial distinction is that the auxiliary random variable $\mathbf{Z}$ defining the ellipse is drawn conditionally on the current state of the Markov chain $\mathbf{x}_i$. Specifically, we assume $(\mathbf{X},\mathbf{Z}) \sim  \mathcal{E}_{2P}(\tilde{\boldsymbol{\mu}}_{\boldsymbol{\gamma}}, \tilde{\boldsymbol{\Sigma}}_{\boldsymbol{\gamma}}, \tilde{g})$, where $\tilde{\boldsymbol{\mu}}_\gamma = (\boldsymbol{\mu}_{\boldsymbol{\gamma}}, \boldsymbol{\mu}_{\boldsymbol{\gamma}})$ and $\tilde{\boldsymbol{\Sigma}}_{\boldsymbol{\gamma}} = I_2 \otimes \boldsymbol{\Sigma}_{\boldsymbol \gamma}$. In this setting, we draw $\mathbf{Z}$ conditional on the current state $\mathbf{X} = \mathbf{x}_i$, so that $\mathbf{Z}\mid \mathbf{X} = \mathbf{x}_i \sim \mathcal{E}_{P}(\boldsymbol{\mu}_{\boldsymbol{\gamma}}, \boldsymbol{\Sigma}_{\boldsymbol{\gamma},\mathbf{x}_i}, g_{\boldsymbol{\gamma},\mathbf{x}_i})$. Consequently, a key consideration when implementing AGESS is the choice of elliptical distribution, as it can affect the integrability of the \textit{transformed likelihood}. In what follows, we focus on two families of elliptical distributions that have convenient conditional distributions:  multivariate Gaussian distributions and multivariate Pearson type VII distributions, the latter constituting a generalization of multivariate t-distributions. In practical applications, we recommend using a multivariate Pearson type VII distribution due to the heavier tails. Beyond choosing the family of elliptical distributions, practitioners must also decide on the adaptation scheme, whether to employ MCMC blocking strategies, whether to transform variables, and whether to mix adaptive kernels with non-adaptive kernels. Section 6 of the Supplementary Materials provides an in-depth discussion of these practical choices and their impact on sampling properties.

\section{Illustrative Examples and Case Studies}
\label{sec: sim_studies}
In this section, we evaluate how AGESS performs relative to widely used alternative MCMC sampling methods---including adaptive random walk \citep[ARW,][]{haario2001adaptive}, elliptical slice sampling \citep[ESS,][]{murray2010elliptical}, generalized elliptical slice sampling \citep[GESS,][]{nishihara2014parallel}, and Hamiltonian Monte Carlo \citep[HMC,][]{betancourt2017conceptual, neal2011mcmc, hoffman2014no}---across a broad range of realistic modeling settings. In these case studies, we do not take into account any structure of the problem and instead treat the adaptive generalized elliptical slice sampler as essentially a \textit{black-box} sampler---constructing a general algorithm for all case studies and just providing the (unnormalized) posterior density for each case study. The general algorithm can be found in Section 5 of the Supplementary Materials. 

To compare the MCMC sampling algorithms, we computed performance metrics based on the multivariate effective sample size \citep{vats2019multivariate, vats2021revisiting} for all converged Markov chains: effective sample size per second, per iteration, and per function evaluation (likelihood or gradient evaluations). To determine whether the Markov chains converged, we calculated the Gelman-Rubin statistic \citep{gelman1992inference} using the $\epsilon = 0.1$ criterion of \citet{vats2021revisiting}. All sampling methods were implemented in compiled languages: with ARW, ESS, GESS, and AGESS implemented in Julia \citep{Julia-2017}, while HMC was conducted using the No-U-Turn Sampler \citep{hoffman2014no} in Stan \citep{carpenter2017stan}. The only exception was the conjugate horseshoe sampler used in Section \ref{sec: HS}, which was conducted using the \textsc{horseshoe} R package \citep{van2016horseshoe}. 
ESS per iteration and per function evaluation are language-independent and directly comparable; ESS per second is roughly comparable since most methods were implemented in compiled languages, but implementation details can still be reflected.

\subsection{Generalized ReLU Regression}
\label{sec: ReLU_Reg}
As illustrated in Section \ref{sec: Mixing_Times}, we prove that the elliptical slice sampler can achieve mixing times that scale logarithmically with dimension when targeting Gaussian target distributions, and, more generally, exhibits extremely efficient sampling when targeting elliptically contoured distributions; see Section 2 of the Supplementary Materials for a case study on the Volcano distribution. The fundamental question becomes: How does the sampling efficiency change as the target distribution becomes less elliptically contoured? To answer this question, we consider Bayesian computation for non-linear regression models of the form $f_Y(y_i \mid g^{-1}(\theta(\mathbf{x}_i)), \tau)$, where $\theta(\mathbf{x}_i) := \max(0, \mathbf{x}_i^{\top}\boldsymbol{\beta})$. Regression models of this type appear in applications such as density discontinuity modeling \citep{tokdar2025density} and in the modeling of nonlinear Hawkes processes \citep{bremaud1996stability, costa2020renewal, bonnet2023inference, sulem2024bayesian}. 
Here, we consider the sub-component of the density discontinuity model proposed by \citet{tokdar2025density}, which takes the following form:
$$Y_i \sim Bernoulli(\Phi(\mu_i)), \quad \Phi(z) = \frac{e^z}{1 + e^z}, \quad \mu_i = \max(0, \mathbf{x}_i^{\top}\boldsymbol{\beta});$$
for $i = 1, \dots, N$, where $\boldsymbol{\beta} \sim \mathcal{N}_{D}(\mathbf{0}, \mathbf{I})$ and $\mathbf{x}_i \in \mathbb{R}^D$. What makes this a particularly interesting case study is that the posterior is non-differentiable and becomes less elliptically contoured as the proportion of $\mathbf{x}_i^{\top}\boldsymbol{\beta}$ such that $\mathbf{x}_i^{\top}\boldsymbol{\beta} < 0$ ($\mu_i = 0$) increases.

\begin{figure}
    \centering
    \includegraphics[width=\linewidth]{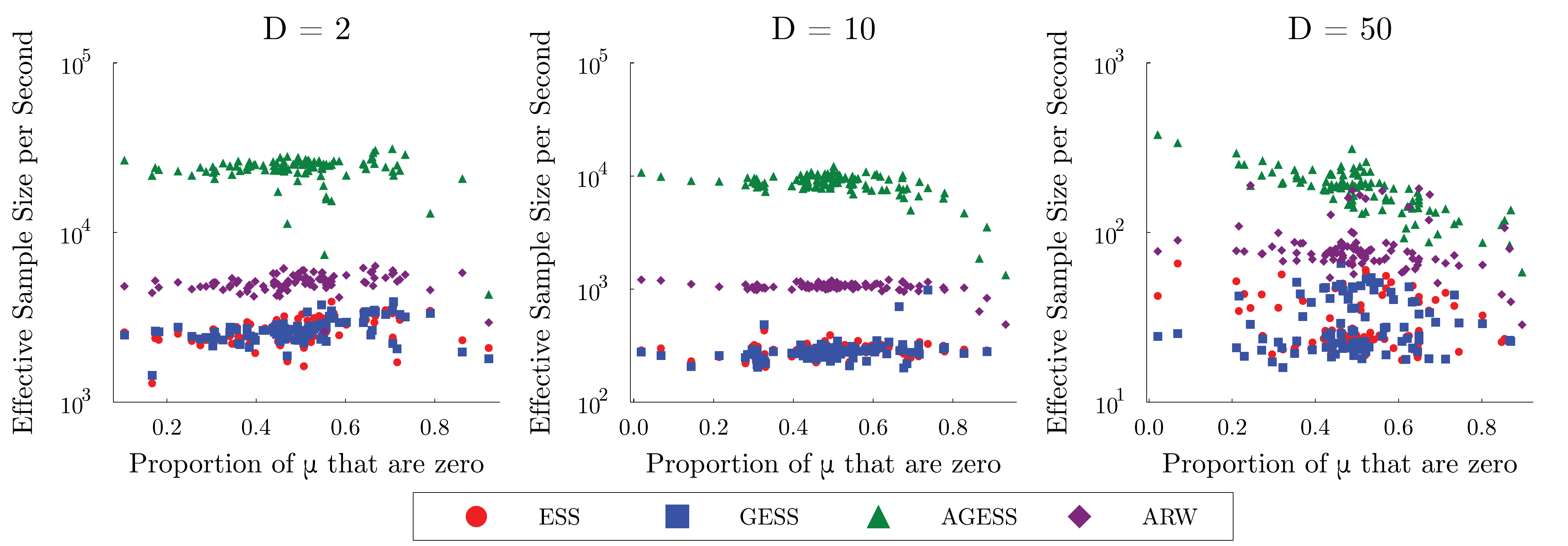}
    \caption{Performance metrics of various samplers when targeting the posterior distribution of $\boldsymbol{\beta}$ in the generalized ReLU regression case study.}
    \label{fig: Constrained_Regression}
\end{figure}

In this case study, we generate 100 datasets with $N =1000$ observations under various numbers of covariates ($D = 2, 10, 50$). The covariate effects were generated from a t-distribution such that $\boldsymbol{\beta} \sim \mathcal{T}_D(\mathbf{0}, \sqrt{2\log(D)}\mathbf{I}, \nu = 6)$ and the design matrix was generated from a normal distribution such that $\mathbf{x}_i \sim \mathcal{N}_D(\mu_x\mathbf{1}, \mathbf{I})$, where $\mu_x \sim \mathcal{N}(0, 0.25)$. Since the target distribution is not differentiable, we compare the performance of AGESS with other gradient-free MCMC algorithms, namely ARW, ESS, and GESS. The Markov chains generated by ARW were run for $30000 \times D$ iterations, while the Markov chains generated by the other sampling schemes were run for $10000 \times D$ iterations. The first $2500 \times D$ iterations of each Markov chain were discarded as burn-in. 

\begin{table}
\centering
\caption{Performance metrics (mean $\pm$ SD) for the generalized ReLU regression case study, by sampler and covariate dimension $D$ ($100$ datasets per dimension). Summary statistics were computed using only the converged chains; see note below.}
\label{tab: relu_summary}
\begin{tabular}{lcccc}
\toprule
Sampler & ESS/sec & ESS/iter & ESS/func. eval. & Comp. Time (sec) \\
\addlinespace
\multicolumn{5}{l}{\textit{$D = 2$}} \\
ESS   & 2679 $\pm$ 452  & 0.294 $\pm$ 0.049 & 0.049 $\pm$ 0.008 & 2.20 $\pm$ 0.23  \\
GESS  & 2669 $\pm$ 420  & 0.296 $\pm$ 0.040 & 0.048 $\pm$ 0.007 & 2.23 $\pm$ 0.19  \\
ARW   & 5125 $\pm$ 582  & 0.118 $\pm$ 0.012 & 0.093 $\pm$ 0.010 & 1.70 $\pm$ 0.09  \\
AGESS & 23790 $\pm$ 4075  & 0.753 $\pm$ 0.110 & 0.440 $\pm$ 0.077 & 0.65 $\pm$ 0.07  \\
\addlinespace
\multicolumn{5}{l}{\textit{$D = 10$}} \\
ESS\textsuperscript{\dag}   & 288.0 $\pm$ 33.9 & 0.046 $\pm$ 0.006 & 0.006 $\pm$ 0.001 & 15.94 $\pm$ 1.02  \\
GESS\textsuperscript{\dag}  & 289.7 $\pm$ 90.1 & 0.048 $\pm$ 0.014 & 0.006 $\pm$ 0.002 & 16.45 $\pm$ 1.12  \\
ARW   & 1052 $\pm$ 94.1  & 0.028 $\pm$ 0.003 & 0.028 $\pm$ 0.003 & 9.80 $\pm$ 0.41  \\
AGESS & 8694 $\pm$ 1707   & 0.508 $\pm$ 0.084 & 0.182 $\pm$ 0.037 & 6.26 $\pm$ 0.43  \\
\addlinespace
\multicolumn{5}{l}{\textit{$D = 50$}} \\
ESS\textsuperscript{\dag}   & 30.3 $\pm$ 11.8 & 0.010 $\pm$ 0.004 & 0.0011 $\pm$ 0.0004 & 157.8 $\pm$ 8.0  \\
GESS\textsuperscript{\dag}  & 30.3 $\pm$ 11.6 & 0.010 $\pm$ 0.004 & 0.0012 $\pm$ 0.0004 & 168.7 $\pm$ 8.7  \\
ARW   & 83.7 $\pm$ 30.4 & 0.006 $\pm$ 0.002 & 0.0056 $\pm$ 0.0020 & 113.7 $\pm$ 8.0  \\
AGESS & 184.3 $\pm$ 53.8 & 0.061 $\pm$ 0.015 & 0.0067 $\pm$ 0.0019 & 186.2 $\pm$ 10.8  \\
\bottomrule
\multicolumn{5}{@{}p{\dimexpr\textwidth-2\tabcolsep}@{}}{\footnotesize \textit{Note} \textsuperscript{\dag}: Number of the 100 Markov chains omitted due to non-convergence: [$P=10$] ESS~$=2$, GESS~$=2$; [$P=50$] ESS~$=7$, GESS~$=5$.}
\end{tabular}
\end{table}

Consistent with previous results, AGESS is extremely efficient when the target distribution is elliptically contoured; achieving gains roughly an order of magnitude larger than any alternative sampling scheme (see Figure \ref{fig: Constrained_Regression}). As the inequality constraint becomes more active and the target distribution becomes less elliptically contoured, we can see that the sampling efficiency degrades. While the sampling efficiency degrades for all samplers (with some ESS and GESS chains failing to converge in higher dimensions), AGESS retains superiority over the alternative sampling schemes. The gains offered by AGESS can be seen across the three performance metrics considered, as illustrated by Table \ref{tab: relu_summary}. Notably, while ARW is the second most efficient method by effective sample size per second and per function evaluation, its effective sample size per iteration is the lowest of all samplers considered: one step of ARW is cheap to compute, but many such steps are required, which can be prohibitive in high-dimensional settings. Overall, this study shows that AGESS remains a strong option even when the target distribution is not elliptically contoured---particularly when it is also non-differentiable, precluding gradient-based methods.


\subsection{Deep Gaussian Process Surrogates}
\label{sec: Deep_GP}
In numerous scientific fields, complex computer simulations have become increasingly prevalent, particularly where obtaining real experimental data is prohibitively costly or challenging \citep{gramacy2020surrogates}. However, these high-fidelity models are typically computationally expensive to simulate from and often depend on a high-dimensional set of input parameters, leading to the use of a \textit{surrogate} model. Recently, the use of deep Gaussian processes \citep{damianou2013deep} have become popular surrogate models for computer simulations due to their flexible nature \citep{montagna2016computer, radaideh2020surrogate, sauer2023non, sauer2023active}. In addition to being practically useful, deep Gaussian process models provide an interesting case study because they often exhibit a challenging multimodal posterior with strong inter-parameter dependencies and regions of high curvature, which causes many sampling methods to perform poorly.

Deep Gaussian process models provide a flexible non-stationary model by using a hierarchical representation of augmented stationary Gaussian processes. Here, we will consider the simple two-layer deep Gaussian process. Let $\mathbf{Y} \in \mathbb{R}^N$ be the outputs of interest, and let $\mathbf{X} \in \mathbb{R}^{N \times D}$ be the inputs to the computer experiment. Fundamentally, the goal is to estimate the function $f: \mathbb{R}^D \rightarrow \mathbb{R}$ where $Y_i = f(\mathbf{x}_i)$. Letting $W(\mathbf{x})$ $(\mathbf{x} \in \mathbb{R}^D)$ be the augmented Gaussian process, with  $\mathbf{W} := (W(\mathbf{x}_1), \dots, W(\mathbf{x}_N)) \in \mathbb{R}^N$, we construct the deep Gaussian process through the following hierarchical representation \citep{montagna2016computer}:
\begin{gather}
    \nonumber \mathbf{Y} \sim \mathcal{N}\left(\mathbf{0}, \tau\left(K_{\theta_y}(\mathbf{X}) + g_y\mathbf{I}_N \right)\right), \quad \mathbf{W} \sim \mathcal{N}\left(\mathbf{0}, K_{\theta_w}(\mathbf{X}) + g_w \mathbf{I}_N\right),\\ \nonumber \tau \sim \text{Inv-Gamma}(\nu/2, \nu/2);
\end{gather}
where $K_{\theta_y}(\mathbf{x}_i, \mathbf{x}_j):= \exp\left( - \sum_{d = 1}^D \left[\left\lVert x_{id} - x_{jd}\right\rVert^2_2 / \theta_{y_d} +  \left\lVert W_{i} - W_{j}\right\rVert ^2 / \theta_{y_{D+1}}\right]\right)$, $K_{\theta_w}(\mathbf{x}_i, \mathbf{x}_j):= \exp\left( - \sum_{d = 1}^D \left\lVert x_{id} - x_{jd}\right\rVert ^2 / \theta_{w_d}\right)$, and $g_y,g_w \in \mathbb{R}_{+}$ are user-specified. When conducting posterior inference, $\tau$ is typically marginalized out of the posterior distribution, leaving us with the desired inference on $\mathbf{W}$, $\{\theta_{y_d}\}_{d=1}^{D+1}$, and $\{\theta_{w_d}\}_{d=1}^{D}$. In this case study, we will consider a one-dimensional input (i.e., $\mathbf{X} \in \mathbb{R}^N$) and use the following hyper-parameters $g_w = 10^{-8}$, $g_y  = 10^{-8}$, and $\nu = 6$. 
\begin{figure}
    \centering
    \includegraphics[width=\linewidth]{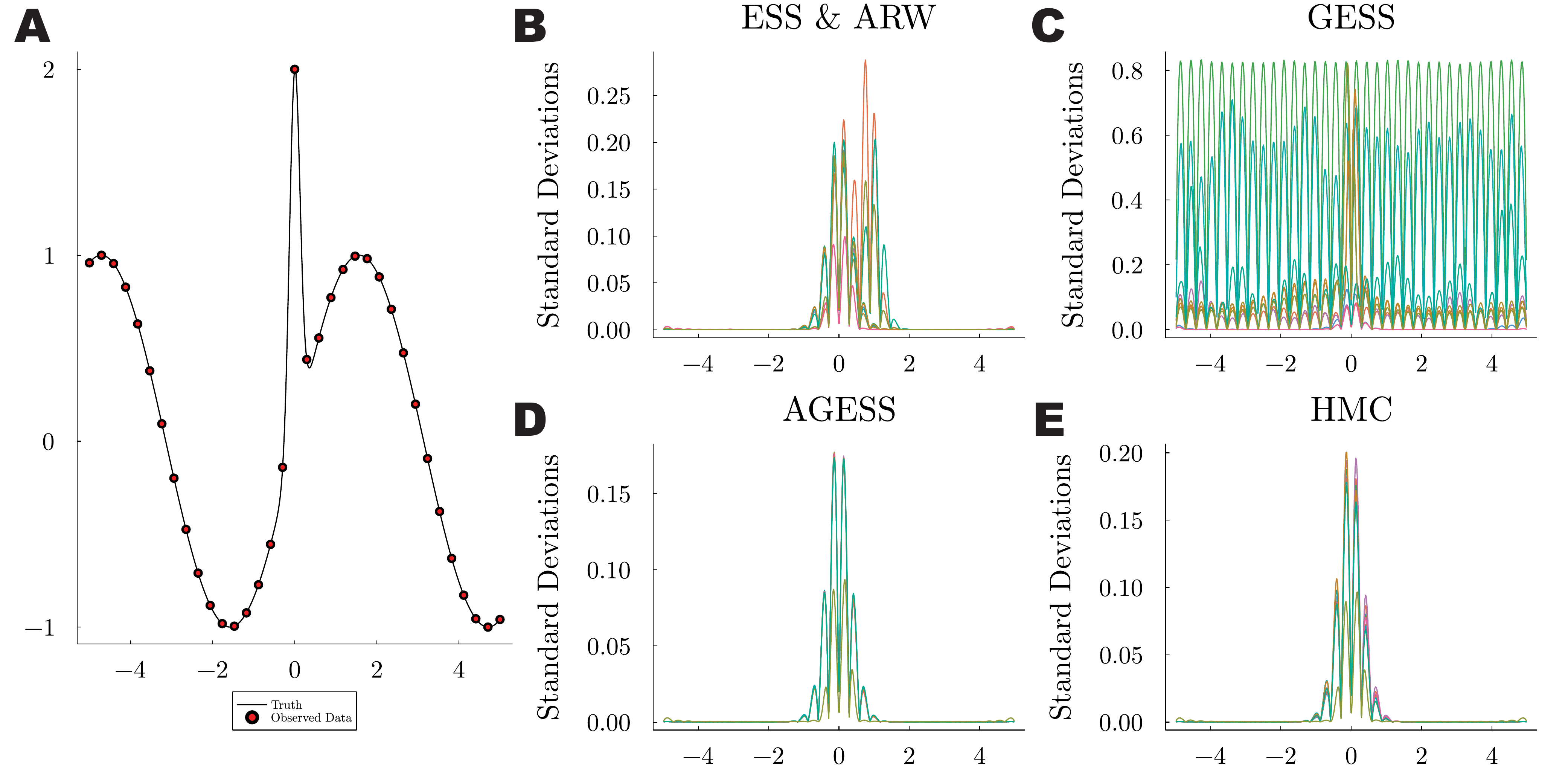}
    \caption{\textbf{Subfigure A} shows the observed data alongside the true underlying function. Visualizations of the standard deviation of the posterior predictive distribution obtained by the various sampling schemes are presented in \textbf{Subfigures B - E}.}   
    \label{fig: DeepGP}
\end{figure}

In this case study, we consider the one-dimensional function used in \citet{montagna2016computer}:
$f(x) = \sin(x) + 2\exp(-30x^2)$. We examine the setting in which the function is observed at $N = 35$ points on a uniform grid over $\Omega = [-5,5]$, as illustrated in Figure \ref{fig: DeepGP}. In addition to comparing AGESS with other general sampling schemes---specifically GESS and HMC---we consider a hybrid scheme from the deep Gaussian process surrogate literature that combines elliptical slice sampling and adaptive random walk \citep[ESS \& ARW,][]{sauer2023active}. Specifically, the hybrid scheme updates the lengthscale parameters $(\{ \theta_{y_1}, \theta_{y_2}, \theta_{w_1}\})$ using an adaptive random walk scheme \citep[ARW,][]{haario2001adaptive} and updates $\mathbf{W}$ using elliptical slice sampling \citep[ESS,][]{murray2010elliptical}. The strong dependence between the lengthscale parameters and the latent vector $\mathbf{W}$, however, gives rise to a multimodal posterior, in which sampling schemes can become trapped in local modes, resulting in poor mixing. Indeed, this dependence is precisely what makes block updates prone to becoming stuck: conditional on one block, the other's distribution can be sharply concentrated around a single mode, leaving it unable to explore the posterior distribution. To reveal whether this occurs in practice, and whether other sampling schemes exhibit similar problems, we randomly initialize the starting states of the lengthscale parameters, while fixing the initial state of $\mathbf{W}$ equal to $\mathbf{X}$ for all sampling schemes. To assess the performance of each sampling method, we run 10 Markov chains per sampler (125,000 iterations for HMC and 250,000 iterations for all other sampling algorithms), discarding the first half of each chain as burn-in.

A key quantity of interest in this setting is the \textit{posterior predictive} distribution of the process, as it quantifies the distribution of the process at unobserved inputs. Figure \ref{fig: DeepGP} illustrates the standard deviation of the predictive distributions across the input space for each sampling scheme considered. Indeed, we can see that the block updates (ESS \& ARW) cause the sampler to get stuck, leading to significantly different estimates of the posterior predictive distribution depending on the initial state of the lengthscale parameters. By contrast, samplers that jointly update all parameters (namely HMC and AGESS) exhibit less dependence on these initial states. Some dependence remains, however, as both AGESS and HMC have one chain that becomes stuck in a local mode under the same initial state---a testament to how challenging this target distribution is. Given the evidence against blocking in this setting, one might expect GESS to be a good alternative to the hybrid scheme (ESS \& ARW). However, in this case, the posterior is quite concentrated, resulting in a severe discrepancy between the prior and the target distribution. This discrepancy produces extremely slow-mixing Markov chains, leading to heavy dependence on the initial states of the chains, as illustrated in Figure \ref{fig: DeepGP}.

\begin{table}
\centering
\caption{Performance metrics (mean $\pm$ SD) for the deep Gaussian process surrogate case study (10 Markov chains). }
\label{tab: dgp_summary}
\begin{tabular}{lcccc}
\toprule
Method & ESS/sec & ESS/iter & ESS/func. eval. & Comp. Time (sec) \\
\addlinespace
AGESS\textsuperscript{\dag} & 300.8 $\pm$ 29.1    & 0.167 $\pm$ 0.013    & 0.0056 $\pm$ 0.0004 & 148.7  $\pm$ 4.8 \\
HMC\textsuperscript{\dag} & 1.91 $\pm$ 0.27  & 0.146 $\pm$ 0.022    & 0.00014 $\pm$ 0.00002 & 9478 $\pm$ 133 \\
ESS \& ARW\textsuperscript{\dag} & 168.2 $\pm$ 33.3    & 0.032 $\pm$ 0.006 & 0.0020 $\pm$ 0.0004 & 48.4 $\pm$ 1.7 \\
\bottomrule
\multicolumn{5}{@{}p{\dimexpr\textwidth-2\tabcolsep}@{}}{\footnotesize \textit{Note} \textsuperscript{\dag}: Number of the 10 Markov chains omitted due to non-convergence: AGESS~$=1$, HMC~$=1$, ESS \& ARW~$=4$.}
\vspace{-1em}
\end{tabular}
\end{table}

While convergence and reliability of the MCMC algorithm are of primary importance, a secondary question still persists: If the Markov chain converged to the correct target distribution, which sampling algorithm is most efficient? To answer this question, we calculated performance metrics related to the multivariate effective sample size \citep{vats2019multivariate, vats2021revisiting}. Since the parameters are not fully identifiable, we will calculate the effective sample size with respect to the posterior predictive mean, evaluated at $20$ roughly uniformly distributed unobserved points within $[-5,5]$. As illustrated in Table \ref{tab: dgp_summary}, AGESS clearly outperforms the other sampling schemes in each of the three main metrics considered. Although HMC is slightly less efficient on a per iteration basis, each iteration of HMC is significantly more expensive---as illustrated by the effective sample size per second and per function evaluation---resulting in prohibitively long computation times in many realistic scenarios. In contrast, each iteration of the hybrid scheme (ESS \& ARW) is cheaper than AGESS; however, each iteration is less efficient, leading to an overall less efficient algorithm. 

 More generally, this simulation study illustrates two separate limitations of existing MCMC methods: gradient-based methods, such as HMC, tend to struggle when the target distribution is highly concentrated and exhibits high curvature, as they often require a large number of leapfrog steps (with small step sizes) to traverse the target distribution, making each iteration prohibitively expensive. Hybrid schemes are a natural alternative when gradient-based procedures are too costly; however, when there is dependence between blocks, block updates are prone to getting stuck in local modes. AGESS is not limited by either of these problems. Its gradient-free nature and adaptation of the ellipse-defining auxiliary variables result in relatively cheap and efficient iterations, while its ability to jointly update all parameters allows it to target distributions with strong inter-parameter dependence---precisely where hybrid sampling schemes struggle. Taken together, these results suggest that AGESS is a compelling option when the target distribution exhibits high curvature and has complex inter-parameter dependence.

\subsection{High-Dimensional Sparse Regression}
\label{sec: HS}
One of the key benefits of elliptical slice samplers is that they are locally adaptive---they can adapt to localized features in the target geometry in a way that many adaptive methods with global tuning parameters cannot. However, the mismatch between the prior and target distributions often precludes the use of elliptical slice samplers in high-dimensional settings. Can adaptation enable elliptical slice samplers to efficiently sample from high-dimensional target distributions with localized features and locally varying curvature? In this section, we address this question by assessing the performance of AGESS as a general sampling scheme in the context of high-dimensional sparse regression using the horseshoe prior \citep{carvalho2009handling, carvalho2010horseshoe}---a popular global-local shrinkage prior \citep{bhadra2019lasso}. The horseshoe prior is a useful case study to address this question, as it often induces challenging posterior distributions with extreme funnel shapes \citep{piironen2017sparsity}. While previous work has proposed replacing the half-Cauchy priors with slightly less heavy-tailed half-$t$ priors to yield a posterior with better geometry \citep{piironen2017sparsity, biswas2022coupling}, it is precisely this more challenging setting that we wish to study here.

Here, we consider the high-dimensional linear regression setting using a horseshoe prior. Specifically, we consider the following model:

\begin{equation}\label{eq: regression_lik}
\begin{gathered}
        Y_i \sim \mathcal{N}(\mathbf{x}_i^{\top}\boldsymbol{\beta}, \sigma^2), \quad
        \beta_j \sim \mathcal{N}(0, \sigma^2\tau^2\lambda_j^2),\\
        p(\sigma^2) \propto \frac{1}{\sigma^2}, \quad \tau \sim C^+(0,1), \quad \lambda_j \sim C^+(0,1);
    \end{gathered}
\end{equation}
for $i = 1, \dots, N$ and $j = 1, \dots, D$, where $C^+$ denotes the half-Cauchy distribution. To compare the various MCMC methods in this case study, we generated 25 datasets; each containing 50 observations ($N = 50$) and 100 covariates ($D = 100$). The design matrix was generated such that $\mathbf{x}_i \sim \mathcal{N}_{100}(\mathbf{0}, \boldsymbol{\Sigma})$, where $\boldsymbol{\Sigma}$ is defined element-wise as $\boldsymbol{\Sigma}_{jk} = 0.8^{|j - k|}$. The covariate effects $\boldsymbol{\beta}^*$ were generated such that $\beta_j^* =0$ with probability $0.95$ and $\beta_j^* = (-1)^jZ_j$  with probability $0.05$ (where $Z_j \sim \mathcal{N}(1, 9)$), with the additional constraint that each dataset contained at least one non-zero covariate effect. The response variable was simulated according to Equation \ref{eq: regression_lik} with $\sigma^2 = 1$, giving a target distribution of dimension $P = 202$.

We compare the performance of AGESS to two alternative general-purpose samplers---GESS and HMC---as well as a bespoke hybrid approach that combines Gibbs sampling and slice sampling \citep[HS,][]{bhattacharya2016fast}, as implemented in the \textsc{horseshoe} R package \citep{van2016horseshoe}. The hybrid approach utilizes the structure of the problem to sample from structured multivariate Gaussian distributions \citep{bhattacharya2016fast}, achieving a computational complexity of $\mathcal{O}(N^2D)$ and thus is particularly well-suited to scenarios where $N \ll D$. In contrast, GESS, AGESS, and HMC are general-purpose samplers that do not leverage any such structural information.
In this case study, GESS and AGESS were run for 1,500,000 iterations, with the initial 250,000 iterations discarded as burn-in. The HS sampling scheme was run for 200,000 iterations, discarding the first 100,000 iterations as burn-in. Finally, HMC was run for 210,000 iterations, with the first 10,000 iterations removed as burn-in.

\begin{figure}
    \centering
    \includegraphics[width =\linewidth]{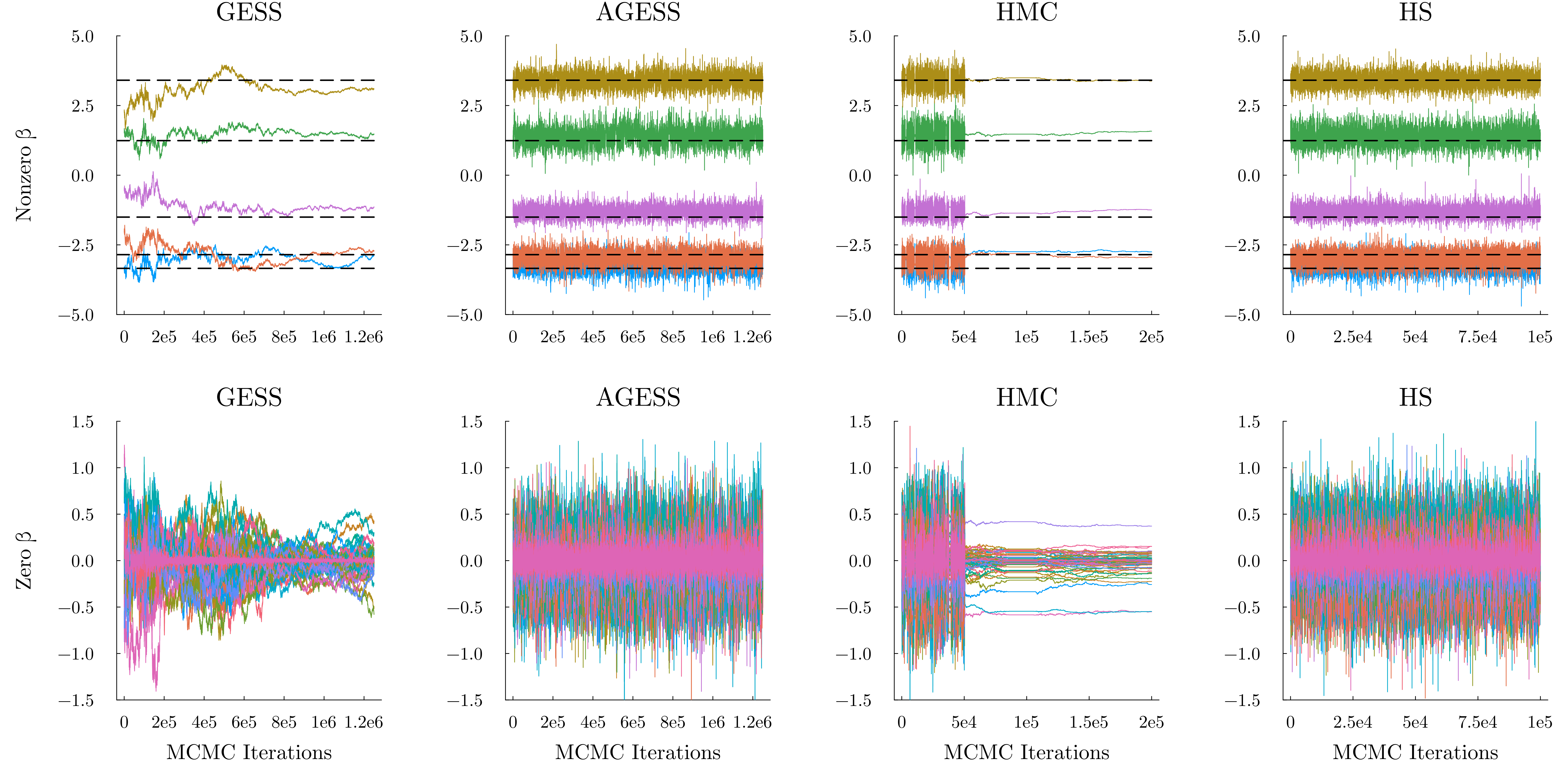}
    \caption{Trace plots of the non-zero coefficients (top row) along with the trace plots of the coefficients that are zero (bottom row) from 1 out of the 25 datasets. The true parameter values can be visualized by the dotted black line.}
    \label{fig: HS_AR1}
\end{figure}

As alluded to earlier, the prior-target mismatch renders GESS ineffective in this high-dimensional setting, as is evident from Figure \ref{fig: HS_AR1}. When viewing the trace plots generated by HMC, we can see that there are periods of high autocorrelation, during which the chain effectively becomes stuck. These are caused by \textit{divergent transitions}, which occur when the approximated Hamiltonian trajectory deviates from the true trajectory. Importantly, divergent transitions are an indication that the Markov chain may no longer be geometrically ergodic \citep{betancourt2017conceptual,livingstone2019geometric}. In practice, this means that the resulting MCMC estimates may be biased and that the Markov chain may fail to fully explore the posterior distribution within finitely many MCMC iterations---the latter potentially leading to overestimates of multivariate effective sample size \citep{vats2021revisiting}. It is precisely for these reasons that it is often suggested to exercise caution when encountering divergent transitions \citep{betancourt2017conceptual}. In this case study, we often found that between 30\% and 60\% of the total MCMC iterations were divergent. In contrast, AGESS is not subject to this problem: it produces a Markov chain with substantially less autocorrelation than GESS, while achieving performance metrics comparable to HMC (Table \ref{tab: HS_summary})---if one were to trust the results obtained by HMC. Unsurprisingly, the bespoke hybrid sampler (HS) performed best in this setting, leveraging the model structure to achieve extremely efficient and computationally cheap sampling.

In general, most settings will not have bespoke samplers that leverage the model structure, leaving users to typically rely on general-purpose samplers. In high-dimensional settings, practitioners commonly use HMC for its scalability and ease of implementation through software such as STAN \citep{carpenter2017stan}. However, as we have illustrated in this case study, HMC can suffer from divergent transitions, rendering its output unreliable. This leads to an important question: When are divergent transitions likely to occur? In cases where the target exhibits locally varying curvature, HMC tends to perform poorly: in areas of high curvature, HMC will likely exhibit divergent transitions, while in areas of low curvature, the step size will be too small leading to inefficient sampling. One solution would be to use Riemann manifold HMC \citep[RMHMC,][]{girolami2011riemann}, which allows the leapfrog integrator to depend on the current location. However, the cost of each leapfrog step becomes $\mathcal{O}(P^3)$, making RMHMC computationally intractable in many high-dimensional settings. Divergent transitions, or more generally loss of geometric ergodicity, can also occur in much simpler settings; such as when the tails whose density decays sufficiently faster or slower than those of a Gaussian distribution \citep{livingstone2019geometric}. In such cases of localized target geometry or non-Gaussian tails, AGESS offers a promising alternative that scales well with dimension.

\begin{table}[t]
\centering
\caption{Performance metrics (mean $\pm$ SD) for the high-dimensional sparse regression case study (25 datasets).}
\label{tab: HS_summary}
\begin{tabular}{lcccc}
\toprule
Method & ESS/sec & ESS/iter & ESS/func.\ eval. & Comp. Time (sec) \\
\addlinespace
AGESS & 18.94 $\pm$ 4.50 & 0.0065 $\pm$ 0.0016 & 0.00017 $\pm$ 0.00004 & 538.6 $\pm$ 23.1 \\
HMC\textsuperscript{\dag *}  & 22.66 $\pm$ 10.09 & 0.1150 $\pm$ 0.0490  & 0.00024 $\pm$ 0.00011 & 1090.0 $\pm$ 498.3 \\
HS    & 3123 $\pm$ 419    & 0.4884 $\pm$ 0.0645  & N/A  & 31.3 $\pm$ 0.9 \\
\bottomrule
\multicolumn{5}{@{}p{\dimexpr\textwidth-2\tabcolsep}@{}}{\footnotesize \textit{Notes}: \textsuperscript{\dag} The effective sample size could not be calculated on 1 out of the 25 simulations for HMC. \textsuperscript{*} HMC exhibited a large number of divergent transitions, which can inflate estimates of ESS; results should be interpreted with caution.}
\end{tabular}
\end{table}


\subsection{Guidelines for Using AGESS}
The illustrative examples considered in this section were carefully chosen to shed light on the performance of AGESS across a wide variety of challenging target distributions. Further examples demonstrating the performance of AGESS on non-convex and multimodal two-dimensional target distributions are provided in Section 7 of the Supplementary Materials. The key takeaway from these studies is that AGESS is particularly beneficial in scenarios where the posterior is non-differentiable, multimodal with strong parameter dependencies, or exhibits locally varying curvature---especially in the absence of a bespoke sampler that leverages the structure of the model. 

\section{Ergodicity of AGESS}
\label{sec: Theory}
In this section, we establish the regularity conditions under which the AGESS algorithm (Algorithm \ref{alg: AGESS}) is ergodic and well-specified (i.e., the number of iterations of the while-loop is finite almost surely). Let $\mu$ be the target distribution, and let $\mathcal{X}$ denote the state space of the Markov chain. We assume $\mathcal{X}$ is open and $\mu \ll \lambda$ where $\lambda$ denotes the Lebesgue measure on $(\mathcal{X},\mathcal{B}(\mathcal{X}))$. Ergodicity requires that each transition kernel in the family of transition kernels considered in the adaptive scheme has $\mu(\cdot)$ as its stationary distribution. Additionally, one must often show that the family of transition kernels is simultaneously strongly aperiodically geometrically ergodic. All proofs are presented in the Supplementary Materials.

Fundamental to the theoretical properties of ESS and AGESS is the inner while-loop of Algorithm \ref{alg: AGESS}, which will be referred to as the \textit{Shrinkage Algorithm} \citep{hasenpflug2025reversibility}; see the Supplementary Materials for a detailed discussion. The Shrinkage Algorithm, denoted $\texttt{shrink}(\theta_{\text{in}},S)$, provides a transition scheme within some set $S \in \mathcal{B}([0, 2\pi))$; starting with some current angle $\theta_{\text{in}} \in S$ and returning a new angle $\theta_{\text{out}} \in S$. 
To analyze the proposed adaptive algorithm, we define the following quantities, which will be referenced throughout the manuscript. Define $\mathcal{X}_{\boldsymbol{\gamma}}(y) := \{\mathbf{x} \in \mathcal{X}  \mid \mathcal{L}^*(\mathbf{x},\boldsymbol{\mu}_{\boldsymbol{\gamma}}, \boldsymbol{\Sigma}_{\boldsymbol{\gamma}}) > y\},$
which is the superlevel set of $\mathcal{L}^*$ with respect to $y$, corresponding to the set of possible moves in one iteration of the elliptical slice sampler for a given $y$. For any $\mathbf{x},\mathbf{z} \in \mathbb{R}^P$ and $\boldsymbol{\gamma} \in \mathcal{Y}$, define the function $p_{\mathbf{x}, \mathbf{z}, \boldsymbol{\gamma}}: [0, 2\pi) \rightarrow \mathbb{R}^P$ as
$p_{\mathbf{x}, \mathbf{z}, \boldsymbol{\gamma}} (\theta) := \left[(\mathbf{x} - \boldsymbol{\mu}_{\boldsymbol{\gamma}})\cos\theta + (\mathbf{z} -\boldsymbol{\mu}_{\boldsymbol{\gamma}})  \sin\theta\right] + \boldsymbol{\mu}_{\boldsymbol{\gamma}},$
and the corresponding pre-image 
$p^{-1}_{\mathbf{x}, \mathbf{z}, \boldsymbol{\gamma}}(A) := \{\theta \in [0,2\pi)\mid p_{\mathbf{x}, \mathbf{z}, \boldsymbol{\gamma}}(\theta) \in A\},$
for some $A \in \mathcal{B}(\mathbb{R}^P)$. Using the Shrinkage Algorithm, we can reformulate the proposed adaptive algorithm as in Algorithm \ref{alg: iteration}.

\begin{algorithm}
\caption{Reformulated iteration of AGESS under fixed $\boldsymbol{\gamma} := (\boldsymbol{\mu}_{\boldsymbol{\gamma}}, \boldsymbol{\Sigma}_{\boldsymbol{\gamma}})$}\label{alg: iteration}
\hspace*{\algorithmicindent} \textbf{Input}: function $\mathcal{L}^*$, family of elliptical distributions $\mathcal{E}$, adaptive parameters $\boldsymbol{\gamma} := (\boldsymbol{\mu}_{\boldsymbol{\gamma}}, \boldsymbol{\Sigma}_{\boldsymbol{\gamma}})$, and current state $\mathbf{x}_i$\\
\hspace*{\algorithmicindent} \textbf{Output}: next state $\mathbf{x}_{i+1}$
\begin{algorithmic}
\State $y \sim \mathcal{U}_{(0, \mathcal{L}^*(\mathbf{x}_{i}, \boldsymbol{\mu}_{\boldsymbol{\gamma}}, \boldsymbol{\Sigma}_{\boldsymbol{\gamma}}))}$
\State $\mathbf{z} \sim \mathcal{E}_{P}(\boldsymbol{\mu}_{\boldsymbol{\gamma}}, \boldsymbol{\Sigma}_{\boldsymbol{\gamma},\mathbf{x}_i}, g_{\boldsymbol{\gamma},\mathbf{x}_i})$ \Comment{Draw $\mathbf{Z}$ conditionally on  $\mathbf{x}_i$}
\State $\theta \gets \texttt{shrink}(0, p^{-1}_{\mathbf{x}_i, \mathbf{z}, \boldsymbol{\gamma}}(\mathcal{X}_{\boldsymbol{\gamma}}(y)))$ \Comment{Shrinkage Algorithm}
\State $\mathbf{x}_{i+1} \gets \left[(\mathbf{x}_{i} - \boldsymbol{\mu}_{\boldsymbol{\gamma}})\cos\theta + (\mathbf{z} - \boldsymbol{\mu}_{\boldsymbol{\gamma}})\sin\theta\right] + \boldsymbol{\mu}_{\boldsymbol{\gamma}}$
\end{algorithmic}
\end{algorithm}

Let $H_{\boldsymbol{\gamma}}(\mathbf{x}, \cdot)$ denote the transition kernel of the proposed adaptive generalized elliptical slice sampler (Algorithm \ref{alg: AGESS}) with a fixed $\boldsymbol{\gamma}:=(\boldsymbol{\mu}_{\boldsymbol{\gamma}}, \boldsymbol{\Sigma}_{\boldsymbol{\gamma}}) \in \mathcal{Y}$. To show that the proposed adaptive algorithm is ergodic, we proceed under the following set of assumptions:
\begin{assumption}[Compact $\mathcal{Y}$]
    \label{assumption1}
    Let $\boldsymbol{\mu}_0,\boldsymbol{\mu}_{\boldsymbol{\gamma}} \in \mathcal{Y}_{\boldsymbol{\mu}}:= \{\boldsymbol{\mu} \in \mathbb{R}^P \mid \lVert\boldsymbol{\mu}\rVert_2 \le R_{\mu}\}$, for some $R_{\mu} > 0$, and $\boldsymbol{\Sigma}_0, \boldsymbol{\Sigma}_{\boldsymbol{\gamma}} \in  \mathcal{Y}_{\boldsymbol{_\Sigma}}:= \left\{\mathbf{A} \in S^P_{++}| \lambda_{\min}(\mathbf{A})\ge k_{\min}, \lambda_{\max}(\mathbf{A}) \le k_{\max}\right\}$ such that $(\boldsymbol{\mu}_{\boldsymbol{\gamma}}, \boldsymbol{\Sigma}_{\boldsymbol{\gamma}}) \in \mathcal{Y}_{\boldsymbol{\mu}} \times \mathcal{Y}_{\boldsymbol{\Sigma}}$, denoted $\boldsymbol{\gamma} \in \mathcal{Y}$.
\end{assumption}

\begin{assumption}[Bounded $\mathcal{L}^*$]
    \label{assumption2}
    $\mathcal{L}^*(\cdot,\boldsymbol{\mu}_{\boldsymbol{\gamma}}, \boldsymbol{\Sigma}_{\boldsymbol{\gamma}})$ is bounded away from $0$ and $\infty$ on any bounded set of $\mathcal{X}$ and all $\boldsymbol{\gamma} \in \mathcal{Y}$.
\end{assumption}

\begin{assumption}[Lower Semi-Continuity of $\mathcal{L}^*$]
    \label{assumption3}
    $\mathcal{L}^*(\cdot,\boldsymbol{\mu}_{\boldsymbol{\gamma}}, \boldsymbol{\Sigma}_{\boldsymbol{\gamma}})$ is lower semi-continuous at every  $\mathbf{x} \in \mathcal{X}$, for all $\boldsymbol{\gamma} \in \mathcal{Y}$.
\end{assumption}

\begin{assumption}[Properties of the Elliptical Distribution]
    \label{assumption4}
   Let $\mathcal{E}$ be an elliptical distribution in the subclass of multivariate Gaussian distributions or symmetric multivariate Pearson type VII distributions \citep{fang2018symmetric}. Thus the continuous functional parameters take one of the following functional forms:
    \allowdisplaybreaks
    \small{
    \begin{align}
        \nonumber \textbf{Multivariate Gaussian:} &\quad g(t) = \exp\left(-0.5t \right) & \\
        \nonumber \textbf{Multivariate Pearson Type VII: } & \quad g(t) = (1 + t/m)^{-M} &  m > 0 , M > P/2,
    \end{align}}
    where $P$ denotes the dimension of the multivariate distribution.
    Let $(\mathbf{X},\mathbf{Z}) \sim \mathcal{E}_{2P}(\tilde{\boldsymbol{\mu}}_{\boldsymbol{\gamma}}, \tilde{\boldsymbol{\Sigma}}_{\boldsymbol{\gamma}}, \tilde{g})$, where $\tilde{\boldsymbol{\mu}}_\gamma = (\boldsymbol{\mu}_{\boldsymbol{\gamma}}, \boldsymbol{\mu}_{\boldsymbol{\gamma}})$, $\tilde{\boldsymbol{\Sigma}}_{\boldsymbol{\gamma}} = I_2 \otimes \boldsymbol{\Sigma}_{\boldsymbol \gamma}$, and $\tilde{g}(t) = (1 + t/m)^{-(M + P/2)}$, such that $\mathbf{X}$ and $\mathbf{Z}$ have marginal distributions $\mathcal{E}_{P}(\boldsymbol{\mu}_{\boldsymbol{\gamma}}, \boldsymbol{\Sigma}_{\boldsymbol{\gamma}}, g)$. Lastly, if $P = 1$ and $\mathcal{E}$ is in the subclass of symmetric multivariate Pearson type VII distributions, let $M > 1$.
\end{assumption}

\begin{assumption}[Elliptical Subcover]
    \label{assumption6}
    \nonumber If $\mathcal{X}$ is not bounded, then there exists $R \in (0, \infty)$, $\alpha \in (0, 1)$, $\xi \in (0 , \sqrt{\alpha})$, $\psi > 0$, and a positive definite matrix $\mathbf{A}$ such that when $\mathbf{x} \in B_R^C(\boldsymbol{\mu}_0, \mathbf{A}) :=\left\{\mathbf{x}\in \mathcal{X} \mid q_{\mathbf{x}}(\boldsymbol{\mu}_0, \mathbf{A}) \ge R \right\}$, where $q_{\mathbf{x}}(\boldsymbol{\mu}_0, \mathbf{A}) := (\mathbf{x} - \boldsymbol{\mu}_{0})^{\top}\mathbf{A}^{-1}(\mathbf{x} - \boldsymbol{\mu}_{0})$, the following holds: 
    \small{
    \begin{align}
        \nonumber \textbf{Elliptical Subcover:} & \quad \left\{\mathbf{y}\in \mathcal{X} \mid q_{\mathbf{y}}(\boldsymbol{\mu}_0, \mathbf{A})<\alpha q_{\mathbf{x}}(\boldsymbol{\mu}_0, \mathbf{A}) \right\} \subseteq\mathcal{X}_{\boldsymbol{\gamma}}(\mathcal{L}^*(\mathbf{x}, \boldsymbol{\mu}_{\boldsymbol{\gamma}}, \boldsymbol{\Sigma}_{\boldsymbol{\gamma}})),\\
        \nonumber \textbf{Diminishing Tails: } & \quad \frac{\max_{\mathbf{x}\in \{ \mathbf{x} \in \mathcal{X} \mid q_{\mathbf{x}}(\boldsymbol{\mu}_0, \mathbf{A}) = R_2\}}\mathcal{L}^*(\mathbf{x}, \boldsymbol{\mu}_{\boldsymbol{\gamma}}, \boldsymbol{\Sigma}_{\boldsymbol{\gamma}})}{\max_{\mathbf{y}\in \{ \mathbf{y} \in \mathcal{X}\mid q_{\mathbf{y}}(\boldsymbol{\mu}_0, \mathbf{A}) = R_1\}}\mathcal{L}^*(\mathbf{y}, \boldsymbol{\mu}_{\boldsymbol{\gamma}}, \boldsymbol{\Sigma}_{\boldsymbol{\gamma}})} \le \left(1 + \left[R_2 - R_1\right]  \right)^{-1},
    \end{align}
    }
    for all $R_2 \ge R_1 \ge R$ and $\boldsymbol{\gamma} \in \mathcal{Y}$, where $\alpha$ satisfies the following requirements:
   \begin{enumerate}
        \item when $\mathcal{E}$ is in the subclass of multivariate Gaussian distributions, $\alpha > 0.75$,
       \item when $\mathcal{E}$ is in the subclass of symmetric multivariate Pearson type VII distributions, $\alpha$ satisfies the following inequality:
       $\left( \frac{1}{\sqrt{\alpha}}  - \sqrt{\alpha} \right)(1 - F_{\alpha}(M, \xi, \psi)) \le \frac{F_1(\alpha, M , \xi, \psi)}{2},$
       where $F_{\alpha}(M,\xi, \psi) := \frac{1}{2\pi} \int_{\Theta_{\alpha}^{\xi}}I_{\frac{g(\alpha, \tilde{\theta}, \xi, \psi)}{1 + g(\alpha, \tilde{\theta}, \xi, \psi)}}(P/2, M - P/2) \text{d}\tilde{\theta}$, $F_1(\alpha, M, \xi, \psi):= \int_{\xi^2}^{\alpha} \frac{1}{2\sqrt{\tilde{\alpha}}} F_{\tilde{\alpha}}(M, \xi, \psi) \text{d}\tilde{\alpha}$, $\Theta_{\alpha}^{\xi} := \left\{\theta \bigm\vert \lvert\cos(\theta)\rvert < \sqrt{\alpha} - \xi\right\}$, and $g(\tilde{\alpha}, \tilde{\theta}, \xi, \psi) := \frac{ \left(\left( \sqrt{\tilde{\alpha}} - \lvert \cos(\tilde{\theta})\rvert\right) - \xi\right)^2}{( 1 + \psi)\left(\frac{k_{\max}\lambda_{\max}(\mathbf{A})}{k_{\min}\lambda_{\min}(\mathbf{A})}\right)\sin^2(\tilde{\theta})}$, 
       where $I_x(\alpha, \beta)$ is the regularized incomplete beta function.
       
   \end{enumerate}
\end{assumption}

Assumptions \labelcref{assumption1,assumption2,assumption3} are fairly typical. 
Assumption \ref{assumption4} restricts the set of elliptical distributions to the family of multivariate Gaussian distributions and the family of symmetric multivariate Pearson type VII distributions \citep{fang2018symmetric}, the latter being a generalization of  multivariate $t$-distributions. Crucially, both these families admit closed-form conditional and marginal distributions from which samples can be easily drawn, and under this assumption the first two moments of the auxiliary variable ($\mathbf{Z}$) are well-defined. These features play a key role in establishing that the family of transition kernels is simultaneously strongly aperiodically geometrically ergodic. 

Assumption \ref{assumption6} is somewhat technical but could be appreciated as follows. The first part implies that there exists a sufficiently large ellipse, centered at the prior mean, such that whenever the current state $\mathbf{x}$ lies outside this large ellipse and $\boldsymbol{\gamma} \in \mathcal{Y}$, the sampler will accept any proposed transition to a state in a smaller ellipse (controlled by $\alpha$) with probability 1.
The required size of the elliptical subcovers, defined by $1/ \alpha$, is relatively straightforward when the elliptical distribution is a multivariate Gaussian distribution, but is rather technical when the elliptical distribution is in the family of multivariate Pearson type VII distributions. In general, $\alpha$ will be closer to $1$ under the multivariate Pearson type VII distribution, and will mainly depend on the parameter $M$ and the choice of $\psi$ and $\xi$, which control the radius of the set $C$ in Proposition \ref{prop: drift}. The second part assumes that the maximum values along the elliptical contours, defined by the matrix $\mathbf{A}$, decay sufficiently fast as the size of the ellipse increases. Importantly, Assumption \ref{assumption6} will play a crucial role in proving that the \textit{geometric drift condition} holds when that state space is not bounded.

Under these assumptions, we can show that the Shrinkage Algorithm is well-specified (i.e., the number of iterations of the while-loop is almost surely finite) and that the set of possible transition angles ($p^{-1}_{\mathbf{x}, \mathbf{z}, \boldsymbol{\gamma}}(\mathcal{X}_{\gamma}(y))$) has positive Lebesgue measure for any $\mathbf{x} \in \mathcal{X}$ and $y \in (0, \mathcal{L}^*(\mathbf{x},\boldsymbol{\mu}_{\boldsymbol{\gamma}}, \boldsymbol{\Sigma}_{\boldsymbol{\gamma}}))$; see the Supplementary Materials. Letting $Q_S(\theta, F)$ denote the transition kernel of the Shrinkage Algorithm ($S \in \mathcal{B}([0, 2\pi))$, $\theta \in S$, and $F \in \mathcal{B}(S)$), we can specify the transition kernel of the proposed adaptive algorithm under a fixed $\boldsymbol{\gamma} \in \mathcal{Y}$. Specifically, for any $\mathbf{x} \in \mathcal{X}$ and $A \in \mathcal{B}(\mathcal{X})$, the transition kernel is defined by:
\setcounter{equation}{2}
\begin{equation}
    \label{eq: trans_Kernel}
    H_{\boldsymbol{\gamma}}(\mathbf{x}, A) = \frac{1}{\mathcal{L}^*} \int_0^{\mathcal{L}^*} \int_{\mathbb{R}^P} Q_{p^{-1}_{\mathbf{x}, \mathbf{z}, \boldsymbol{\gamma}}(\mathcal{X}_{\boldsymbol{\gamma}}(y))}(0, p^{-1}_{\mathbf{x}, \mathbf{z}, \boldsymbol{\gamma}}(\mathcal{X}_{\boldsymbol{\gamma}}(y) \cap A))\mathcal{E}^{\boldsymbol{\gamma}}_{\mathbf{Z}|\mathbf{x}}(\text{d}\mathbf{z})\text{d}y,
\end{equation}
where $\mathcal{E}^{\boldsymbol{\gamma}}_{\mathbf{Z}|\mathbf{x}}$ denotes the conditional probability distribution of $\mathbf{Z}$ given $\mathbf{X} = \mathbf{x}$ and $\mathcal{L}^*:= \mathcal{L}^*(\mathbf{x},\boldsymbol{\mu}_{\boldsymbol{\gamma}}, \boldsymbol{\Sigma}_{\boldsymbol{\gamma}})$. Using the specified transition kernel, we show that $H_{\boldsymbol{\gamma}}$ is reversible with respect to $\mu$ for all $\boldsymbol{\gamma} \in \mathcal{Y}$.

\begin{theorem}[Reversibility]
    \label{thm: stationary}
    Suppose Assumptions \labelcref{assumption1,assumption2,assumption3,assumption4} hold. Then for $A,B \in \mathcal{B}(\mathcal{X})$ and $\boldsymbol{\gamma} \in \mathcal{Y}$, we have $\int_{B} H_{\boldsymbol{\gamma}}(\mathbf{x}, A) \mu(\text{d}\mathbf{x}) = \int_{A} H_{\boldsymbol{\gamma}}(\mathbf{x}, B) \mu(\text{d}\mathbf{x}).$
\end{theorem} 

Theorem \ref{thm: stationary} implies that for every $\boldsymbol{\gamma} \in \mathcal{Y}$, $\mu$ is stationary for the transition kernel $H_{\boldsymbol{\gamma}}$. Since the AirMCMC scheme \citep{chimisov2018air} updates the adaptive parameters with increasing rarity, the diminishing adaptation condition is satisfied. To establish ergodicity of the adaptive algorithm, it is sufficient to show that the family of transition kernels is \textit{simultaneously strongly aperiodically geometrically ergodic} \citep{roberts2007coupling}.


 To show that the family of transition kernels is simultaneously strongly aperiodically geometrically ergodic, we will build on the results of \citet{natarovskii2021geometric} who showed geometric convergence of the elliptical slice sampler. We will first start by showing that the minorization condition holds for general open and bounded sets in $\mathcal{B}(\mathcal{X})$ in Proposition \ref{prop: minorization}, and then refine the possible \textit{small sets}, $C$, under which the geometric drift condition holds in Proposition \ref{prop: drift}. Proving that the geometric drift condition holds is not straightforward, as the transition kernel (Equation \ref{eq: trans_Kernel}) is quite complex. Utilizing Assumption \ref{assumption6}, we can (1) lower bound the probability of transitioning to within some smaller ellipse by the probability of transitioning to that set in the first iteration of the while-loop using the \textit{elliptical subcover} property and (2) upper bound the probability of moving outside a covering set using the \textit{diminishing tails} property. 

\begin{proposition}[Strongly Aperiodic Minorization Condition]
    \label{prop: minorization}
    Suppose Assumptions \labelcref{assumption1,assumption2,assumption3,assumption4} hold. Let $\boldsymbol{\gamma} \in \mathcal{Y}$ and let $C$ be an open and bounded set in $\mathcal{B}(\mathcal{X})$. Then there exists a $\delta > 0$ and a probability measure on $C$, $\nu_{\boldsymbol{\gamma}}(\cdot)$, such that $H_{\boldsymbol{\gamma}}(\mathbf{x}, \cdot) \ge \delta \nu_{\boldsymbol{\gamma}}(\cdot)$ for all $\mathbf{x} \in C$.
\end{proposition}

\begin{proposition}[Geometric Drift Condition]
    \label{prop: drift}
    Suppose Assumptions \labelcref{assumption1,assumption2,assumption3,assumption4,assumption6} hold. Define $V(\mathbf{x}):= 1 + \left[\left(\mathbf{x} - \boldsymbol{\mu}_0 \right)^{\top} \mathbf{A}^{-1}\left(\mathbf{x} - \boldsymbol{\mu}_0 \right)\right]^{1/2}$ where $\mathbf{A}$ is defined as in Assumption \ref{assumption6} if $\mathcal{X}$ is not bounded, otherwise let $\mathbf{A} = \mathbf{I}$.  Then, there exist $\phi < 1$, $b < \infty$, and a set $C= B_{\tilde{R}}(\boldsymbol{\mu}_0, \mathbf{A}) := \left\{\mathbf{x}\in \mathcal{X} \mid q_{\mathbf{x}}(\boldsymbol{\mu}_0, \mathbf{A}) < \tilde{R} \right\}$ ($\tilde{R} > 0$), such that $H_{\boldsymbol{\gamma}}V(\mathbf{x}) \le \phi V(\mathbf{x}) + b \mathbbm{1}_{C}(\mathbf{x})$ for all $\mathbf{x} \in \mathcal{X}$.
\end{proposition}

As illustrated by Propositions \labelcref{prop: minorization,prop: drift}, the family of Markov chain transition kernels considered in this adaptive scheme is simultaneously strongly aperiodically geometrically ergodic. Thus, there exist constants $K< \infty$ and $\rho < 1$ that depend on $b$, $\delta$, $\tilde{R}$, and $\phi$, such that $\lVert H^{n}_{\boldsymbol{\gamma}}(\mathbf{x}, \cdot) - \mu(\cdot)\rVert \le K V(\mathbf{x}) \rho^{n}$ for all $\mathbf{x} \in \mathcal{X}$ and $\boldsymbol{\gamma} \in \mathcal{Y}$ \citep{roberts2007coupling}. 
The values of $b$, $\delta$, $\tilde{R}$, and $\phi$ found in Propositions \labelcref{prop: minorization,prop: drift} are highly dependent on the values of $\xi$ and $\psi$ in Assumption \ref{assumption6}, with smaller values of $\xi$ and $\psi$ leading to slower upper bounds for the rate of convergence.  However, to show that the adaptive algorithm is ergodic, it is sufficient that the family of Markov chain transition kernels is simultaneously strongly aperiodically geometrically ergodic for any $b < \infty$, $\tilde{R} < \infty$, $\delta > 0$, and $\phi< 1$, meaning that we can choose $\xi$ and $\psi$ arbitrarily small. 

\begin{theorem}[Ergodicity]
    \label{thm: ergodicitiy}
    Suppose Assumptions \labelcref{assumption1,assumption2,assumption3,assumption4,assumption6} hold. Then the adaptive scheme proposed in Algorithm \ref{alg: AGESS} is ergodic.
\end{theorem} 

Theorem \ref{thm: ergodicitiy} illustrates that Assumptions \labelcref{assumption1,assumption2,assumption3,assumption4,assumption6} are sufficient to show that the adaptive algorithm is ergodic. Therefore, the distribution of the adaptive algorithm converges in total variation to the target distribution, which means that the adaptive algorithm is a valid method for drawing samples from the target distribution.

\section{Discussion}
\label{sec: discussion}

Adaptive generalized elliptical slice sampling is a promising gradient-free MCMC method that is capable of handling high-dimensional multimodal target distributions with strong dependencies among parameters. Although the method involves tuning adaptive parameters, we demonstrate that a general adaptation strategy is effective for a wide variety of target distributions---including non-convex, non-differentiable, non-elliptical, multimodal, and/or high-dimensional target distributions---thereby supporting its use as a \textit{black-box} sampler. A related approach proposed by \citet{schar2024parallel} adapts elliptical slice sampling via a learned affine transformation; however, their proof of ergodicity (Theorem 4.2) requires adaptation to stop after finitely many updates. Instead, AGESS establishes ergodicity under a schedule that adapts indefinitely at a diminishing rate, and further accommodates a broader class of elliptical distributions for the auxiliary variable, allowing practitioners to target heavy-tailed distributions by pairing AGESS with a heavy-tailed elliptical distribution in place of a Gaussian. We conclude the manuscript by discussing the limitations of our theoretical analysis of AGESS.


As shown in Propositions \labelcref{prop: minorization,prop: drift}, the family of transition kernels in this adaptive framework is simultaneously strongly aperiodically geometrically ergodic, meaning that for fixed $\boldsymbol{\gamma} \in \mathcal{Y}$, $H_{\boldsymbol{\gamma}}$ converges to $\mu$ at a geometric rate. 
A natural question is whether the adaptive scheme itself converges at a geometric rate, and whether this rate can be bounded. Recently, \citet{brown2024upper} provided a way to find upper and lower bounds on the convergence rate of adaptive sampling schemes, but obtaining such bounds requires stronger assumptions, including a sufficiently fast decay in the adaptation rate and a \textit{simultaneous subgeometric drift condition}. While the bounds on the rate of convergence would likely depend on the choices of $\xi$ and $\psi$ in Assumption \ref{assumption6}, it remains an open question whether one would be able to obtain meaningful upper and lower bounds on the convergence rate of the proposed adaptive algorithm. In fact, for the simple Gaussian target distribution examined in Section \ref{sec: Mixing_Times}, we were unable to obtain sharp convergence-rate bounds using standard drift and minorization techniques \citep{rosenthal1995minorization, meyn1994computable}, and therefore we adopted a more direct approach instead. Nevertheless, since adaptation occurs increasingly rarely and the family of transition kernels is simultaneously strongly aperiodically geometrically ergodic, we can apply the results of \citet{hofstadler2026almost} (Corollary 4.10) to obtain bounds on the almost sure convergence rate of MCMC estimates for the expectation of suitably regular functions under the target distribution. Taken together with the strong results across our case studies, these findings showcase the utility of adaptive generalized elliptical slice sampling across a broad range of Bayesian computational problems.

The Julia package \textsc{AdaptEllipticalSliceSampler.jl} is available on GitHub, enabling the use of adaptive generalized elliptical slice sampling for a broad range of applications. Tutorials covering the case studies presented in this manuscript are available in the software documentation. While we have only studied the sampling properties on a limited set of potential target distributions, the software package allows readers to test the sampler on their own problems by simply providing a function that efficiently evaluates the log target density. 

\section{Software}
The code associated with this manuscript can be found as follows:\\
\small{
    \textbf{Case Studies: \quad} \texttt{https://github.com/ndmarco/AGESS}\\
    \textbf{Julia Package: \quad} \texttt{https://github.com/ndmarco/AdaptEllipticalSliceSampler.jl}}

\section{Disclosure}
The authors report that there are no competing interests to declare.

\section{Supplementary Materials}
The Supplementary Materials contains (1) a review of elliptical distributions, (2) a detailed discussion on the mixing rates of the elliptical slice sampler, (3) a detailed discussion on the Shrinkage Algorithm, (4) proofs of all theorems and propositions found in the paper, (5) implementation details used in the case studies, (6) a discussion on practical considerations when using AGESS, and (7) additional case studies and use cases of AGESS.

\section{Acknowledgments}
The authors thank Filippo Ascolani, Sifan Liu, and Alexander Fisher for their helpful feedback. The authors gratefully acknowledge funding from NIH awards R01 DC013096 and R01 DC016363.

\bibliographystyle{abbrvnat}
\bibliography{AGESS}

\end{document}


\title{\bf Supplementary Materials for ``Adaptive Generalized Elliptical Slice Sampling''}
\author{Nicholas Marco\thanks{Corresponding author: nicholas.marco@duke.edu}\hspace{.2cm}\\
    Department of Statistical Science, Duke University\\
    and\\
    Surya T. Tokdar\\
    Department of Statistical Science, Duke University}
\maketitle
\bigskip

\maketitle

\begin{abstract}
    Supplementary Materials for the Manuscript ``Adaptive Generalized Elliptical Slice Sampling''. Section \ref{sec: elliptical} contains a review of the properties of Elliptical distributions. Section \ref{sec: Mixing_Times} contains a detailed discussion on the mixing time of the elliptical slice sampler when the target distribution is Gaussian. Section \ref{sec: proofs} contains the proofs of all theorems and propositions in the main manuscript. Section \ref{sec: algorithm} contains the algorithm used for all case studies in the main manuscript. Section \ref{sec: Practical} contains a detailed discussion on the practical considerations necessary to implement the adaptive algorithm. Section \ref{sec: add_case} contains additional case studies and potential applications of AGESS.
\end{abstract}

\vfill

\newpage
\onehalfspacing

\section{Review of Elliptical Distributions}
\label{sec: elliptical}

We will start with a review of elliptical distributions, and will primarily focus on two subclasses of elliptical distributions: multivariate Gaussian distributions and symmetric multivariate Pearson type VII distributions. Let $\mathcal{E}_P(\cdot; \boldsymbol{\mu}, \boldsymbol{\Sigma}, g)$ denote a $P$-dimensional elliptical distribution \citep{gomez2003survey,frahm2004generalized, fang2018symmetric} with a median vector $\boldsymbol{\mu}$, a positive-definite scale matrix $\boldsymbol{\Sigma}$, and a continuous non-negative functional parameter $g$ such that $\int_0^\infty t^{\frac{P}{2}-1}g(t) dt < \infty$. We will assume that the probability density function of the elliptical distribution exists, leading to $\text{pr}(\mathbf{X}\in \mathbf{A}) = \int_\mathbf{A}k g\left( (\mathbf{x} - \boldsymbol{\mu})^{\top} \boldsymbol{\Sigma}^{-1}(\mathbf{x} - \boldsymbol{\mu})\right)\lambda(d\mathbf{x})$ for some constant $k:= \frac{\Gamma(P/2)}{\pi^{P/2}\int_0^\infty t^{\frac{P}{2}-1}g(t) dt}$ ($\mathbf{X} \sim \mathcal{E}_P( \boldsymbol{\mu}, \boldsymbol{\Sigma}, g)$). When working with elliptical distributions, the following properties are often useful:

\begin{property}[Stochastic Representation \citep{cambanis1981theory}]
$\mathbf{X} \sim \mathcal{E}_P( \boldsymbol{\mu}, \boldsymbol{\Sigma}, g)$  with $\text{rank}(\Sigma) = K$ if and only if $$\mathbf{X} \overset{d}{=} \boldsymbol{\mu} + \mathcal{R}\boldsymbol{\Lambda}\mathcal{U}^{(K)},$$
where $\mathcal{U}^{(K)}$ is a $K$-dimensional random vector that is uniformly distributed on the $(K-1)$-dimensional sphere ($\mathcal{S}^{(K-1)}$), $\mathcal{R}$ is a non-negative random variable that is independent of $\mathcal{U}^{(K)}$, and $\boldsymbol{\Lambda}$ is the Cholesky root of $\boldsymbol{\Sigma}$.
\end{property}

\begin{property}[Moments \citep{frahm2004generalized}]
Let $\mathbf{X} \sim \mathcal{E}_P( \boldsymbol{\mu}, \boldsymbol{\Sigma}, g)$. Suppose $E(\mathcal{R}^2)$ is finite in the corresponding stochastic representation. Then we have
\begin{align}
    \nonumber E(\mathbf{X}) &= \boldsymbol{\mu},\\
    \nonumber \text{cov}(\mathbf{X}) &= \frac{E(\mathcal{R}^2)}{K} \boldsymbol{\Sigma}.
\end{align}
\end{property}

\begin{property}[Affine Transformations \citep{frahm2004generalized}]
\label{prop: affine_transformation}
Let $\mathbf{X} \sim \mathcal{E}_P( \boldsymbol{\mu}, \boldsymbol{\Sigma}, g)$, $\mathbf{B} \in \mathbb{R}^{K \times P}$, and $\mathbf{b} \in \mathbb{R}^K$. Then  we have
$$\mathbf{b} + \mathbf{B}\mathbf{X} \sim \mathcal{E}_K( \mathbf{B}\boldsymbol{\mu} + \mathbf{b}, \mathbf{B}\boldsymbol{\Sigma}\mathbf{B}^{\top}, g).$$
    
\end{property}

\begin{property}[Conditional and Marginal Distributions \citep{gomez2003survey}]
\label{prop: Conditional_Marginal}
Let $(\mathbf{X},\mathbf{Z}) \sim \mathcal{E}_{N}(\boldsymbol{\mu}, \boldsymbol{\Sigma}, g)$, where $\boldsymbol{\mu} = (\boldsymbol{\mu}_{x}, \boldsymbol{\mu}_{z})$ and $\boldsymbol{\Sigma} = \begin{bmatrix}
\boldsymbol{\Sigma}_{xx} & \boldsymbol{\Sigma}_{xz}\\
\boldsymbol{\Sigma}_{zx} & \boldsymbol{\Sigma}_{zz}
\end{bmatrix}$ ($\boldsymbol{\Sigma}$ is full rank, $\mathbf{X} \in \mathbb{R}^P$, and $\mathbf{Z} \in \mathbb{R}^{N-P}$). Then we have 
$$\mathbf{X} \sim \mathcal{E}_P(\boldsymbol{\mu}_x, \boldsymbol{\Sigma}_{xx}, g_x),$$
where $g_x(t) := \int_0^\infty w^{\frac{N-P}{2}-1}g(t + w) \text{d}w$ and
$$\mathbf{X} \mid \mathbf{Z} = \mathbf{z} \sim \mathcal{E}_P(\boldsymbol{\mu}_x + \boldsymbol{\Sigma}_{xz}\boldsymbol{\Sigma}_{zz}^{-1}(\mathbf{z} - \boldsymbol{\mu}_z), \boldsymbol{\Sigma}_{xx} - \boldsymbol{\Sigma}_{xz}\boldsymbol{\Sigma}_{zz}^{-1}\boldsymbol{\Sigma}_{zx}, \tilde{g}_{\mathbf{z}}),$$
where $\tilde{g}_{\mathbf{z}}(t):= g\left(t + (\mathbf{z} - \boldsymbol{\mu}_z)^{\top}\boldsymbol{\Sigma}_{zz}^{-1}(\mathbf{z} - \boldsymbol{\mu}_z)\right)$.
\end{property}

\begin{property}[Moments of Conditional Distribution \citep{gomez2003survey}]
\label{prop: Moments_Conditional}
Let $(\mathbf{X},\mathbf{Z}) \sim \mathcal{E}_{N}(\boldsymbol{\mu}, \boldsymbol{\Sigma}, g)$,  as above. Suppose $E(\mathcal{R}^2)$ is finite. Then the first two moments of the conditional distribution exist and can be expressed as
\begin{align}
    \nonumber E(\mathbf{X} \mid \mathbf{Z} = \mathbf{z}) &= \boldsymbol{\mu}_x + \boldsymbol{\Sigma}_{xz}\boldsymbol{\Sigma}_{zz}^{-1}(\mathbf{z} - \boldsymbol{\mu}_z), \\
    \nonumber \text{cov}(\mathbf{X} \mid \mathbf{Z} = \mathbf{z}) &= \frac{\int_0^\infty t^{\frac{P}{2}}g\left(t + (\mathbf{z} - \boldsymbol{\mu}_z)^{\top}\boldsymbol{\Sigma}_{zz}^{-1}(\mathbf{z} - \boldsymbol{\mu}_z)\right)\text{d}t}{P\int_0^\infty t^{\frac{P}{2} - 1}g\left(t + (\mathbf{z} - \boldsymbol{\mu}_z)^{\top}\boldsymbol{\Sigma}_{zz}^{-1}(\mathbf{z} - \boldsymbol{\mu}_z)\right)\text{d}t}\left(\boldsymbol{\Sigma}_{xx} - \boldsymbol{\Sigma}_{xz}\boldsymbol{\Sigma}_{zz}^{-1}\boldsymbol{\Sigma}_{zx}\right) .
\end{align}
\end{property}

As illustrated throughout the manuscript, the conditional distributions of elliptical distributions are crucial to the adaptive generalized algorithm. However, the general form of the conditional distributions are hard to work with, so we will focus on two subclasses of elliptical distributions:  Multivariate Gaussian distributions and Symmetric Multivariate Pearson Type VII distributions.

\subsection{Multivariate Gaussian Distributions}
\label{sec: Mult_Gaussian}
Multivariate Gaussian distributions \citep{fang2018symmetric} are elliptical distributions such that
$$g(t) = \exp{\left( -0.5t\right)}.$$

 Letting $(\mathbf{X},\mathbf{Z}) \sim \mathcal{N}_{N}(\boldsymbol{\mu}, \boldsymbol{\Sigma})$ as in Property \ref{prop: Conditional_Marginal}, that is, the joint distribution of $\mathbf{X}$ and $\mathbf{Z}$ is a Multivariate Gaussian distribution, we have
\begin{equation}
    \label{eq: var_gaussian}
    (\mathbf{X} \mid \mathbf{Z} = \mathbf{z}) \sim \mathcal{N}_{P}\left(\boldsymbol{\mu}_x + \boldsymbol{\Sigma}_{xz}\boldsymbol{\Sigma}_{zz}^{-1}(\mathbf{z} - \boldsymbol{\mu}_z), \boldsymbol{\Sigma}_{xx} - \boldsymbol{\Sigma}_{xz}\boldsymbol{\Sigma}_{zz}^{-1}\boldsymbol{\Sigma}_{zx} \right) .
\end{equation}
Moreover, the conditional distribution has the following stochastic representation
\begin{equation}
    \label{eq: stochastic_gaussian}
    (\mathbf{X} \mid \mathbf{Z} = \mathbf{z})\overset{d}{=} \left( \boldsymbol{\mu}_x + \boldsymbol{\Sigma}_{xz}\boldsymbol{\Sigma}_{zz}^{-1}(\mathbf{z} - \boldsymbol{\mu}_z)\right) + \mathcal{R} \boldsymbol{\Lambda} \mathcal{U}^{(P)}, 
\end{equation}
where $\boldsymbol{\Lambda}$ is the Cholesky root of $\boldsymbol{\Sigma}_{xx} - \boldsymbol{\Sigma}_{xz}\boldsymbol{\Sigma}_{zz}^{-1}\boldsymbol{\Sigma}_{zx}$, and $\mathcal{R}$ is a random variable independent of $\mathcal{U}^{(P)}$, with $\mathcal{R}^2 \sim \text{Ga}(P/2, 1/2)$.

\subsection{Symmetric Multivariate Pearson Type VII Distributions}
Symmetric multivariate Pearson type VII distributions \citep{fang2018symmetric} are elliptical distributions such that
$$g(t) = \left(1 + t/m \right)^{-M},$$
where $m > 0$ and $M > P/2$. Notice that the multivariate $t$-distribution is a special case of symmetric multivariate Pearson type VII distributions, specifically when $m = \nu$ (where $\nu$ is the degrees of freedom) and $M = \frac{P + \nu}{2}$. Using Theorem 3.8 in \citet{fang2018symmetric}, we have that the conditional distribution has the following stochastic representation
\begin{equation}
    \label{eq: stochastic_pearson}
    \left(\mathbf{X} \mid \mathbf{Z} = \mathbf{z}\right) \overset{d}{=} \left(\boldsymbol{\mu}_x + \boldsymbol{\Sigma}_{xz}\boldsymbol{\Sigma}_{zz}^{-1}(\mathbf{z} - \boldsymbol{\mu}_z) \right) + \mathcal{R} \boldsymbol{\Lambda} \mathcal{U}^{(P)},
\end{equation}
where $(\mathbf{X},\mathbf{Z}) \sim SMPVII_{N}(\boldsymbol{\mu}, \boldsymbol{\Sigma}, g)$ as in Property \ref{prop: Conditional_Marginal} and $\boldsymbol{\Lambda}$ is the Cholesky root of $\boldsymbol{\Sigma}_{xx} - \boldsymbol{\Sigma}_{xz}\boldsymbol{\Sigma}_{zz}^{-1}\boldsymbol{\Sigma}_{zx}$. Under this stochastic representation, we have $\mathcal{R}^2 / (m + q) \sim BeII(P/2, M - P/2)$ where $q:= (\mathbf{z} - \boldsymbol{\mu}_z)^{\top}\boldsymbol{\Sigma}_{zz}^{-1}(\mathbf{z} - \boldsymbol{\mu}_z)$ and $BeII$ is the Beta type II (Beta prime) distribution. Using the stochastic representation, we have
\begin{align}
    \nonumber \text{cov}(\mathbf{X} \mid \mathbf{Z} = \mathbf{z}) & = E\left[\left(\mathcal{R} \Lambda \mathcal{U}^{(P)}\right)\left(\mathcal{R} \Lambda \mathcal{U}^{(P)}\right)^{\top} \right] \\
    \nonumber & = \frac{E(\mathcal{R}^2)}{P} \left(\boldsymbol{\Sigma}_{xx} - \boldsymbol{\Sigma}_{xz}\boldsymbol{\Sigma}_{zz}^{-1}\boldsymbol{\Sigma}_{zx} \right).
\end{align}
If $M - P /2 > 1$, then we have that $E(\mathcal{R}^2)$ exists and
\begin{equation}
    \label{eq: var_pearson}
    \text{cov}(\mathbf{X} \mid \mathbf{Z} = \mathbf{z}) = \left(\frac{m + (\mathbf{z} - \boldsymbol{\mu}_z)^{\top}\boldsymbol{\Sigma}_{zz}^{-1}(\mathbf{z} - \boldsymbol{\mu}_z)}{2M - P - 2}\right)\left(\boldsymbol{\Sigma}_{xx} - \boldsymbol{\Sigma}_{xz}\boldsymbol{\Sigma}_{zz}^{-1}\boldsymbol{\Sigma}_{zx} \right).
\end{equation}
Note that when we are considering a $t$-distribution, $M = \frac{N + \nu}{2}$ and $m = \nu$, leading to
\begin{equation}
    \label{eq: var_t}
    \text{cov}(\mathbf{X} \mid \mathbf{Z} = \mathbf{z}) = \left(\frac{\nu + (\mathbf{z} - \boldsymbol{\mu}_z)^{\top}\boldsymbol{\Sigma}_{zz}^{-1}(\mathbf{z} - \boldsymbol{\mu}_z)}{N-P + \nu- 2}\right)\left(\boldsymbol{\Sigma}_{xx} - \boldsymbol{\Sigma}_{xz}\boldsymbol{\Sigma}_{zz}^{-1}\boldsymbol{\Sigma}_{zx} \right).
\end{equation}
The marginal distribution in this setting can be obtained from Theorem 3.7 in \citet{fang2018symmetric} and can be expressed as
$$ \mathbf{X} \sim SMPVII_P(\boldsymbol{\mu}_x, \boldsymbol{\Sigma}_{xx}, \tilde{g}),$$ where $\tilde{g}(t) = (1 + t/m)^{-\left( M  - 0.5(N - P)\right)}$.
Additional properties of the family of symmetric multivariate Pearson type VII distributions can be found in \citet{fang2018symmetric}.

\section{Theoretical Properties of an Optimal Elliptical Slice Sampler}
\label{sec: Mixing_Times}

Consider the case where the target distribution $\mu = \mathcal{N}(\mathbf{0}, \boldsymbol{\Sigma})$ corresponds to the prior distribution $\pi_0$. Thus, we have
$$\mu(\text{d}\mathbf{x}) = \mathcal{L}(\mathbf{x})\pi_0(\text{d}\mathbf{x}),$$
where $\mathcal{L}(\mathbf{x}) = 1$. In this setting, observe that we always accept the first proposed angle on the slice, since $\mathcal{L}$ is constant. Thus, given an initial state $\mathbf{x}_0$, the elliptical slice sampler produces the following Markov chain:
$$\mathbf{x}_n = \mathbf{x}_{n-1}\cos(\theta_n) + \mathbf{z}_n \sin(\theta_n), \quad n = 1,2,\dots,$$
where $(\mathbf{z}_n, \theta_n) \overset{\text{iid}}{\sim} \mathcal{N}(\mathbf{0}, \boldsymbol{\Sigma}) \times \mathcal{U}_{[0, 2\pi]}$. By a recursion argument, the Markov chain chain can be written as
\begin{equation}
    \label{eq: simple_ESS}
    \mathbf{x}_n = \mathbf{x}_0 \prod_{j = 1}^n \cos(\theta_j) + \sum_{k=1}^n \mathbf{z}_k\sin(\theta_k)\left[\prod_{j= k+1}^n \cos(\theta_j) \right].
\end{equation}

Under this simplified setting, we would like to study the efficiency of elliptical slice sampling. Specifically, we will study the convergence rate of the Markov chain and the autocorrelation of the Markov chain. 
We will start by first bounding the rate of convergence of the Markov chain.

\begin{proposition}
    \label{prop: log_p_convergence}
    Consider the $P$-dimensional target distribution $\mu = \mathcal{N}(\mathbf{0},\boldsymbol{\Sigma})$. Let $\mathbf{x}_0$ be the initial state of the Markov chain, and let $H^n(\mathbf{x}_0, \cdot)$ be the $n$-step transition kernel of the elliptical slice sampler with $\pi_0 = \mathcal{N}(\mathbf{0},\boldsymbol{\Sigma})$. The KL divergence between $\mu(\cdot)$ and $H^n(\mathbf{x}_0, \cdot)$ for $n \ge 3$ can be bounded as follows:
    \begin{align}
        \nonumber D_{\text{KL}}\infdiv{\mu(\cdot)}{H^n(\mathbf{x}_0, \cdot)}  & \le \left(\mathbf{x}_0^\intercal\boldsymbol{\Sigma}^{-1}\mathbf{x}_0 + P\right)\left(2^{-(n+1)} + \pi^{-n/2}\right).
    \end{align}
\end{proposition}
\begin{proof}
    Define $\mathcal{G}_n:= \sigma(\theta_1, \dots, \theta_n)$ and $\psi_n := \prod_{j=1}^n \cos(\theta_j)$. Conditional on $\mathcal{G}_n$,  $\mathbf{X}_n$ has a Gaussian distribution with 
    $$E\left[ \mathbf{X}_n \mid \mathcal{G}_n\right] = \mathbf{x}_0\psi_n, \quad \text{cov}\left[\mathbf{X}_n \mid \mathcal{G}_n\right] = \boldsymbol{\Sigma} \left(\sum_{k=1}^n \sin^2(\theta_k)\left[\prod_{j= k+1}^n \cos^2(\theta_j) \right]\right) = \boldsymbol{\Sigma}(1 - \psi_n^2).$$
    Therefore, conditional on $\mathcal{G}_n$, the KL divergence of the target $\mu(\cdot)$ from $H^n(\mathbf{x}_0, \cdot)$ is
    \begin{equation}\label{eq: KL_div_cond}
    \begin{aligned}
    D_{\text{KL}}\infdiv{\mu(\cdot)}{H^n(\mathbf{x}_0, \cdot\mid \mathcal{G}_n)} & = D_{\text{KL}}\infdiv{\mathcal{N}(\mathbf{0}, \boldsymbol{\Sigma})}{\mathcal{N}(\mathbf{x}_0\psi_n, \boldsymbol{\Sigma}(1 - \psi_n^2)} \\
    & =\frac{P}{2}\log(1 - \psi_n^2) + \frac{\mathbf{x}_0^\intercal\boldsymbol{\Sigma}^{-1}\mathbf{x}_0 + P}{2}\left( \frac{\psi^2_n}{1 - \psi^2_n} \right).
    \end{aligned}
    \end{equation}
    By convexity of the KL divergence and Jensen's inequality, we have
    \begin{equation} \label{eq: KL_div}
        \begin{aligned}
            D_{\text{KL}}\infdiv{\mu(\cdot)}{H^n(\mathbf{x}_0, \cdot)} & \le E\left[D_{\text{KL}}\infdiv{\mathcal{N}(\mathbf{0}, \boldsymbol{\Sigma})}{\mathcal{N}(\mathbf{x}_0\psi_n, \boldsymbol{\Sigma}(1 - \psi_n^2)}\right] \\
    & =\frac{P}{2}E\left[\log(1 - \psi^2_n)\right] + \frac{\mathbf{x}_0^\intercal\boldsymbol{\Sigma}^{-1}\mathbf{x}_0 + P}{2}E\left[ \frac{\psi^2_n}{1 - \psi^2_n} \right].
        \end{aligned}
    \end{equation}
    Letting $a_k:= \frac{1}{2\pi} \int_{0}^{2\pi}\cos^{2k}(\theta) \text{d}\theta = \frac{(2k)!}{2^{2k}(k!)^2} \le (\pi k)^{-1/2}$ for every integer $k\ge 1$, we have $E(\psi^{2k}_n) = a_k^n$. Using the power series of $\log(1 - x^2)$, we have 
    \begin{equation}\label{eq: power_log}
        E\left[\log(1 - \psi^2_n)\right]  = - E\left[\sum_{k=1}^{\infty} \frac{\psi^{2k}_n}{k}\right] = - \sum_{k=1}^{\infty}\frac{a_k^n}{k}
    \end{equation}
    where the second equality follows from the monotone convergence theorem. Similarly, we have
    \begin{equation}\label{eq: power_ratio}
    E\left[ \frac{\psi^2_n}{1 - \psi^2_n} \right] = E\left[\sum_{k=1}^{\infty} \psi^{2k}_n\right] = \sum_{k=1}^{\infty} a_k^n.
    \end{equation}
    Using the series representations in Equations \labelcref{eq: power_log,eq: power_ratio}, we have
    \begin{equation} \label{eq: KL_div_series}
        \begin{aligned}
            D_{\text{KL}}\infdiv{\mu(\cdot)}{H^n(\mathbf{x}_0, \cdot)} \le \frac{1}{2}\left[ \mathbf{x}_0^\intercal\boldsymbol{\Sigma}^{-1}\mathbf{x}_0\left(\sum_{k=1}^\infty a_k^n\right) + P \left(\sum_{k=1}^\infty a_k^n\left(1 - \frac{1}{k}\right)\right)\right].
        \end{aligned}
    \end{equation}
    Using the fact that $a_k^n \le (\pi k)^{-n/2}$, for $n \ge 3$, we can obtain the following bounds on the series:
    \begin{align}\label{eq: bd_series1}
        \nonumber \sum_{k=1}^\infty a_k^n &\le 2^{-n} + \int_{1}^{\infty}(\pi x)^{-n/2}\text{d}x \\
        & = 2^{-n} + \frac{\pi^{-n/2}}{n/2 - 1}.
    \end{align}
    Since $(1 - \frac{1}{k}) \le 1$ for all integers $k \ge 1$, we have that Equation \ref{eq: bd_series1} also bounds $\sum_{k=1}^\infty a_k^n\left(1 - \frac{1}{k}\right)$. Using this along with the assumption that $n \ge 3$ gives us the following:
    \begin{equation}\label{eq: KL_bound2}
    \begin{aligned}
        D_{\text{KL}}\infdiv{\mu(\cdot)}{H^n(\mathbf{x}_0, \cdot)} & \le  \left(\mathbf{x}_0^\intercal\boldsymbol{\Sigma}^{-1}\mathbf{x}_0 + P\right)\left(2^{-(n+1)} + \frac{\pi^{-n/2}}{n -2 }\right)\\
        & \le \left(\mathbf{x}_0^\intercal\boldsymbol{\Sigma}^{-1}\mathbf{x}_0 + P\right)\left(2^{-(n+1)} + \pi^{-n/2}\right).
    \end{aligned}
    \end{equation}

\end{proof}

In practice, we often look at measurements of the effective sample size as a proxy for the mixing rate. When considering the multivariate effective sample size \citep{vats2019multivariate}, the autocorrelation of the Markov chain plays a key role in determining the effective sample size. Specifically, the multivariate effective sample size is defined as
$$\text{mESS} = N\left(\frac{\left|\boldsymbol{\Lambda}\right|}{\left|\boldsymbol{\Sigma}\right|}\right)^{1/P}$$
where $N$ is the total number of iterations of the Markov chain. Here, $\boldsymbol{\Lambda}$ is the covariance of the target distribution and $\boldsymbol{\Sigma}$ is the Monte Carlo asymptotic covariance matrix. Specifically, we have $\boldsymbol{\Sigma}  =  \boldsymbol{\Lambda} + \sum_{k=1}^{\infty} \boldsymbol{\Gamma}_k + \boldsymbol{\Gamma}_k^\intercal$, where $\boldsymbol{\Gamma}_k = \text{cov}\left(\mathbf{X}_0, \mathbf{X}_k\right)$ is the $k$-lag autocovariance matrix. In the scenario where the target distribution and the prior distribution coincide and are both isotropic normal distributions, we can show that the multivariate effective sample size of the constructed Markov chain is equal to the length of the Markov chain; illustrating that the elliptical slice sampler generates uncorrelated samples in this scenario.

\begin{lemma}
    \label{prop: mESS}
    Consider the Markov chain described in Equation \ref{eq: simple_ESS}. Suppose $\mathbf{X}_0 \sim \mu = \mathcal{N}(\mathbf{0}, \tilde{\boldsymbol{\Sigma}})$. The multivariate effective sample size for a chain of length $N$ is $N$.
\end{lemma}
\begin{proof}
    By the definition of multivariate effective sample size, we have 
   $$\text{mESS} = N\left(\frac{\left|\boldsymbol{\Lambda}\right|}{\left|\boldsymbol{\Sigma}\right|}\right)^{1/P}$$
where $\boldsymbol{\Sigma}  =  \boldsymbol{\Lambda} + \sum_{k=1}^{\infty} \boldsymbol{\Gamma}_k + \boldsymbol{\Gamma}_k^\intercal$ and $\boldsymbol{\Gamma}_k = \text{cov}\left(\mathbf{X}_0, \mathbf{X}_k\right)$. Here, we have $\boldsymbol{\Lambda} = \tilde{\boldsymbol{\Sigma}}$. Define $\mathcal{G}_n:= \sigma(\theta_1, \dots, \theta_n)$ and $\psi_n := \prod_{j=1}^n \cos(\theta_j)$. As in the proof of Proposition \ref{prop: log_p_convergence}, we have 
$$E\left[\mathbf{X}_n \mid \mathbf{X}_0 = \mathbf{x}_0, \mathcal{G}_n \right] = \mathbf{x}_0\psi_n, \quad \text{cov}\left[\mathbf{X}_n \mid \mathbf{X}_0 = \mathbf{x}_0, \mathcal{G}_n \right] = \tilde{\boldsymbol{\Sigma}}(1 - \psi_n^2).$$
Using the law of total covariance, we have  
$$\boldsymbol{\Gamma}_n :=\text{cov}(\mathbf{X}_0, \mathbf{X}_n) = \text{cov}\left(\mathbf{X}_0, E\left[\mathbf{X}_n| \mathbf{X}_0\right]\right).$$
Notice that 
$$E\left[\mathbf{X}_n| \mathbf{X}_0 = \mathbf{x}_0\right] = E\left[E\left(\mathbf{X}_n| \mathbf{X}_0= \mathbf{x}_0, \mathcal{G}_n\right)\right] = \mathbf{x}_0 E(\psi_n) = \mathbf{x}_0\left(\frac{1}{2\pi} \int_{0}^{2\pi} \cos(\theta) \text{d}\theta\right)^n =\mathbf{0}.$$
Thus, we have $\boldsymbol{\Gamma}_n = \text{cov}(\mathbf{X}_0, \mathbf{0}) = \mathbf{O}$ for all integers $n \ge 1$.
Therefore, we have  
$$\left(\frac{\left|\boldsymbol{\Lambda}\right|}{\left|\boldsymbol{\Sigma}\right|}\right)^{1/P} = \left(\frac{|\tilde{\boldsymbol{\Sigma}}|}{|\tilde{\boldsymbol{\Sigma}}|}\right)^{1/P} = 1.$$
This directly gives us $\text{mESS} = N$.
\end{proof}

From Proposition \ref{prop: log_p_convergence} and Lemma \ref{prop: mESS}, we can see that elliptical slice sampling exhibits extremely fast mixing times and dimension-independent multivariate effective sample size calculations when the target and prior distributions coincide and are both Gaussian distributions. Under an isotropic Gaussian setting, we will show that the mixing time increases substantially when the prior distribution and target distributions differ. 

\begin{proposition}
 \label{prop: sub_opt_bound}
    Consider an elliptical slice sampler where the target distribution is $\mu = \mathcal{N}(\mathbf{0}, \sigma^2\mathbf{I}_P$) and the prior is $\pi_0 = \mathcal{N}(\mathbf{0}, (1 + \alpha)\sigma^2\mathbf{I}_P)$ for some $\alpha > 0$. For any $\epsilon \in (0,1)$ and $\alpha > 0$, there exists $P_\alpha \in \mathbb{N}$ such that with $N = \left\lfloor \frac{\sqrt{P(1 + \alpha)}}{4(\log(P))^2} \right\rfloor$,$$\norm{H^N(\mathbf{0}, \cdot) - \mu(\cdot)}_{TV} \ge 1 - \epsilon \quad \forall P > P_\alpha.$$    
\end{proposition}
\begin{proof}
    Consider an elliptical slice sampler where $\mu = \mathcal{N}(0, \sigma^2\mathbf{I}_P)$ and $\pi_0 = \mathcal{N}(0, (1 + \alpha)\sigma^2\mathbf{I}_P)$ for some $\alpha > 0$. In this setup, we have $\mathcal{L}(\mathbf{x}) \propto \exp\left(-\frac{\norm{\mathbf{x}}^2}{2\beta} \right)$ where $\beta := \sigma^2(1 + \alpha^{-1})$. Consequently, with $\mathbf{z} \sim \pi_0$, $y \sim \mathcal{U}_{[0,1]}$, the slice $S(y; \mathbf{x}, \mathbf{z})$ is
    $$S(y; \mathbf{x}, \mathbf{z}) = \left\{\theta \in [0,2\pi) \mid \norm{\mathbf{x}_{\theta}}^2 \le \norm{\mathbf{x}}^2 + 2\beta \log(1/y)\right\},$$
    where $\mathbf{x}_{\theta} = \mathbf{x}\cos(\theta) + \mathbf{z}\sin(\theta)$. Under the choice of $\pi_0$, we have $\frac{\norm{\mathbf{z}}^2}{(1 + \alpha)\sigma^2} \sim \chi^2_P$. Additionally, for fixed $\mathbf{x} \in \mathbb{R}^P$, we can express
    $$\mathbf{x}^{\intercal}\mathbf{z} = \frac{\norm{\mathbf{x}}\norm{\mathbf{z}}}{\sqrt{P}}\eta,$$
    for a single random variable $\eta$ which is independent of $\norm{\mathbf{z}}$ (and whose distribution does not depend on $\mathbf{x}$), and $\eta^2/P \sim \text{Beta}\left( \frac{1}{2}, \frac{P-1}{2}\right)$. Define $\xi = \frac{\norm{\mathbf{z}}}{\sqrt{(1 + \alpha)}\sigma} - \sqrt{P}$. A standard argument involving the central limit theorem and Slutsky's theorem shows $(\xi, \eta) \xrightarrow{d} \mathcal{N}(0, 1/2) \times \mathcal{N}(0,1)$. In addition, applying Chernoff bounds (Lemmas \labelcref{lemma: gamma_tail,lemma: beta_tail}) we have $\text{pr}\left(\frac{\norm{\mathbf{z}}^2}{(1 + \alpha)\sigma^2} > tP \right) \le \exp \left\{-\frac{P}{2}(t-1-\log t) \right\}$ if $t > 1$, $\text{pr}\left(\frac{\norm{\mathbf{z}}^2}{(1 + \alpha)\sigma^2} < tP \right) \le \exp \left\{-\frac{P}{2}(t-1-\log t) \right\}$ if $0 < t < 1$, and $\text{pr}( |\eta| > t) \le \exp\left\{ \frac{1}{2}\left[\log(t^2) - (P-1)\log\left( \frac{P-1}{P-t^2}\right) \right]\right\}$ if $1 < t< \sqrt{P}$.

    Since $\norm{\mathbf{x}_{\theta}}^2 = \norm{\mathbf{x}}^2\cos^2(\theta) + \norm{\mathbf{z}}^2\sin^2(\theta) + \mathbf{x}^\intercal \mathbf{z}\sin(2\theta)$, the slice inclusion event $\{\theta \in S(y; \mathbf{x}, \mathbf{z})\}$ is measurable with respect to the $\sigma$-algebra generated by $(\xi, \eta, y)$. Take $\delta_{\alpha, P}:= \frac{(\log(P))^2}{\sqrt{P}(1 + \alpha)}$. Define the events
    $$\mathcal{E}_{P,\mathbf{x}, \alpha} = \left \{|\xi| > \log P \right\} \cup \left\{S(y;\mathbf{x}, \mathbf{z}) \text{ includes a $\theta$ with } |\sin(\theta)| > \delta_{\alpha,P}  \right\},$$
    and let $\pi_P = \sup_{\norm{\mathbf{x}} \le \sigma\sqrt{P}}\text{pr}(\mathcal{E}_{P,\mathbf{x}, \alpha})$. Notice that if $\norm{\mathbf{x}} \le \sigma\sqrt{P}$ then
    \begin{align}
        \nonumber \norm{\mathbf{x}_\theta}^2 - \norm{\mathbf{x}}^2 = & (\norm{\mathbf{z}}^2 - \norm{\mathbf{x}}^2) \sin^2(\theta) + \mathbf{x}^{\intercal}\mathbf{z}\sin(2\theta) \\
        \nonumber \ge & \sigma^2\left[(1 + \alpha)\left(\sqrt{P} + \xi\right)^2 - P\right] \sin^2(\theta) - 2\sigma^2\sqrt{1 + \alpha}\left(\xi + \sqrt{P} \right)|\eta\sin(\theta)|\\
        \nonumber = & \sigma^2\left[\alpha P\sin^2(\theta) + 2(1 + \alpha)\sqrt{P}\sin^2(\theta)\left(\xi + \frac{\xi^2}{2\sqrt{P}} - \frac{|\eta| + \xi|\eta|}{|\sin(\theta)|\sqrt{1 + \alpha}} \right) \right]\\
        \nonumber \ge & \sigma^2\left[\alpha P\sin^2(\theta) + 2(1 + \alpha)\sqrt{P}\sin^2(\theta)\left(\xi - \frac{|\eta| + \xi|\eta|}{\delta_{\alpha,P}\sqrt{1 + \alpha}} \right) \right]\\
        \nonumber \ge & \sigma^2\left[\alpha P\sin^2(\theta) - \sqrt{P}\sin^2(\theta)\phi\right],
        \end{align}
        where $\phi = 2(1 + \alpha)\left(|\xi|  + \frac{|\eta| + |\xi\eta|}{\delta_{\alpha,P}\sqrt{1 + \alpha}}\right)$. Consequently, we have 
        \begin{align}
            \nonumber \pi_P & \le \sup_{\norm{\mathbf{x}} \le \sigma\sqrt{P}} \text{pr}\left[\inf_{\theta : |\sin(\theta)| > \delta_{\alpha,P}} \left(\norm{\mathbf{x}_{\theta}}^2 - \norm{\mathbf{x}}^2\right) \le 2 \beta \log(1/y)\right] + \text{pr}\left[|\xi| > \log(P) \right] \\
            \nonumber & \le \text{pr}\left[\inf_{\theta : |\sin(\theta)| > \delta_{\alpha,P}} \sin^2(\theta)\left(\alpha P - \sqrt{P}\phi\right) \le 2 (1 + \alpha^{-1}) \log(1/y)\right] + \text{pr}\left[\xi^2 > (\log(P))^2 \right].
        \end{align}
        Notice that if $\phi \le \alpha \sqrt{P}$, then $\inf_{\theta : |\sin(\theta)| > \delta_{\alpha,P}} \sin^2(\theta)\left(\alpha P - \sqrt{P}\phi\right) = \delta_{\alpha,P}^2 \left(\alpha P - \sqrt{P}\phi\right)$. Consequently, using the union bound, we have
        \begin{align}
            \nonumber \text{pr}&\left[\inf_{\theta : |\sin(\theta)| > \delta_{\alpha,P}} \sin^2(\theta)\left(\alpha P - \sqrt{P}\phi\right) \le 2 (1 + \alpha^{-1}) \log(1/y)\right]\\
            \nonumber &\le  \text{pr}\left[\delta_{\alpha,P}^2 \left(\alpha P - \sqrt{P}\phi\right) \le  2 (1 + \alpha^{-1}) \log(1/y)\right] + \text{pr}\left[\phi >\alpha \sqrt{P} \right]\\
            \nonumber &  \le \text{pr}\left[\phi > \frac{1}{2} \alpha\sqrt{P}\right] + \text{pr} \left[2(1 + \alpha^{-1})\log(1/y) > \frac{1}{2}\alpha P\delta^2_{\alpha, P} \right] + \text{pr}\left[\phi >\alpha \sqrt{P} \right]\\
            \nonumber & \le 2\text{pr}\left[\phi > \frac{1}{2} \alpha\sqrt{P}\right] + \text{pr} \left[\log(1/y) > \frac{\alpha^2}{(1+\alpha)^{3}}\frac{(\log(P))^4}{4} \right]\\
            \label{eq: inf_prob_bound} & \le 2\text{pr}\left[ \xi^2 > \frac{1}{4}\left(\sqrt{\frac{(\log(P))^2\alpha}{4(1 + \alpha)}} - 1\right)^2\right] + 2\text{pr}\left[ \eta^2 > \frac{\alpha(\log(P))^2}{16(1 + \alpha)^2}\right]  + \text{pr} \left[\log(1/y) > \frac{\alpha^2}{(1+\alpha)^{3}}\frac{(\log(P))^4}{4} \right],
        \end{align}
        for large $P$, where the last inequality can be realized due to the fact that for $b \in \mathbb{R}_+$ there exists a $P_b \in \mathbb{N}$ such that 
        $$\text{pr}\left( \phi > 2b(1 + \alpha)\sqrt{P}\right) \le \text{pr}\left( \xi^2 > \frac{1}{4}\left[\log(P) \sqrt{b} - 1\right]^2\right) + \text{pr}\left(\eta^2 > \frac{b}{4(1 + \alpha)} (\log(P))^2\right) \quad \forall P > P_b.$$ 
        Since $\log(1/y) \sim Ex(1)$, we have 
        \begin{equation}
            \label{eq: exponential_bound}
            \text{pr} \left[\log(1/y) > \frac{\alpha^2}{(1+\alpha)^{3}}\frac{(\log(P))^4}{4} \right] \le \exp\left(-\frac{\alpha^2}{(1+\alpha)^{3}}\frac{(\log(P))^4}{4}  \right).
        \end{equation}
        Using the Chernoff bound on $|\eta|$, and letting $\epsilon = \frac{\alpha(\log(P))^2}{16(1 + \alpha)^2}$, we can bound the second term of Equation \ref{eq: inf_prob_bound} by
        \begin{align}
            \label{eq: chernoff_eta}
            \nonumber 2\text{pr}\left[ \eta^2 > \frac{\alpha(\log(P))^2}{16(1 + \alpha)^2}\right] &\le 2\exp\left\{ \frac{1}{2}\left[\log\left(\epsilon \right) - (P-1)\log\left( \frac{P-1}{P-\epsilon}\right) \right]\right\}\\
              &= 2\exp\left\{-\frac{1}{2}\left[(P-1)\log\left( \frac{P-1}{P-\mathcal{O}(\log(P)^2)}\right) - \smallO(\log(P)^2 )\right]\right\}.
        \end{align}
        Using the Chernoff bound for $\frac{\norm{\mathbf{z}}^2}{(1 + \alpha)\sigma^2}$, and letting $\epsilon = \frac{1}{4}\left(\sqrt{\frac{\log(P)^2\alpha}{4(1 + \alpha)}} - 1\right)^2$, we have
        {\small
        \begin{align}
           \nonumber 2\text{pr}\left[ \xi^2 > \frac{1}{4}\left(\sqrt{\frac{\log(P)^2\alpha}{4(1 + \alpha)}} - 1\right)^2\right]  = & 2\left[\text{pr}\left(\frac{\norm{\mathbf{z}}^2}{(1 + \alpha)\sigma^2}\ge P\left(1 + \sqrt{\frac{\epsilon}{P}}\right)^2 \right) + \text{pr}\left(\frac{\norm{\mathbf{z}}^2}{(1 + \alpha)\sigma^2}\le P\left(1 - \sqrt{\frac{\epsilon}{P}}\right)^2 \right) \right] \\
           \nonumber \le & 2\left[\exp\left\{-\frac{P}{2}\left(\frac{\epsilon}{P} + 2\left(\sqrt{\frac{\epsilon}{P}} - \log\left(1 + \sqrt{\frac{\epsilon}{P}}\right)\right) \right)\right\}\right. \\
           \nonumber & \;\;\; + \left.\exp\left\{-\frac{P}{2}\left(\frac{\epsilon}{P} - 2\left(\sqrt{\frac{\epsilon}{P}} + \log\left(1 - \sqrt{\frac{\epsilon}{P}}\right)\right) \right)\right\}\right]\\
           \nonumber \le & 4\exp\left\{- \frac{\epsilon}{2}\right\} \\
           \nonumber = & 4\exp\left\{-\frac{1}{8}\left(\sqrt{\frac{\log(P)^2\alpha}{4(1 + \alpha)}} - 1\right)^2 \right\}\\
          \label{eq: chernoff_xi} = & 4P^{-\left(\frac{\alpha}{(1 + \alpha)32}\log(P)\right) + \mathcal{O}(1)},
        \end{align}}
        where we use the fact that $x - \log(1 + x) > 0$ for $x > 0$ and $x + \log(1 -x) < 0$ for $0 < x< 1$. Similarly, we can find $\text{pr}[\xi^2 > (\log(P))^2] \le 2P^{-0.5\log(P)}$.
        Consequently, for fixed $\alpha > 0$, we can see that $\lim_{P \rightarrow \infty} \sqrt{P}\pi_P = 0$.

        Let us now consider the Markov chain $\mathbf{x}_n \sim H(\mathbf{x}_{n-1}, \cdot)$ starting at $\mathbf{x}_0 = \mathbf{0}$. For $n \ge 1$, let $\mathcal{E}_n = \mathcal{E}_{P, \mathbf{x}_{n-1}, \alpha}$. Take $N = \left\lfloor \frac{\sqrt{P(1 + \alpha)}}{4(\log(P))^2} \right\rfloor$ and define $\Omega_N = \cap_{n \le N}\mathcal{E}^c_n$. We make the following claim:
        \begin{equation}
            \label{eq: claim_MCMC}
            \text{On $\Omega_N$, $\norm{\mathbf{x}_n} \le \frac{2n\sigma(\log(P))^2}{\sqrt{1 + \alpha}}$ for each $n \le N$.}
        \end{equation}
        The claim is trivially true for $n = 0$. By induction, we have
        $$\norm{\mathbf{x}_n} \le \norm{\mathbf{x}_{n-1}} + \norm{\mathbf{z}_n}|\sin\theta_n| \le \frac{2(n-1)\sigma (\log(P))^2}{\sqrt{1 + \alpha}} + 2\sigma\sqrt{1+ \alpha}\sqrt{P}\left(\frac{(\log(P))^2}{\sqrt{P}(1 + \alpha)}\right) = \frac{2\sigma n(\log(P))^2}{\sqrt{1 + \alpha}},$$
        for each $n \ge 1$. Consequently on $\Omega_N$, $\norm{\mathbf{x}_N} \le \sigma\sqrt{P}/2$, and hence
        $$\text{pr}(\norm{\mathbf{x}_N} > \sigma\sqrt{P}/2 \mid \mathbf{x}_0 = \mathbf{0}) \le \text{pr}(\Omega_N^C) \le 
        \sum_{n \le N}\text{pr}(\mathcal{E}_n) \le N \pi_P \rightarrow 0,
        $$
        as $P \rightarrow \infty$. However, for any $\mathbf{y} \sim \mu$, $\text{pr}(\norm{\mathbf{y}} > \sigma\sqrt{P}/2) = 1 - \text{pr}(\norm{\mathbf{y}}^2 \le \sigma^2P/4) \ge 1 - e^{-P/2 (-\frac{3}{4}-\log(1/4))} \rightarrow 1$ as $P \rightarrow \infty$.
        Therefore, for any $\epsilon \in (0,1)$ and $\alpha > 0$,  one can find a $P_\alpha$ such that with $N = \left\lfloor \frac{\sqrt{P(1 + \alpha)}}{4(\log(P))^2} \right\rfloor$,
        $$\norm{H^N(\mathbf{0}, \cdot) - \mu}_{TV} \ge 1 - \epsilon \quad \forall P > P_{\alpha}.$$
\end{proof}

\begin{lemma}[Gamma tail bound]
    \label{lemma: gamma_tail}
    Suppose $X \sim \text{Gamma}(r, \lambda)$ where $\mu = E(X) = r/\lambda$. Take $g(t) = r(t - 1 - \log t)$, where $t > 0$. Then $\text{pr}(X > t\mu) \le \exp(-g(t))$ if $t > 1$, and $\text{pr}(X < t\mu) \le \exp(-g(t))$ if $t < 1$.
\end{lemma}
\begin{proof}
    By Markov's inequality, for any $x > 0$,
    $$\text{pr}(X > x) \le \inf_{s > 0} e^{-sx}E\left( e^{sX}\right) = \inf_{s > 0} e^{-sx}(1 - s/\lambda)^{-r} = \inf_{s > 0} e^{-h(s)},$$
    where $h(s) = sx + r\log(1 - s/\lambda)$. Notice that this is uniquely maximized at $\hat{s} = \max\left(0, \lambda - \frac{r}{x}\right)$. If $x = t\mu > \mu$, then $\hat{s} = \lambda(t-1)/t$ and $h(\hat{s}) = g(t)$. Thus, we have $\text{pr}(X > t\mu) \le e^{-g(t)}$. A similar calculation gives the lower tail bound. 
\end{proof}

\begin{lemma}[Beta tail bound]
    \label{lemma: beta_tail}
    Suppose $X \sim \text{Beta}(a,b)$, where $\mu = E(X) = a/(b +a)$. Then for any $x \in (\mu,1)$, 
    $$\text{pr}(X > x) \le \exp\left[-(a + b)\left\{\mu\log\left(\frac{\mu}{x}\right) + (1-\mu)\log\left(\frac{1 - \mu}{1-x}\right)\right\} \right].$$
\end{lemma}
\begin{proof}
    Using properties of the Beta distribution, we can write $X = \frac{U}{U+V}$, where $U \sim \text{Gamma}(a, 1)$, $V\sim \text{Gamma}(b, 1)$, and $U$ and $V$ are independent. Therefore,
    \begin{align}
        \nonumber \text{pr}(X > x) & = \text{pr}\left[(1-x)U - xV > 0 \right] & \\
        \nonumber & \le \inf_{s > 0}\left\{E\left(\exp\left[s(1-x)U - sxV\right] \right)\right\}& \text{[Markov's inequality]}\\
        \nonumber & = \inf_{s > 0}\left\{E\left(\exp\left[s(1-x)U\right]\right) E\left(\exp\left[- sxV\right] \right)  \right\} & \text{[Independence]}\\
        \nonumber & = \inf_{s > 0}\left\{ \left[1 - s(1-x)\right]^{-a}\left[1 + sx\right]^{-b}\right\} & \\
        \nonumber & = \inf_{s > 0} e^{-h(s)},
    \end{align}
    with $h(s) = a\log\left(1 - s(1-x) \right) + b\log(1 + sx)$. If $x > \mu$, then $h(s)$ is uniquely maximized at $\hat{s} = \frac{x - \mu}{x(1-x)}$. Noticing that $h(\hat{s}) = a\log\left(\frac{\mu}{x}\right) + b \log\left(\frac{1 - \mu}{1 -x} \right)$ completes the result.
\end{proof}

\subsection{The Volcano Distribution}
\begin{figure}
    \centering
    \includegraphics[width=1.0\linewidth]{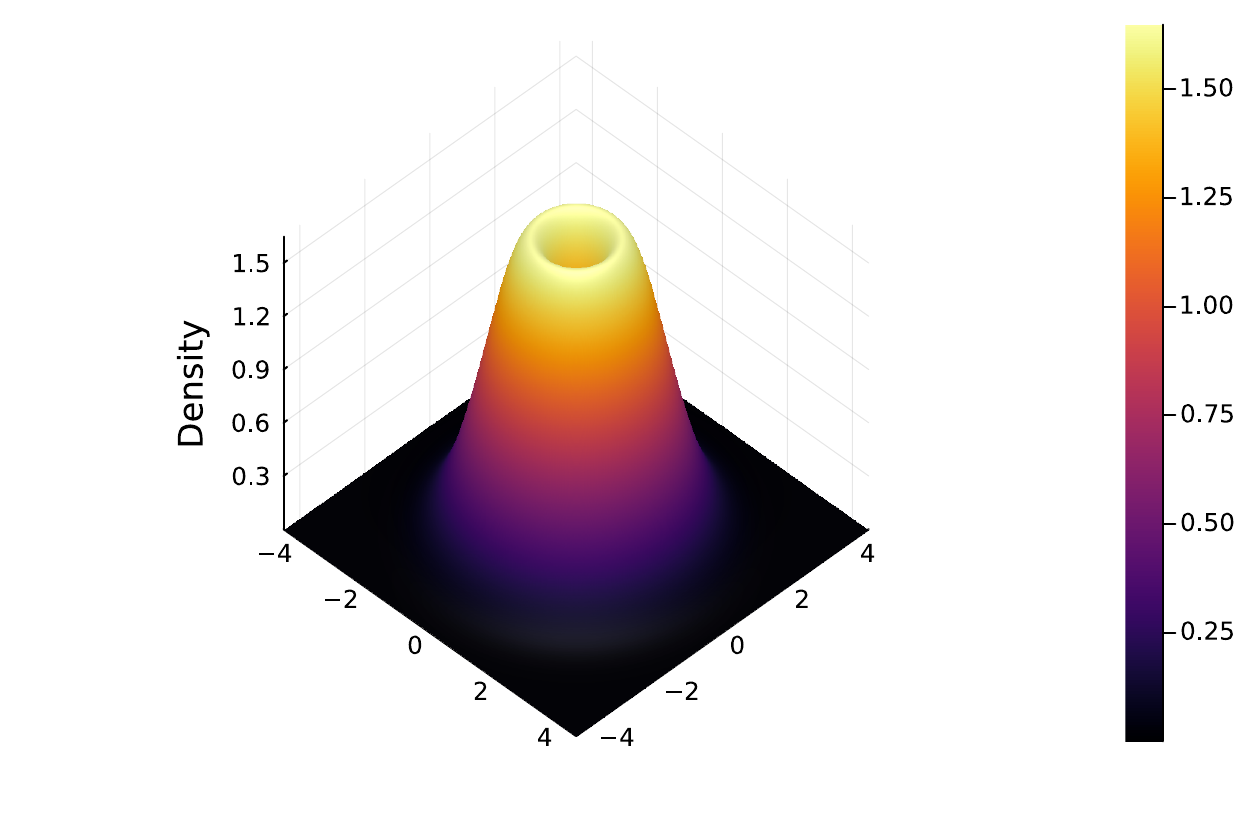}
    \caption{Visualization of the Volcano distribution.}
    \label{fig: volcano_density}
\end{figure}

In the work of \citet{natarovskii2021geometric}, the authors investigate the sampling efficiency of the elliptical slice sampler on the \textit{Volcano} distribution; see Figure \ref{fig: volcano_density}. In particular, they demonstrate that the elliptical slice sampler attains an effective sample size that appears to be dimension independent. As illustrated earlier in the section, we showed that the elliptical slice sampler can exhibit an effective sample size that is not dependent on the dimension of the target distribution, yet the mixing time is $\mathcal{O}(\log(P))$. Therefore, although the mixing times are likely not independent of the dimension, this suggests that fast mixing times (i.e., $\mathcal{O}(\log(P))$) may extend beyond Gaussian target distributions and even beyond monotonically decreasing elliptically contoured distributions.

\begin{figure}
    \centering
    \includegraphics[width=\linewidth]{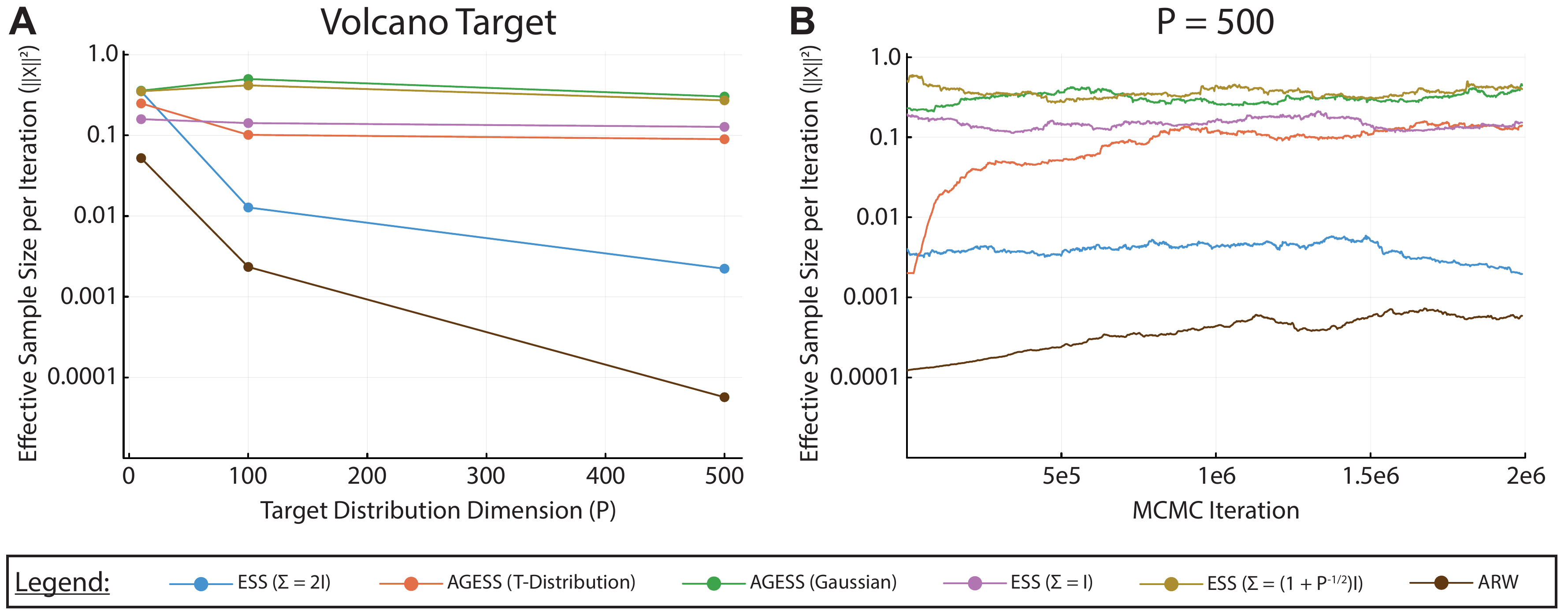}
    \caption{Sampling performance of the various MCMC algorithms when targeting the Volcano distribution. \textbf{Subfigure A} shows the effective sample size per iteration of $\norm{\mathbf{x}}^2$, computed using the final 40\% of iterations to ensure that adaptive methods have had sufficient time to adapt to the target distribution. \textbf{Subfigure B} illustrates how the adaptive schemes are able to adapt to the target distribution and achieve higher effective sample size per iteration as the Markov chain runs.}
    \label{fig: volcano_results}
\end{figure}

The Volcano distribution \citep{natarovskii2021geometric} is an elliptically-contoured distribution that is specified as follows:
$$\mu(\mathbf{x}) \propto \exp\left\{\norm{\mathbf{x}} - \frac{1}{2}\norm{\mathbf{x}}^2 \right\}.$$
Here, we will compare the effective sample size estimates across the following MCMC schemes: (1) adaptive random walk \citep[ARW,][]{haario2001adaptive}, (2) an optimal elliptical slice sampler ($\boldsymbol{\Sigma} = (1 + P^{-1/2})\mathbf{I}_P$), (3) the elliptical slice sampler used in \citet{natarovskii2021geometric} ($\boldsymbol{\Sigma} = \mathbf{I}_P$), (4) an adaptive generalized elliptical slice sampler using a T-distribution, (5) a modified adaptive generalized elliptical slice sampler using a Normal distribution with structured covariance structure ($\boldsymbol{\Sigma} = \hat{\sigma}^2\mathbf{I}_P$, where $\hat{\sigma}$ is adaptively learned), and (6) a sub-optimal elliptical slice sampler ($\boldsymbol{\Sigma} = 2\mathbf{I}_P$). To evaluate the sampling performance of each MCMC scheme, we estimated the effective sample size of the quantity $\norm{\mathbf{x}}^2$. The results of this case study can be visualized in Figure \ref{fig: volcano_results}.

Figure \ref{fig: volcano_results} supports the results found by \citet{natarovskii2021geometric}, showing that ESS can achieve an effective sample size that appears to be independent of the dimension of the target distribution. Similarly, we can see that both variants of AGESS (Structured Normal and T-distribution) exhibited essentially dimension-independent effective sample size calculations, despite initialization with a poor covariance structure ($\boldsymbol{\Sigma} = 2\mathbf{I}_P$). Without adaptation, we can see that the sampling performance decreased substantially as the dimension of the target distribution increased. Specifically, we can see that the effective sample size per iteration was roughly two orders of magnitude smaller at $P = 500$ compared to $P= 10$, indicating that we would need to run the Markov chain for substantially more iterations to achieve a comparable effective sample size. Overall, these findings suggest that rapid mixing with the elliptical slice sampler is probably attainable for a far broader range of target distributions than just Gaussian ones, although establishing theoretical bounds on mixing times for this broader class of targets remains an open research problem.

\section{Properties of the Shrinkage Algorithm}
\label{sec: shrinkage}
As discussed in the main manuscript, the Shrinkage Algorithm \citep{hasenpflug2025reversibility} plays a pivotal role in the adaptive generalized elliptical slice sampler. The Shrinkage Algorithm (Algorithm \ref{alg: shrinkage}) provides a transition scheme within some set $S \in \mathcal{B}([0, 2\pi))$; starting with some current angle $\theta_{\text{in}} \in S$ and returning a new angle $\theta_{\text{out}} \in S$.

\begin{algorithm}
\caption{Shrinkage Algorithm \citep{hasenpflug2025reversibility} -- shrink$(\theta_{in}, S)$}
\label{alg: shrinkage}
\hspace*{\algorithmicindent} \textbf{Input}: a current state $\theta_{\text{in}} \in S$ and a set $S \in \mathcal{B}([0,2\pi))$\\
\hspace*{\algorithmicindent} \textbf{Output}: a next state $\theta_{\text{out}} \in S$
\begin{algorithmic}
\State $i \gets 1$
\State $\theta_i \sim \mathcal{U}_{[0, 2\pi)}$
\State $\theta_i^{\text{min}} \gets \theta_i$
\State $\theta_i^{\text{max}} \gets \theta_i$
\While{$\theta_i \not\in S$}
\vspace{-1.5em}
    \If{$\theta_i \in J(\theta_i^{\text{min}}, \theta_{\text{in}})$}  \Comment{$\begin{gathered}
       J(\alpha, \beta) := \begin{cases} 
      [0,\beta) \cup [\alpha, 2\pi) & \alpha > \beta \\
      [\alpha, \beta) & \alpha < \beta  \\
      \varnothing & \alpha =  \beta 
   \end{cases}
    \end{gathered}$}
    \vspace{-1.5em}
    \State $\theta_{i+1}^{\text{min}} \gets \theta_i$
    \State $\theta_{i+1}^{\text{max}} \gets \theta_{i}^{\text{max}}$
    \Else
    \State $\theta_{i+1}^{\text{min}} \gets \theta_{i}^{\text{min}}$
    \State $\theta_{i+1}^{\text{max}} \gets \theta_{i}$
    \EndIf
    \vspace{-1.5em}
    \State $\theta_{i+1} \sim \mathcal{U}_{I(\theta_{i+1}^{\text{min}},\theta_{i+1}^{\text{max}})}$ \Comment{$\begin{gathered}
        I(\alpha, \beta) := \begin{cases} 
      [0,\beta) \cup [\alpha, 2\pi) & \alpha > \beta \\
      [\alpha, \beta) & \alpha < \beta  \\
      [0, 2\pi) & \alpha =  \beta 
   \end{cases}
    \end{gathered}$}
   \vspace{-1.5em}
    \State $i \gets i + 1$
\EndWhile
\State $\theta_{\text{out}} \gets \theta_i$
\end{algorithmic}
\end{algorithm}

We proceed by reviewing properties of the Shrinkage Algorithm (Algorithm \ref{alg: shrinkage}), which will play an important role in showing that the procedure is well-specified and that each transition kernel in the family of transition kernels has $\mu(\cdot)$ as its stationary distribution.  
\begin{definition}[Open on the Circle]
    A set $S \in \mathcal{B}([0,2\pi))$ is called open on the circle if for all $\theta \in S$, there exists a $\epsilon > 0$ such that $(\theta - \epsilon \text{ mod } 2\pi, \theta + \epsilon \text{ mod } 2\pi) \subseteq S.$
    \label{def: open}
\end{definition}
Letting $S \in \mathcal{B}([0, 2\pi))$, $\theta \in S$, and $F \in \mathcal{B}(S)$, we can define the transition kernel of the Shrinkage Algorithm  as $Q_S(\theta, F) := \text{pr}(\theta_{\text{out}} \in F, i < \infty|\theta_{\text{in}} = \theta)$. \cite{hasenpflug2025reversibility} provides the following properties of the Shrinkage Algorithm:

 \begin{property}[Corollary 2.7 \citep{hasenpflug2025reversibility}]
    \label{prop: hasenpflung1}
     Let $S \in \mathcal{B}([0, 2\pi))$ be open on the circle and non-empty. Then, for any $\theta \in S$, we have $Q_S(\theta, S) = 1$.
 \end{property}
 
 \begin{property}[Theorem 2.10 \citep{hasenpflug2025reversibility}]
    \label{prop: hasenpflung2}
     If $S \in \mathcal{B}([0,2\pi))$ is open on the circle and non-empty, then the shrinkage kernel $Q_S$ is reversible with respect to the uniform distribution on $S$, $\mathcal{U}_S$. Specifically, for $F,G \in \mathcal{B}([0,2\pi))$, we have
    $\int_G Q_S(\theta,F)\mathcal{U}_S(\text{d}\theta) = \int_F Q_S(\theta,G)\mathcal{U}_S(\text{d}\theta).$
\end{property}

\begin{property}[Lemma 2.12 \cite{hasenpflug2025reversibility}]
    \label{prop: hasenpflung3}
    Let $g_\theta: [0,2\pi) \rightarrow [0,2\pi)$ such that $g_\theta(\alpha) := (\theta - \alpha) \text{ mod } 2\pi$. For $S \in \mathcal{B}([0,2\pi))$ such that $S$ is open on the circle and non-empty, we have
    $Q_{g^{-1}_\theta(S)}(g^{-1}_\theta(\alpha), g^{-1}_\theta(B)) = Q_S(\alpha,B)$ for $\alpha \in S, B \in \mathcal{B}(S).$
\end{property}

As illustrated in Algorithm 2 of the main manuscript, we are primarily interested in studying the properties of the Shrinkage Algorithm when transitioning on the set $p^{-1}_{\mathbf{x}, \mathbf{z}, \boldsymbol{\gamma}}(\mathcal{X}_{\boldsymbol{\gamma}}(y))$, namely the collection of all admissible transition angles given $\mathbf{x}$, $\mathbf{z}$, $\boldsymbol{\gamma}$, and $y$.

\begin{suppproposition}
     \label{prop: open}
    Let $\mathbf{x} \in \mathcal{X}$, $\mathbf{z} \in \mathbb{R}^P$, and $\boldsymbol{\gamma} \in \mathcal{Y}$. Under Assumption \labelcref{assumption2,assumption3}, $p^{-1}_{\mathbf{x}, \mathbf{z}, \boldsymbol{\gamma}}(\mathcal{X}_{\boldsymbol{\gamma}}(y))$ is open on the circle and non-empty for $y \in (0, \mathcal{L}^*(\mathbf{x},\boldsymbol{\mu}_{\boldsymbol{\gamma}}, \boldsymbol{\Sigma}_{\boldsymbol{\gamma}}))$.
\end{suppproposition}

\begin{proof}
Let $y\in (0, \mathcal{L}^*(\mathbf{x},\boldsymbol{\mu}_{\boldsymbol{\gamma}}, \boldsymbol{\Sigma}_{\boldsymbol{\gamma}}))$ and fix $\mathbf{x} \in \mathcal{X}$, $\mathbf{z} \in \mathbb{R}^P$. Since $\mathcal{L}^*(\mathbf{x},\boldsymbol{\mu}_{\boldsymbol{\gamma}}, \boldsymbol{\Sigma}_{\boldsymbol{\gamma}}) > 0$ by Assumption 2, we know that $0 \in p^{-1}_{\mathbf{x}, \mathbf{z}, \boldsymbol{\gamma}}(\mathcal{X}_{\gamma}(y))$, showing that the set is non-empty. Since $\mathcal{L}^*(\mathbf{x},\boldsymbol{\mu}_{\boldsymbol{\gamma}}, \boldsymbol{\Sigma}_{\boldsymbol{\gamma}})$ is lower-semicontinuous, we have that $\mathcal{X}_{\gamma}(y)$ is an open set by definition. Since $p_{\mathbf{x}, \mathbf{z}, \boldsymbol{\gamma}} (\theta)$ is a continuous function, and  $\mathcal{X}_{\gamma}(y)$ is open, we have that $p^{-1}_{\mathbf{x}, \mathbf{z}, \boldsymbol{\gamma}}(\mathcal{X}_{\gamma}(y))$ is open.
\end{proof}

 Applying the results of Property \ref{prop: hasenpflung1} to $p^{-1}_{\mathbf{x}, \mathbf{z}, \boldsymbol{\gamma}}(\mathcal{X}_{\boldsymbol{\gamma}}(y))$, we have for any $\theta \in p^{-1}_{\mathbf{x}, \mathbf{z}, \boldsymbol{\gamma}}(\mathcal{X}_{\gamma}(y))$, $$Q_{p^{-1}_{\mathbf{x}, \mathbf{z}, \boldsymbol{\gamma}}(\mathcal{X}_{\gamma}(y))}(\theta, p^{-1}_{\mathbf{x}, \mathbf{z}, \boldsymbol{\gamma}}(\mathcal{X}_{\gamma}(y)) =1.$$ Thus, the stopping times of the Shrinkage Algorithm, corresponding to the number of iterations of the while-loop of Algorithm \ref{alg: shrinkage}, are almost surely finite; leading to the notion of a well-specified algorithm. Moreover, from Proposition \ref{prop: open} and Property \ref{prop: hasenpflung1}, we have that for any $y \in (0, \mathcal{L}^*(\mathbf{x},\boldsymbol{\mu}_{\boldsymbol{\gamma}}, \boldsymbol{\Sigma}_{\boldsymbol{\gamma}}))$ the set of possible transition angles has positive Lebesgue measure (i.e., $\lambda\left(p^{-1}_{\mathbf{x}, \mathbf{z}, \boldsymbol{\gamma}}(\mathcal{X}_{\gamma}(y)) \right) > 0$). These properties hold for all $\mathbf{x} \in \mathcal{X}$, which will be used to show that the transition kernel satisfies detailed balance.

\section{Proof of Theorems and Propositions}
\label{sec: proofs}
In this section, we provide proofs to all Propositions and Theorems in the main manuscript. As in the main manuscript, we will continue under the following assumptions:
\begin{assumption}[Compact $\mathcal{Y}$]
    \label{assumption1}
    Let $\boldsymbol{\mu}_0,\boldsymbol{\mu}_{\boldsymbol{\gamma}} \in \mathcal{Y}_{\boldsymbol{\mu}}:= \{\boldsymbol{\mu} \in \mathbb{R}^P \mid \lVert\boldsymbol{\mu}\rVert_2 \le R_{\mu}\}$, for some $R_{\mu} > 0$, and $\boldsymbol{\Sigma}_0, \boldsymbol{\Sigma}_{\boldsymbol{\gamma}} \in  \mathcal{Y}_{\boldsymbol{_\Sigma}}:= \left\{\mathbf{A} \in S^P_{++}| \lambda_{\min}(\mathbf{A})\ge k_{\min}, \lambda_{\max}(\mathbf{A}) \le k_{\max}\right\}$ such that $(\boldsymbol{\mu}_{\boldsymbol{\gamma}}, \boldsymbol{\Sigma}_{\boldsymbol{\gamma}}) \in \mathcal{Y}_{\boldsymbol{\mu}} \times \mathcal{Y}_{\boldsymbol{\Sigma}}$, denoted $\boldsymbol{\gamma} \in \mathcal{Y}$.
\end{assumption}
\begin{assumption}[Bounded $\mathcal{L}^*$]
    \label{assumption2}
    $\mathcal{L}^*(\cdot,\boldsymbol{\mu}_{\boldsymbol{\gamma}}, \boldsymbol{\Sigma}_{\boldsymbol{\gamma}})$ is bounded away from $0$ and $\infty$ on any bounded set of $\mathcal{X}$ and all $\boldsymbol{\gamma} \in \mathcal{Y}$.
\end{assumption}

\begin{assumption}[Lower Semi-Continuity of $\mathcal{L}^*$]
    \label{assumption3}
    $\mathcal{L}^*(\cdot,\boldsymbol{\mu}_{\boldsymbol{\gamma}}, \boldsymbol{\Sigma}_{\boldsymbol{\gamma}})$ is lower semi-continuous at every  $\mathbf{x} \in \mathcal{X}$, for all $\boldsymbol{\gamma} \in \mathcal{Y}$.
\end{assumption}

\begin{assumption}[Properties of the Elliptical Distribution]
    \label{assumption4}
    Let $\mathcal{E}$ be an elliptical distribution in the subclass of multivariate Gaussian distributions or symmetric multivariate Pearson type VII distributions \citep{fang2018symmetric}. Thus the continuous functional parameters take one of the following functional forms:
    \allowdisplaybreaks
    \small{
    \begin{align}
        \nonumber \textbf{Multivariate Gaussian:} &\quad g(t) = \exp\left(-0.5t \right) & \\
        \nonumber \textbf{Multivariate Pearson Type VII: } & \quad g(t) = (1 + t/m)^{-M} &  m > 0 , M > P/2,
    \end{align}}
    where $P$ denotes the dimension of the multivariate distribution.
    Let $(\mathbf{X},\mathbf{Z}) \sim \mathcal{E}_{2P}(\tilde{\boldsymbol{\mu}}_{\boldsymbol{\gamma}}, \tilde{\boldsymbol{\Sigma}}_{\boldsymbol{\gamma}}, \tilde{g})$, where $\tilde{\boldsymbol{\mu}}_\gamma = (\boldsymbol{\mu}_{\boldsymbol{\gamma}}, \boldsymbol{\mu}_{\boldsymbol{\gamma}})$, $\tilde{\boldsymbol{\Sigma}}_{\boldsymbol{\gamma}} = I_2 \otimes \boldsymbol{\Sigma}_{\boldsymbol \gamma}$, and $\tilde{g}(t) = (1 + t/m)^{-(M + P/2)}$, such that $\mathbf{X}$ and $\mathbf{Z}$ have marginal distributions $\mathcal{E}_{P}(\boldsymbol{\mu}_{\boldsymbol{\gamma}}, \boldsymbol{\Sigma}_{\boldsymbol{\gamma}}, g)$. Lastly, if $P = 1$ and $\mathcal{E}$ is in the subclass of symmetric multivariate Pearson type VII distributions, let $M > 1$.
\end{assumption}
\begin{assumption}[Elliptical Subcover]
    \label{assumption6}
     If $\mathcal{X}$ is not bounded, then there exists $R \in (0, \infty)$, $\alpha \in (0, 1)$, $\xi \in (0 , \sqrt{\alpha})$, $\psi > 0$, and a positive definite matrix $\mathbf{A}$ such that when $\mathbf{x} \in B_R^C(\boldsymbol{\mu}_0, \mathbf{A}) :=\left\{\mathbf{x}\in \mathcal{X} \mid q_{\mathbf{x}}(\boldsymbol{\mu}_0, \mathbf{A}) \ge R \right\}$, where $q_{\mathbf{x}}(\boldsymbol{\mu}_0, \mathbf{A}) := (\mathbf{x} - \boldsymbol{\mu}_{0})^{\top}\mathbf{A}^{-1}(\mathbf{x} - \boldsymbol{\mu}_{0})$, the following holds: 
    \begin{align}
        \nonumber \textbf{Elliptical Subcover:} & \;\; \left\{\mathbf{y}\in \mathcal{X} \mid q_{\mathbf{y}}(\boldsymbol{\mu}_0, \mathbf{A})<\alpha q_{\mathbf{x}}(\boldsymbol{\mu}_0, \mathbf{A}) \right\} \subseteq\mathcal{X}_{\boldsymbol{\gamma}}(\mathcal{L}^*(\mathbf{x}, \boldsymbol{\mu}_{\boldsymbol{\gamma}}, \boldsymbol{\Sigma}_{\boldsymbol{\gamma}})),\\
        \nonumber \textbf{Diminishing Tails: } & \;\; \frac{\max_{\mathbf{x}\in \{ \mathbf{x} \in \mathcal{X} \mid q_{\mathbf{x}}(\boldsymbol{\mu}_0, \mathbf{A}) = R_2\}}\mathcal{L}^*(\mathbf{x}, \boldsymbol{\mu}_{\boldsymbol{\gamma}}, \boldsymbol{\Sigma}_{\boldsymbol{\gamma}})}{\max_{\mathbf{y}\in \{ \mathbf{y} \in \mathcal{X}\mid q_{\mathbf{y}}(\boldsymbol{\mu}_0, \mathbf{A}) = R_1\}}\mathcal{L}^*(\mathbf{y}, \boldsymbol{\mu}_{\boldsymbol{\gamma}}, \boldsymbol{\Sigma}_{\boldsymbol{\gamma}})} \le \left(1 + \left[R_2 - R_1\right]  \right)^{-1},
    \end{align}
    for all $R_2 \ge R_1 \ge R$ and $\boldsymbol{\gamma} \in \mathcal{Y}$, where $\alpha$ satisfies the following requirements:
   \begin{enumerate}
        \item when $\mathcal{E}$ is in the subclass of multivariate Gaussian distributions, $\alpha > 0.75$.
       \item when $\mathcal{E}$ is in the subclass of symmetric multivariate Pearson type VII distributions, $\alpha$ satisfies the following inequality:
       $$\left( \frac{1}{\sqrt{\alpha}}  - \sqrt{\alpha} \right)(1 - F_{\alpha}(M, \xi, \psi)) \le \frac{F_1(\alpha, M , \xi, \psi)}{2},$$
       where $F_{\alpha}(M,\xi, \psi) := \frac{1}{2\pi} \int_{\Theta_{\alpha}^{\xi}}I_{\frac{g(\alpha, \tilde{\theta}, \xi, \psi)}{1 + g(\alpha, \tilde{\theta}, \xi, \psi)}}(P/2, M - P/2) \text{d}\tilde{\theta}$, $F_1(\alpha, M, \xi, \psi):= \int_{\xi^2}^{\alpha} \frac{1}{2\sqrt{\tilde{\alpha}}} F_{\tilde{\alpha}}(M, \xi, \psi) \text{d}\tilde{\alpha}$, $\Theta_{\alpha}^{\xi} := \left\{\theta \bigm\vert \lvert\cos(\theta)\rvert < \sqrt{\alpha} - \xi\right\}$, and $g(\tilde{\alpha}, \tilde{\theta}, \xi, \psi) := \frac{ \left(\left( \sqrt{\tilde{\alpha}} - \lvert \cos(\tilde{\theta})\rvert\right) - \xi\right)^2}{( 1 + \psi)\left(\frac{k_{\max}\lambda_{\max}(\mathbf{A})}{k_{\min}\lambda_{\min}(\mathbf{A})}\right)\sin^2(\tilde{\theta})}$, 
       where $I_x(\alpha, \beta)$ is the regularized incomplete beta function.
   \end{enumerate}
\end{assumption}



\subsection{Theorem 1}
\label{sec: reversibility}
Before proving Theorem 1 from the main manuscript, we will first prove two lemmas (Lemmas \labelcref{prop: transformation,prop: rev_prop}) that will be helpful in proving that $H_{\boldsymbol{\gamma}}$ is reversible with respect to $\mu$.

\begin{lemma}
\label{prop: transformation}
Let $\mathbf{W} := (\mathbf{X},\mathbf{Y}) \sim  \mathcal{E}_{2P}(\tilde{\boldsymbol{\mu}}_{\boldsymbol{\gamma}}, \tilde{\boldsymbol{\Sigma}}_{\boldsymbol{\gamma}}, \tilde{g})$ where $\tilde{\boldsymbol{\mu}}_\gamma = (\boldsymbol{\mu}_{\boldsymbol{\gamma}}, \boldsymbol{\mu}_{\boldsymbol{\gamma}})$ and $\tilde{\boldsymbol{\Sigma}}_{\boldsymbol{\gamma}} = \begin{bmatrix}
\boldsymbol{\Sigma}_{\boldsymbol{\gamma}} & \mathbf{0}\\
\mathbf{0} & \boldsymbol{\Sigma}_{\boldsymbol{\gamma}}
\end{bmatrix}$, such that $\mathbf{X}$ and $\mathbf{Y}$ have marginal distributions $\mathcal{E}_{P}(\boldsymbol{\mu}_{\boldsymbol{\gamma}}, \boldsymbol{\Sigma}_{\boldsymbol{\gamma}}, g)$. Define the transformation $T^{\theta}_{\boldsymbol{\gamma}}:\mathbb{R}^P \times \mathbb{R}^P \rightarrow \mathbb{R}^P \times \mathbb{R}^P$ such that
 $$T^{\theta}_{\boldsymbol{\gamma}}(\mathbf{x}, \mathbf{y}):= \left([(\mathbf{x} - \boldsymbol{\mu}_{\boldsymbol{\gamma}})\cos\theta + (\boldsymbol{y}- \boldsymbol{\mu}_{\boldsymbol{\gamma}})\sin\theta] + \boldsymbol{\mu}_{\boldsymbol{\gamma}}, \left[(\mathbf{x} - \boldsymbol{\mu}_{\boldsymbol{\gamma}})\sin\theta- (\boldsymbol{y}- \boldsymbol{\mu}_{\boldsymbol{\gamma}})\cos\theta\right] + \boldsymbol{\mu}_{\boldsymbol{\gamma}} \right).$$ 
 Then for a function $f:[0,2\pi)\times \mathbb{R}^P \times \mathbb{R}^P \rightarrow \mathbb{R}$ such that $E_{\mathbf{W}}(f(\theta,\mathbf{X},\mathbf{Y}))$ exists, we have
 $$\int_{\mathbb{R}^P\times \mathbb{R}^P}f(\theta, \mathbf{x},\mathbf{y})\text{pr}(\mathbf{X} \in d\mathbf{x},\mathbf{Y} \in d\mathbf{y}) =\int_{\mathbb{R}^P\times \mathbb{R}^P} f(\theta, T^{\theta}_{\boldsymbol{\gamma}}(\mathbf{x}, \mathbf{y}))\text{pr}(\mathbf{X} \in d\mathbf{x},\mathbf{Y} \in d\mathbf{y}),$$
 or equivalently,
 $$E_{\mathbf{W}}\left[f(\theta, \mathbf{X},\mathbf{Y})\right] = E_{\mathbf{W}}\left[f(\theta, T^{\theta}_{\boldsymbol{\gamma}}(\mathbf{X}, \mathbf{Y}))\right]$$
 for $\theta \in [0,2\pi)$.
\end{lemma}

 \begin{proof}
     Let $\mathbf{W}^\theta = (\mathbf{X}^\theta, \mathbf{Y}^\theta) =  T_{\gamma}^{\theta}(\mathbf{X}, \mathbf{Y})$. Thus we have that
     \begin{align}
         \nonumber\int_{\mathbb{R}^P\times \mathbb{R}^P}f(\theta, T^{\theta}_{\boldsymbol{\gamma}}(\mathbf{x}, \mathbf{y}))\text{pr}(\mathbf{X} \in d\mathbf{x},\mathbf{Y} \in d\mathbf{y}) & = \nonumber\int_{\mathbb{R}^P\times \mathbb{R}^P}f(\theta, \mathbf{x}^\theta, \mathbf{y}^\theta)\text{pr}(\mathbf{X}^\theta \in d\mathbf{x}^\theta,\mathbf{Y}^\theta \in d\mathbf{y}^\theta)\\
         \nonumber & = \int_{\mathbb{R}^P\times \mathbb{R}^P}f(\theta, \mathbf{w}^\theta)\text{pr}(\mathbf{W}^\theta \in d\mathbf{w}^\theta).
     \end{align}
     Thus, we would like to show $\text{pr}(\mathbf{W} \in \mathbf{w}) = \text{pr}(\mathbf{W}^\theta \in \mathbf{w})$ for all $\mathbf{w} \in \mathcal{B}(\mathbb{R}^{2P})$. Notice that $\boldsymbol{W}^\theta = T_{\gamma}^{\theta} \circ \mathbf{W} = \mathbf{R}_\theta(\mathbf{W} - \tilde{\boldsymbol{\mu}}_\gamma) + \tilde{\boldsymbol{\mu}}_\gamma$, where $$\mathbf{R}_\theta = \begin{bmatrix}
         \cos \theta & \sin \theta \\
         \sin \theta & -\cos \theta
     \end{bmatrix} \otimes \mathbf{I}_{P}.$$
     Using Property \ref{prop: affine_transformation}, we have that
     $\mathbf{W} - \tilde{\boldsymbol{\mu}}_\gamma \sim \mathcal{E}_{2P}(\mathbf{0},  \tilde{\boldsymbol{\Sigma}}_{\boldsymbol{\gamma}}, \tilde{g})$, $\mathbf{R}_\theta(\mathbf{W} - \tilde{\boldsymbol{\mu}}_\gamma) \sim \mathcal{E}_{2P}(\mathbf{0},\mathbf{R}_\theta\tilde{\boldsymbol{\Sigma}}_{\boldsymbol{\gamma}}\mathbf{R}_\theta^\intercal = \tilde{\boldsymbol{\Sigma}}_{\boldsymbol{\gamma}}, \tilde{g})$, and finally $\mathbf{W}^\theta :=\mathbf{R}_\theta(\mathbf{W} - \tilde{\boldsymbol{\mu}}_\gamma) + \tilde{\boldsymbol{\mu}}_\gamma \sim \mathcal{E}_{2P}(\tilde{\boldsymbol{\mu}}_\gamma, \tilde{\boldsymbol{\Sigma}}_{\boldsymbol{\gamma}}, \tilde{g})$. 
     Thus we have that $\text{pr}(\mathbf{W} \in \mathbf{w}) = \text{pr}(\mathbf{W}^\theta \in \mathbf{w})$ for all $\mathbf{w} \in \mathcal{B}(\mathbb{R}^{2P})$, giving us 
     $$\int_{\mathbb{R}^P\times \mathbb{R}^P}f(\theta, \mathbf{x},\mathbf{y})\text{pr}(\mathbf{X} \in d\mathbf{x},\mathbf{Y} \in d\mathbf{y}) = \int_{\mathbb{R}^P\times \mathbb{R}^P}f(\theta, T^{\theta}_{\boldsymbol{\gamma}}(\mathbf{x}, \mathbf{y}))\text{pr}(\mathbf{X} \in d\mathbf{x},\mathbf{Y} \in d\mathbf{y}),$$
    for $\theta \in [0,2\pi)$.
 \end{proof}

\begin{lemma}
\label{prop: rev_prop}
    Let $A \in \mathcal{B}([0,2\pi))$. Then we have that
    \begin{footnotesize}
        \begin{align}
    \nonumber &\int_{p^{-1}_{T^{\theta}_{\boldsymbol{\gamma}}(\mathbf{x}, \mathbf{z}), \boldsymbol{\gamma}}(\mathcal{X}_{\boldsymbol{\gamma}}(y) \cap A)} Q_{p^{-1}_{T^{\theta}_{\boldsymbol{\gamma}}(\mathbf{x}, \mathbf{z}), \boldsymbol{\gamma}}(\mathcal{X}_{\boldsymbol{\gamma}}(y))}(0, \text{d}\alpha) = \int_{p^{-1}_{\mathbf{x}, \mathbf{z}, \boldsymbol{\gamma}}(\mathcal{X}_{\boldsymbol{\gamma}}(y) \cap A)} Q_{p^{-1}_{\mathbf{x}, \mathbf{z}, \boldsymbol{\gamma}}(\mathcal{X}_{\boldsymbol{\gamma}}(y))}(\theta, \text{d}\alpha).
    \end{align}
    \end{footnotesize}
    
\end{lemma}
\begin{proof}
    Letting $g_\theta(\alpha):= (\theta - \alpha) \mod 2\pi$, we have
    \begin{align}
    \nonumber &\int_{p^{-1}_{T^{\theta}_{\boldsymbol{\gamma}}(\mathbf{x}, \mathbf{z}), \boldsymbol{\gamma}}(\mathcal{X}_{\boldsymbol{\gamma}}(y) \cap A)} Q_{p^{-1}_{T^{\theta}_{\boldsymbol{\gamma}}(\mathbf{x}, \mathbf{z}), \boldsymbol{\gamma}}(\mathcal{X}_{\boldsymbol{\gamma}}(y))}(0, \text{d}\alpha)\\  &\nonumber = \int_0^{2\pi} \mathbbm{1}_{p^{-1}_{T^{\theta}_{\boldsymbol{\gamma}}(\mathbf{x}, \mathbf{z}), \boldsymbol{\gamma}}(\mathcal{X}_{\boldsymbol{\gamma}}(y) \cap A)}(\alpha)Q_{g_{\theta}^{-1}\left(p^{-1}_{\mathbf{x}, \mathbf{z}, \boldsymbol{\gamma}}(\mathcal{X}_{\boldsymbol{\gamma}}(y))\right)}\left(g^{-1}_{\theta}(\theta), \text{d}\alpha\right) \\
    &\nonumber = \int_0^{2\pi} \mathbbm{1}_{p^{-1}_{\mathbf{x}, \mathbf{z}, \boldsymbol{\gamma}}(\mathcal{X}_{\boldsymbol{\gamma}}(y) \cap A)}(g_{\theta}(\alpha))Q_{g_{\theta}^{-1}\left(p^{-1}_{\mathbf{x}, \mathbf{z}, \boldsymbol{\gamma}}(\mathcal{X}_{\boldsymbol{\gamma}}(y))\right)}\left(g^{-1}_{\theta}(\theta), \text{d}\alpha\right),
    \end{align}
    where we used the fact that (1) $\alpha \in p^{-1}_{T^{\theta}_{\boldsymbol{\gamma}}(\mathbf{x}, \mathbf{z}), \boldsymbol{\gamma}}(B) \iff g_\theta(\alpha) \in p^{-1}_{\mathbf{x}, \mathbf{z}, \boldsymbol{\gamma}}(B)$  and (2)  $g_{\theta}^{-1}\left(p^{-1}_{\mathbf{x}, \mathbf{z}, \boldsymbol{\gamma}}(B)\right) = p^{-1}_{T^{\theta}_{\boldsymbol{\gamma}}(\mathbf{x}, \mathbf{z}), \boldsymbol{\gamma}}(B)$ for $B \in \mathcal{B}(\mathbb{R}^P)$ . Using Property \ref{prop: hasenpflung3}, we have
    \begin{align}
    &\nonumber = \int_0^{2\pi} \mathbbm{1}_{p^{-1}_{\mathbf{x}, \mathbf{z}, \boldsymbol{\gamma}}(\mathcal{X}_{\boldsymbol{\gamma}}(y) \cap A)}(g_{\theta}(\alpha))Q_{g_{\theta}^{-1}\left(p^{-1}_{\mathbf{x}, \mathbf{z}, \boldsymbol{\gamma}}(\mathcal{X}_{\boldsymbol{\gamma}}(y))\right)}\left(g^{-1}_{\theta}(\theta), \text{d}\alpha\right)\\
    \nonumber &= \int_0^{2\pi} \mathbbm{1}_{p^{-1}_{\mathbf{x}, \mathbf{z}, \boldsymbol{\gamma}}(\mathcal{X}_{\boldsymbol{\gamma}}(y) \cap A)}(\alpha)Q_{p^{-1}_{\mathbf{x}, \mathbf{z}, \boldsymbol{\gamma}}(\mathcal{X}_{\boldsymbol{\gamma}}(y))}\left(\theta, \text{d}\alpha\right)\\
  \nonumber & =\int_{p^{-1}_{\mathbf{x}, \mathbf{z}, \boldsymbol{\gamma}}(\mathcal{X}_{\boldsymbol{\gamma}}(y) \cap A)} Q_{p^{-1}_{\mathbf{x}, \mathbf{z}, \boldsymbol{\gamma}}(\mathcal{X}_{\boldsymbol{\gamma}}(y))}(\theta, \text{d}\alpha).
    \end{align}
\end{proof}

\begin{theorem}[Reversibility]
    \label{thm: reversibility}
    Suppose Assumptions \labelcref{assumption1,assumption2,assumption3,assumption4} hold. Then for $A,B \in \mathcal{B}(\mathcal{X})$ and $\boldsymbol{\gamma} \in \mathcal{Y}$, the following holds: $$\int_{B} H_{\boldsymbol{\gamma}}(\mathbf{x}, A) \mu(\text{d}\mathbf{x}) = \int_{A} H_{\boldsymbol{\gamma}}(\mathbf{x}, B) \mu(\text{d}\mathbf{x}).$$
\end{theorem} 

\begin{proof}
    Fix $\boldsymbol{\gamma} \in \mathcal{Y}$. Let $A,B \in \mathcal{B}(\mathcal{X})$ and let $\mathcal{L}^*  := \mathcal{L}^*(\mathbf{x},\boldsymbol{\mu}_{\boldsymbol{\gamma}}, \boldsymbol{\Sigma}_{\boldsymbol{\gamma}})$ and $\mathcal{E}^{\boldsymbol{\gamma}}_{\mathbf{X}}(\text{d}\mathbf{x}) \propto g\left( (\mathbf{x} - \boldsymbol{\mu}_{\boldsymbol{\gamma}})^{\top} \boldsymbol{\Sigma}_{\boldsymbol{\gamma}}^{-1}(\mathbf{x} - \boldsymbol{\mu}_{\boldsymbol{\gamma}})\right)\lambda(d\mathbf{x})$. Using the fact that $\mu(\mathbf{x}) =\frac{1}{Z} \mathcal{E}_P(\mathbf{x}; \boldsymbol{\mu}_{\boldsymbol{\gamma}}, \boldsymbol{\Sigma}_{\boldsymbol{\gamma}}, g) \mathcal{L}^*(\mathbf{x},\boldsymbol{\mu}_{\boldsymbol{\gamma}}, \boldsymbol{\Sigma}_{\boldsymbol{\gamma}})$ (Equation 1 in the main manuscript), we have
    \begin{footnotesize}
    \begin{align}
        &\nonumber\int_{B} H_{\gamma}(\mathbf{x}, A) \mu(\text{d} \mathbf{x})\\
        &\nonumber = \frac{1}{Z}\int_B \frac{1}{\mathcal{L}^*} \int^{\mathcal{L}^*}_0 \int_{\mathbb{R}^P} Q_{p^{-1}_{\mathbf{x}, \mathbf{z}, \boldsymbol{\gamma}}(\mathcal{X}_{\boldsymbol{\gamma}}(y))}\left(0, p^{-1}_{\mathbf{x}, \mathbf{z}, \boldsymbol{\gamma}}\left(\mathcal{X}_{\boldsymbol{\gamma}}(y)\cap A\right)\right)\mathcal{E}^{\boldsymbol{\gamma}}_{\mathbf{Z}|\mathbf{x}}(\text{d}\mathbf{z})\text{d}y\mu(\text{d}\mathbf{x})\\
        &\nonumber = \frac{1}{Z}\int_B  \int^{\mathcal{L}^*}_0 \int_{\mathbb{R}^P} Q_{p^{-1}_{\mathbf{x}, \mathbf{z}, \boldsymbol{\gamma}}(\mathcal{X}_{\boldsymbol{\gamma}}(y))}\left(0, p^{-1}_{\mathbf{x}, \mathbf{z}, \boldsymbol{\gamma}}\left(\mathcal{X}_{\boldsymbol{\gamma}}(y)\cap A\right)\right)\mathcal{E}^{\boldsymbol{\gamma}}_{\mathbf{Z}|\mathbf{x}}(\text{d}\mathbf{z})\text{d}y\mathcal{E}^{\boldsymbol{\gamma}}_{\mathbf{X}}(\text{d}\mathbf{x})\\
        &\nonumber = \frac{1}{Z}\int_{B} \int^{\mathcal{L}^*}_0 \int_{\mathbb{R}^P} \int_{p^{-1}_{\mathbf{x}, \mathbf{z}, \boldsymbol{\gamma}}(\mathcal{X}_{\boldsymbol{\gamma}}(y) \cap A)} Q_{p^{-1}_{\mathbf{x}, \mathbf{z}, \boldsymbol{\gamma}}(\mathcal{X}_{\boldsymbol{\gamma}}(y))}\left(0, \text{d}\alpha \right)\mathcal{E}^{\boldsymbol{\gamma}}_{\mathbf{Z}|\mathbf{x}}(\text{d}\mathbf{z})\text{d}y\mathcal{E}^{\boldsymbol{\gamma}}_{\mathbf{X}}(\text{d}\mathbf{x}).
    \end{align}
    \end{footnotesize}
    From Proposition \ref{prop: open}, we know that $p^{-1}_{\mathbf{x}, \mathbf{z}, \boldsymbol{\gamma}}(\mathcal{X}_{\boldsymbol{\gamma}}(y))$ is open and $\lambda(p^{-1}_{\mathbf{x}, \mathbf{z}, \boldsymbol{\gamma}}(\mathcal{X}_{\boldsymbol{\gamma}}(y))) > 0$, leading to  $\int_{0}^{2\pi}\frac{\mathbbm{1}_{p^{-1}_{\mathbf{x}, \mathbf{z}, \boldsymbol{\gamma}}(\mathcal{X}_{\boldsymbol{\gamma}}(y))}(\theta)}{\lambda\left(p^{-1}_{\mathbf{x}, \mathbf{z}, \boldsymbol{\gamma}}(\mathcal{X}_{\boldsymbol{\gamma}}(y))\right)} \text{d}\theta = 1$. Thus have
    \begin{footnotesize}
    \begin{align}
    & \nonumber \frac{1}{Z}\int_B  \int^{\mathcal{L}^*}_0 \int_{\mathbb{R}^P} \int_{p^{-1}_{\mathbf{x}, \mathbf{z}, \boldsymbol{\gamma}}(\mathcal{X}_{\boldsymbol{\gamma}}(y) \cap A)} Q_{p^{-1}_{\mathbf{x}, \mathbf{z}, \boldsymbol{\gamma}}(\mathcal{X}_{\boldsymbol{\gamma}}(y))}\left(0, \text{d}\alpha \right)\mathcal{E}^{\boldsymbol{\gamma}}_{\mathbf{Z}|\mathbf{x}}(\text{d}\mathbf{z})\text{d}y\mathcal{E}^{\boldsymbol{\gamma}}_{\mathbf{X}}(\text{d}\mathbf{x})\\
        &\nonumber = \frac{1}{Z}\int_{B}  \int^{\mathcal{L}^*}_0 \int_{\mathbb{R}^P} \int_{0}^{2\pi}\frac{\mathbbm{1}_{p^{-1}_{\mathbf{x}, \mathbf{z}, \boldsymbol{\gamma}}(\mathcal{X}_{\boldsymbol{\gamma}}(y))}(\theta)}{\lambda\left(p^{-1}_{\mathbf{x}, \mathbf{z}, \boldsymbol{\gamma}}(\mathcal{X}_{\boldsymbol{\gamma}}(y))\right)}\int_{p^{-1}_{\mathbf{x}, \mathbf{z}, \boldsymbol{\gamma}}(\mathcal{X}_{\boldsymbol{\gamma}}(y) \cap A)} Q_{p^{-1}_{\mathbf{x}, \mathbf{z}, \boldsymbol{\gamma}}(\mathcal{X}_{\boldsymbol{\gamma}}(y))}\left(0, \text{d}\alpha \right)\text{d}\theta\mathcal{E}^{\boldsymbol{\gamma}}_{\mathbf{Z}|\mathbf{x}}(\text{d}\mathbf{z})\text{d}y\mathcal{E}^{\boldsymbol{\gamma}}_{\mathbf{X}}(\text{d}\mathbf{x})\\
        &\nonumber = \frac{1}{Z}  \int^{\infty}_0  \int_{0}^{2\pi}\int_{\mathbb{R}^P}\int_{\mathbb{R}^P}\frac{\mathbbm{1}_{B}(\mathbf{x})\mathbbm{1}_{\mathcal{X}_{\boldsymbol{\gamma}}(y)}(\mathbf{x})\mathbbm{1}_{p^{-1}_{\mathbf{x}, \mathbf{z}, \boldsymbol{\gamma}}(\mathcal{X}_{\boldsymbol{\gamma}}(y))}(\theta)}{\lambda\left(p^{-1}_{\mathbf{x}, \mathbf{z}, \boldsymbol{\gamma}}(\mathcal{X}_{\boldsymbol{\gamma}}(y))\right)}\int_{p^{-1}_{\mathbf{x}, \mathbf{z}, \boldsymbol{\gamma}}(\mathcal{X}_{\boldsymbol{\gamma}}(y) \cap A)}  Q_{p^{-1}_{\mathbf{x}, \mathbf{z}, \boldsymbol{\gamma}}(\mathcal{X}_{\boldsymbol{\gamma}}(y))}\left(0, \text{d}\alpha \right)\mathcal{E}^{\boldsymbol{\gamma}}_{\mathbf{Z}|\mathbf{x}}(\text{d}\mathbf{z})\mathcal{E}^{\boldsymbol{\gamma}}_{\mathbf{X}}(\text{d}\mathbf{x})\text{d}\theta\text{d}y.
    \end{align}
    \end{footnotesize}
    Letting 
    \begin{footnotesize}
        $$f(\theta, \mathbf{x}, \mathbf{z}):= \frac{\mathbbm{1}_{B}(\mathbf{x})\mathbbm{1}_{\mathcal{X}_{\boldsymbol{\gamma}}(y)}(\mathbf{x})\mathbbm{1}_{p^{-1}_{\mathbf{x}, \mathbf{z}, \boldsymbol{\gamma}}(\mathcal{X}_{\boldsymbol{\gamma}}(y))}(\theta)}{\lambda(p^{-1}_{\mathbf{x}, \mathbf{z}, \boldsymbol{\gamma}}(\mathcal{X}_{\boldsymbol{\gamma}}(y))}\int_{p^{-1}_{\mathbf{x}, \mathbf{z}, \boldsymbol{\gamma}}(\mathcal{X}_{\boldsymbol{\gamma}}(y) \cap A)} Q_{p^{-1}_{\mathbf{x}, \mathbf{z}, \boldsymbol{\gamma}}(\mathcal{X}_{\boldsymbol{\gamma}}(y))}\left(0, \text{d}\alpha \right),$$    
    \end{footnotesize}
    and using Lemma \ref{prop: transformation}, we have
        \begin{align}
        &\nonumber \frac{1}{Z}  \int^{\infty}_0  \int_{0}^{2\pi}\int_{\mathbb{R}^P}\int_{\mathbb{R}^P}f(\theta, \mathbf{x}, \mathbf{z})\mathcal{E}^{\boldsymbol{\gamma}}_{\mathbf{Z}|\mathbf{x}}(\text{d}\mathbf{z})\mathcal{E}^{\boldsymbol{\gamma}}_{\mathbf{X}}(\text{d}\mathbf{x})\text{d}\theta\text{d}y \\
        &\nonumber = \frac{1}{Z}  \int^{\infty}_0  \int_{0}^{2\pi}\int_{\mathbb{R}^P}\int_{\mathbb{R}^P}f\left(\theta, T^{\theta}_{\boldsymbol{\gamma}}(\mathbf{x}, \mathbf{z})\right)\mathcal{E}^{\boldsymbol{\gamma}}_{\mathbf{Z}|\mathbf{x}}(\text{d}\mathbf{z})\mathcal{E}^{\boldsymbol{\gamma}}_{\mathbf{X}}(\text{d}\mathbf{x})\text{d}\theta\text{d}y.
    \end{align}

    Letting $g_\theta(\alpha):= (\theta - \alpha) \mod 2\pi$, we will use the following properties:
    \begin{enumerate}
        \item $p_{T^{\theta}_{\boldsymbol{\gamma}}(\mathbf{x}, \mathbf{z}), \boldsymbol{\gamma}}(\alpha) = p_{\mathbf{x}, \mathbf{z}, \boldsymbol{\gamma}}(g_\theta(\alpha))$ for all $\alpha \in [0,2\pi)$,
        \item $\alpha \in p^{-1}_{T^{\theta}_{\boldsymbol{\gamma}}(\mathbf{x}, \mathbf{z}), \boldsymbol{\gamma}}(\mathcal{X}_{\boldsymbol{\gamma}}(y)) \iff g_\theta(\alpha) \in p^{-1}_{\mathbf{x}, \mathbf{z}, \boldsymbol{\gamma}}(\mathcal{X}_{\boldsymbol{\gamma}}(y))$,
        \item $p^{-1}_{T^{\theta}_{\boldsymbol{\gamma}}(\mathbf{x}, \mathbf{z}), \boldsymbol{\gamma}}(\mathcal{X}_{\boldsymbol{\gamma}}(y)) = g^{-1}_{\theta}\left(p^{-1}_{\mathbf{x}, \mathbf{z}, \boldsymbol{\gamma}}(\mathcal{X}_{\boldsymbol{\gamma}}(y)) \right)$,
        \item $\lambda\left( p^{-1}_{T^{\theta}_{\boldsymbol{\gamma}}(\mathbf{x}, \mathbf{z}), \boldsymbol{\gamma}}(\mathcal{X}_{\boldsymbol{\gamma}}(y))\right) = \lambda\left(p^{-1}_{\mathbf{x}, \mathbf{z}, \boldsymbol{\gamma}}(\mathcal{X}_{\boldsymbol{\gamma}}(y))\right)$,
        \item $\mathbbm{1}_{p^{-1}_{T^{\theta}_{\boldsymbol{\gamma}}(\mathbf{x}, \mathbf{z}), \boldsymbol{\gamma}}(\mathcal{X}_{\boldsymbol{\gamma}}(y))}(\theta) = \mathbbm{1}_{p^{-1}_{\mathbf{x}, \mathbf{z}, \boldsymbol{\gamma}}(\mathcal{X}_{\boldsymbol{\gamma}}(y))}(0) = \mathbbm{1}_{\mathcal{X}_{\boldsymbol{\gamma}}(y)}(\mathbf{x})$.
    \end{enumerate}
    Using these properties along with Lemma \ref{prop: rev_prop}, we have 

    \begin{footnotesize}
        \begin{align}
            \nonumber f\left(\theta, T^{\theta}_{\boldsymbol{\gamma}}(\mathbf{x}, \mathbf{z})\right) &= \frac{\mathbbm{1}_{B} \left(p_{\mathbf{x}, \mathbf{z}, \boldsymbol{\gamma}}(\theta)\right)\mathbbm{1}_{\mathcal{X}_{\boldsymbol{\gamma}}(y)}\left(p_{\mathbf{x}, \mathbf{z}, \boldsymbol{\gamma}}(\theta)\right)\mathbbm{1}_{p^{-1}_{T^{\theta}_{\boldsymbol{\gamma}}(\mathbf{x}, \mathbf{z}), \boldsymbol{\gamma}}(\mathcal{X}_{\boldsymbol{\gamma}}(y))}(\theta)}{\lambda\left(p^{-1}_{T^{\theta}_{\boldsymbol{\gamma}}(\mathbf{x}, \mathbf{z}), \boldsymbol{\gamma}}(\mathcal{X}_{\boldsymbol{\gamma}}(y))\right)}\int_{p^{-1}_{\mathbf{x}, \mathbf{z}, \boldsymbol{\gamma}}(\mathcal{X}_{\boldsymbol{\gamma}}(y)\cap A)}Q_{p^{-1}_{\mathbf{x}, \mathbf{z}, \boldsymbol{\gamma}}(\mathcal{X}_{\boldsymbol{\gamma}}(y))}(\theta, \text{d}\alpha)\\
            \nonumber &=  \frac{\mathbbm{1}_{\mathcal{X}_{\boldsymbol{\gamma}}(y) \cap B}\left(p_{\mathbf{x}, \mathbf{z}, \boldsymbol{\gamma}}(\theta)\right)\mathbbm{1}_{\mathcal{X}_{\boldsymbol{\gamma}}(y)}(\mathbf{x})}{\lambda\left(p^{-1}_{\mathbf{x}, \mathbf{z}, \boldsymbol{\gamma}}(\mathcal{X}_{\boldsymbol{\gamma}}(y))\right)}\int_{p^{-1}_{\mathbf{x}, \mathbf{z}, \boldsymbol{\gamma}}(\mathcal{X}_{\boldsymbol{\gamma}}(y)\cap A)}Q_{p^{-1}_{\mathbf{x}, \mathbf{z}, \boldsymbol{\gamma}}(\mathcal{X}_{\boldsymbol{\gamma}}(y))}(\theta, \text{d}\alpha).
        \end{align}
    \end{footnotesize}
    Thus we have
    \begin{footnotesize}
    \begin{align}
         \nonumber & \frac{1}{Z}  \int^{\infty}_0  \int_{0}^{2\pi}\int_{\mathbb{R}^P}\int_{\mathbb{R}^P}f\left(\theta, T^{\theta}_{\boldsymbol{\gamma}}(\mathbf{x}, \mathbf{z})\right)\mathcal{E}^{\boldsymbol{\gamma}}_{\mathbf{Z}|\mathbf{x}}(\text{d}\mathbf{z})\mathcal{E}^{\boldsymbol{\gamma}}_{\mathbf{X}}(\text{d}\mathbf{x})\text{d}\theta\text{d}y \\
         \nonumber & = \frac{1}{Z}  \int^{\infty}_0  \int_{0}^{2\pi}\int_{\mathbb{R}^P}\int_{\mathbb{R}^P}\frac{\mathbbm{1}_{\mathcal{X}_{\boldsymbol{\gamma}}(y) \cap B}\left(p_{\mathbf{x}, \mathbf{z}, \boldsymbol{\gamma}}(\theta)\right)\mathbbm{1}_{\mathcal{X}_{\boldsymbol{\gamma}}(y)}(\mathbf{x})}{\lambda\left(p^{-1}_{\mathbf{x}, \mathbf{z}, \boldsymbol{\gamma}}(\mathcal{X}_{\boldsymbol{\gamma}}(y))\right)}Q_{p^{-1}_{\mathbf{x}, \mathbf{z}, \boldsymbol{\gamma}}(\mathcal{X}_{\boldsymbol{\gamma}}(y))}\left(\theta, p^{-1}_{\mathbf{x}, \mathbf{z}, \boldsymbol{\gamma}}\left(\mathcal{X}_{\boldsymbol{\gamma}}(y)\cap A\right)\right)\mathcal{E}^{\boldsymbol{\gamma}}_{\mathbf{Z}|\mathbf{x}}(\text{d}\mathbf{z})\mathcal{E}^{\boldsymbol{\gamma}}_{\mathbf{X}}(\text{d}\mathbf{x})\text{d}\theta \text{d}y\\
         \nonumber & = \frac{1}{Z}  \int^{\infty}_0  \int_{\mathbb{R}^P}\int_{\mathbb{R}^P}\int_{0}^{2\pi}\frac{\mathbbm{1}_{p^{-1}_{\mathbf{x}, \mathbf{z}, \boldsymbol{\gamma}}(\mathcal{X}_{\boldsymbol{\gamma}}(y) \cap B)}\left(\theta\right)\mathbbm{1}_{\mathcal{X}_{\boldsymbol{\gamma}}(y)}(\mathbf{x})}{\lambda\left(p^{-1}_{\mathbf{x}, \mathbf{z}, \boldsymbol{\gamma}}(\mathcal{X}_{\boldsymbol{\gamma}}(y))\right)}Q_{p^{-1}_{\mathbf{x}, \mathbf{z}, \boldsymbol{\gamma}}(\mathcal{X}_{\boldsymbol{\gamma}}(y))}\left(\theta, p^{-1}_{\mathbf{x}, \mathbf{z}, \boldsymbol{\gamma}}\left(\mathcal{X}_{\boldsymbol{\gamma}}(y)\cap A\right)\right)\text{d}\theta \mathcal{E}^{\boldsymbol{\gamma}}_{\mathbf{Z}|\mathbf{x}}(\text{d}\mathbf{z})\mathcal{E}^{\boldsymbol{\gamma}}_{\mathbf{X}}(\text{d}\mathbf{x}) \text{d}y\\
         & = \nonumber \frac{1}{Z}  \int^{\infty}_0  \int_{\mathbb{R}^P}\int_{\mathbb{R}^P} \mathbbm{1}_{\mathcal{X}_{\boldsymbol{\gamma}}(y)}(\mathbf{x})\int_{p^{-1}_{\mathbf{x}, \mathbf{z}, \boldsymbol{\gamma}}(\mathcal{X}_{\boldsymbol{\gamma}}(y) \cap B)}Q_{p^{-1}_{\mathbf{x}, \mathbf{z}, \boldsymbol{\gamma}}(\mathcal{X}_{\boldsymbol{\gamma}}(y))}\left(\theta, p^{-1}_{\mathbf{x}, \mathbf{z}, \boldsymbol{\gamma}}\left(\mathcal{X}_{\boldsymbol{\gamma}}(y)\cap A\right)\right)\mathcal{U}_{p^{-1}_{\mathbf{x}, \mathbf{z}, \boldsymbol{\gamma}}(\mathcal{X}_{\boldsymbol{\gamma}}(y))}(\text{d}\theta)\mathcal{E}^{\boldsymbol{\gamma}}_{\mathbf{Z}|\mathbf{x}}(\text{d}\mathbf{z})\mathcal{E}^{\boldsymbol{\gamma}}_{\mathbf{X}}(\text{d}\mathbf{x}) \text{d}y.
    \end{align}
    \end{footnotesize}
    Letting 
    $$\tilde{T}(A, B):=\int_{p^{-1}_{\mathbf{x}, \mathbf{z}, \boldsymbol{\gamma}}(\mathcal{X}_{\boldsymbol{\gamma}}(y) \cap B)}Q_{p^{-1}_{\mathbf{x}, \mathbf{z}, \boldsymbol{\gamma}}(\mathcal{X}_{\boldsymbol{\gamma}}(y))}\left(\theta, p^{-1}_{\mathbf{x}, \mathbf{z}, \boldsymbol{\gamma}}\left(\mathcal{X}_{\boldsymbol{\gamma}}(y)\cap A\right)\right)\mathcal{U}_{p^{-1}_{\mathbf{x}, \mathbf{z}, \boldsymbol{\gamma}}(\mathcal{X}_{\boldsymbol{\gamma}}(y))}(\text{d}\theta),$$ we have that for $A, B \in \mathcal{B}(\mathcal{X})$, we have
    \begin{equation}
        \nonumber\int_{B} H_{\gamma}(\mathbf{x}, A) \mu(\text{d} x) = \frac{1}{Z}  \int^{\infty}_0  \int_{\mathbb{R}^P}\int_{\mathbb{R}^P} \mathbbm{1}_{\mathcal{X}_{\boldsymbol{\gamma}}(y)}(\mathbf{x}) \tilde{T}(A, B)\mathcal{E}^{\boldsymbol{\gamma}}_{\mathbf{Z}|\mathbf{x}}(\text{d}\mathbf{z})\mathcal{E}^{\boldsymbol{\gamma}}_{\mathbf{X}}(\text{d}\mathbf{x}) \text{d}y
    \end{equation}
    Using Property \ref{prop: hasenpflung2} with Proposition \ref{prop: open}, we have that $\tilde{T}(A, B) = \tilde{T}(B,A)$. Thus we have 
    \begin{align}
    \nonumber \int_{B} H_{\gamma}(\mathbf{x}, A) \mu(\text{d} x) &= \frac{1}{Z}  \int^{\infty}_0  \int_{\mathbb{R}^P}\int_{\mathbb{R}^P} \mathbbm{1}_{\mathcal{X}_{\boldsymbol{\gamma}}(y)}(\mathbf{x}) \tilde{T}(A, B)\mathcal{E}^{\boldsymbol{\gamma}}_{\mathbf{Z}|\mathbf{x}}(\text{d}\mathbf{z})\mathcal{E}^{\boldsymbol{\gamma}}_{\mathbf{X}}(\text{d}\mathbf{x}) \text{d}y\\
    \nonumber & = \frac{1}{Z}  \int^{\infty}_0  \int_{\mathbb{R}^P}\int_{\mathbb{R}^P} \mathbbm{1}_{\mathcal{X}_{\boldsymbol{\gamma}}(y)}(\mathbf{x}) \tilde{T}(B, A)\mathcal{E}^{\boldsymbol{\gamma}}_{\mathbf{Z}|\mathbf{x}}(\text{d}\mathbf{z})\mathcal{E}^{\boldsymbol{\gamma}}_{\mathbf{X}}(\text{d}\mathbf{x}) \text{d}y\\
    \nonumber & = \int_{A} H_{\gamma}(\mathbf{x}, B) \mu(\text{d} x),
    \end{align}
    showing that $H_{\boldsymbol{\gamma}}$ is reversible with respect to $\mu$.
\end{proof}

\subsection{Proposition 3}
\begin{proposition}
    \label{prop: minorization}
    Suppose Assumptions \labelcref{assumption1,assumption2,assumption3,assumption4} hold. Let $\boldsymbol{\gamma} \in \mathcal{Y}$ and let $C$ be an open and bounded set in $\mathcal{B}(\mathcal{X})$. Then there exists a $\delta > 0$ and a probability measure on $C$, $\nu_{\boldsymbol{\gamma}}(\cdot)$, such that $H_{\boldsymbol{\gamma}}(\mathbf{x}, \cdot) \ge \delta \nu_{\boldsymbol{\gamma}}(\cdot)$ for all $\mathbf{x} \in C$.
\end{proposition}
\begin{proof}
    Let $A \in \mathcal{B}(C)$. Recall that $H_{\boldsymbol{\gamma}}(\mathbf{x}, A)$ is defined as 
    $$H_{\boldsymbol{\gamma}}(\mathbf{x}, A) = \frac{1}{\mathcal{L}^*(\mathbf{x}, \boldsymbol{\mu}_{\boldsymbol{\gamma}}, \boldsymbol{\Sigma}_{\boldsymbol{\gamma}})} \int_0^{\mathcal{L}^*(\mathbf{x}, \boldsymbol{\mu}_{\boldsymbol{\gamma}}, \boldsymbol{\Sigma}_{\boldsymbol{\gamma}})} \int_{\mathbb{R}^P} Q_{p^{-1}_{\mathbf{x}, \mathbf{z}, \boldsymbol{\gamma}}(\mathcal{X}_{\boldsymbol{\gamma}}(y))}(0, p^{-1}_{\mathbf{x}, \mathbf{z}, \boldsymbol{\gamma}}(\mathcal{X}_{\boldsymbol{\gamma}}(y) \cap A))\mathcal{E}_{\mathbf{Z}|\mathbf{x}}(\text{d}\mathbf{z})\text{d}y.$$
    Following the argument used in \citet{natarovskii2021geometric}, given a fixed $y \in (0, \mathcal{L}^*(\mathbf{x}, \boldsymbol{\mu}_{\boldsymbol{\gamma}}, \boldsymbol{\Sigma}_{\boldsymbol{\gamma}}))$ and $\mathbf{z} \in \mathbb{R}^P$ the probability of transitioning from $\mathbf{x}$ to $A$, i.e. $Q_{p^{-1}_{\mathbf{x}, \mathbf{z}, \boldsymbol{\gamma}}(\mathcal{X}_{\boldsymbol{\gamma}}(y))}(0, p^{-1}_{\mathbf{x}, \mathbf{z}, \boldsymbol{\gamma}}(\mathcal{X}_{\boldsymbol{\gamma}}(y) \cap A))$, is greater than or equal to the probability of transitioning from $\mathbf{x}$ to $A$ in the first iteration of the while loop in Algorithm 1 of the main manuscript, i.e. $\frac{1}{2\pi} \int_0^{2\pi} \mathbbm{1}_{A \cap \mathcal{X}_{\boldsymbol{\gamma}}(y)}(p_{\mathbf{x},\mathbf{z},\boldsymbol{\gamma}}(\theta)) \text{d}\theta$. Thus, we have 
    \begin{align}
        \nonumber H_{\boldsymbol{\gamma}}(\mathbf{x}, A) &\ge \frac{1}{2\pi\mathcal{L}^*(\mathbf{x}, \boldsymbol{\mu}_{\boldsymbol{\gamma}}, \boldsymbol{\Sigma}_{\boldsymbol{\gamma}})} \int_0^{\mathcal{L}^*(\mathbf{x}, \boldsymbol{\mu}_{\boldsymbol{\gamma}}, \boldsymbol{\Sigma}_{\boldsymbol{\gamma}})} \int_{\mathbb{R}^P} \int_0^{2\pi} \mathbbm{1}_{A \cap \mathcal{X}_{\boldsymbol{\gamma}}(y)}(p_{\mathbf{x},\mathbf{z},\boldsymbol{\gamma}}(\theta)) \text{d}\theta\mathcal{E}_{\mathbf{Z}|\mathbf{x}}(\text{d}\mathbf{z})\text{d}y\\
        \nonumber & = \frac{1}{2\pi\mathcal{L}^*(\mathbf{x}, \boldsymbol{\mu}_{\boldsymbol{\gamma}}, \boldsymbol{\Sigma}_{\boldsymbol{\gamma}})} \int_0^{\mathcal{L}^*(\mathbf{x}, \boldsymbol{\mu}_{\boldsymbol{\gamma}}, \boldsymbol{\Sigma}_{\boldsymbol{\gamma}})}\int_0^{2\pi} E_{\mathbf{Z} \mid \mathbf{x}}\left[\mathbbm{1}_{A \cap \mathcal{X}_{\boldsymbol{\gamma}}(y)}(p_{\mathbf{x},\mathbf{z},\boldsymbol{\gamma}}(\theta)) \right]\text{d}\theta\text{d}y.
    \end{align}
    Letting $\tilde{\mathbf{X}}:= p_{\mathbf{x},\mathbf{Z}_{\mathbf{x}},\boldsymbol{\gamma}}(\theta)$ (where $\mathbf{Z}_{\mathbf{x}} := \mathbf{Z} \mid \mathbf{X} = \mathbf{x}$), for fixed $\theta \in [0, 2\pi)$, we have $\tilde{\mathbf{X}} \sim \mathcal{E}_P\left(\boldsymbol{\mu}_{\mathbf{x},\boldsymbol{\gamma}, \theta}, \boldsymbol{\Sigma}_{\mathbf{x},\boldsymbol{\gamma}, \theta}, g_{\mathbf{x}, \boldsymbol{\gamma}} \right) := \mathcal{E}_{\mathbf{x},\boldsymbol{\gamma}, \theta}$ where $\boldsymbol{\mu}_{\mathbf{x},\boldsymbol{\gamma}, \theta} := (\mathbf{x} - \boldsymbol{\mu}_{\boldsymbol{\gamma}})\cos \theta + \boldsymbol{\mu}_{\boldsymbol{\gamma}}$ and $\boldsymbol{\Sigma}_{\mathbf{x},\boldsymbol{\gamma}, \theta}:= \sin^2\theta \boldsymbol{\Sigma}_{\mathbf{x, \boldsymbol{\gamma}}}$, where $\boldsymbol{\Sigma}_{\mathbf{x}, \boldsymbol{\gamma}}$ is proportional to the conditional covariance (Equations \labelcref{eq: var_gaussian,eq: var_pearson}) and $ g_{\mathbf{x}, \boldsymbol{\gamma}}$ is the functional parameter of the corresponding conditional distribution (Property \ref{prop: Conditional_Marginal}). Thus, we have
    \begin{align}
        \nonumber H_{\boldsymbol{\gamma}}(\mathbf{x}, A) &\ge  \frac{1}{2\pi\mathcal{L}^*(\mathbf{x}, \boldsymbol{\mu}_{\boldsymbol{\gamma}}, \boldsymbol{\Sigma}_{\boldsymbol{\gamma}})} \int_0^{\mathcal{L}^*(\mathbf{x}, \boldsymbol{\mu}_{\boldsymbol{\gamma}}, \boldsymbol{\Sigma}_{\boldsymbol{\gamma}})}\int_0^{2\pi} E_{\mathbf{Z} \mid \mathbf{x}}\left[\mathbbm{1}_{A \cap \mathcal{X}_{\boldsymbol{\gamma}}(y)}(p_{\mathbf{x},\mathbf{z},\boldsymbol{\gamma}}(\theta)) \right]\text{d}\theta\text{d}y \\
        \nonumber & = \frac{1}{2\pi\mathcal{L}^*(\mathbf{x}, \boldsymbol{\mu}_{\boldsymbol{\gamma}}, \boldsymbol{\Sigma}_{\boldsymbol{\gamma}})} \int_0^{\mathcal{L}^*(\mathbf{x}, \boldsymbol{\mu}_{\boldsymbol{\gamma}}, \boldsymbol{\Sigma}_{\boldsymbol{\gamma}})}\int_{0}^{2\pi}\int_{\mathbb{R}^P} \mathbbm{1}_{A \cap\mathcal{X}_{\boldsymbol{\gamma}}(y)}(\tilde{\mathbf{x}}) \mathcal{E}_{\mathbf{x},\boldsymbol{\gamma}, \theta}(\text{d}\tilde{\mathbf{x}})\text{d}\theta\text{d}y \\
        \nonumber & = \frac{1}{2\pi\mathcal{L}^*(\mathbf{x}, \boldsymbol{\mu}_{\boldsymbol{\gamma}}, \boldsymbol{\Sigma}_{\boldsymbol{\gamma}})} \int_0^{\mathcal{L}^*(\mathbf{x}, \boldsymbol{\mu}_{\boldsymbol{\gamma}}, \boldsymbol{\Sigma}_{\boldsymbol{\gamma}})}\int_{0}^{2\pi}\int_{A} \mathbbm{1}_{\mathcal{X}_{\boldsymbol{\gamma}}(y)}(\tilde{\mathbf{x}}) \mathcal{E}_{\mathbf{x},\boldsymbol{\gamma}, \theta}(\text{d}\tilde{\mathbf{x}})\text{d}\theta\text{d}y.
    \end{align}
    Using the fact that $\mathbbm{1}_{\mathcal{X}_{\boldsymbol{\gamma}}(y)}(\tilde{\mathbf{x}}) = \mathbbm{1}_{(0, \mathcal{L}^*(\tilde{\mathbf{x}}, \boldsymbol{\mu}_{\boldsymbol{\gamma}}, \boldsymbol{\Sigma}_{\boldsymbol{\gamma}}))}(y)$, we have
    \begin{align}
        \nonumber H_{\boldsymbol{\gamma}}(\mathbf{x}, A) &\ge   \frac{1}{2\pi\mathcal{L}^*(\mathbf{x}, \boldsymbol{\mu}_{\boldsymbol{\gamma}}, \boldsymbol{\Sigma}_{\boldsymbol{\gamma}})} \int_0^{\mathcal{L}^*(\mathbf{x}, \boldsymbol{\mu}_{\boldsymbol{\gamma}}, \boldsymbol{\Sigma}_{\boldsymbol{\gamma}})}\int_{0}^{2\pi}\int_{A} \mathbbm{1}_{ \mathcal{X}_{\boldsymbol{\gamma}}(y)}(\tilde{\mathbf{x}}) \mathcal{E}_{\mathbf{x},\boldsymbol{\gamma}, \theta}(\text{d}\tilde{\mathbf{x}})\text{d}\theta\text{d}y \\
        \nonumber & = \frac{1}{2\pi\mathcal{L}^*(\mathbf{x}, \boldsymbol{\mu}_{\boldsymbol{\gamma}}, \boldsymbol{\Sigma}_{\boldsymbol{\gamma}})} \int_{0}^{2\pi}\int_{A} \int_0^{\mathcal{L}^*(\mathbf{x}, \boldsymbol{\mu}_{\boldsymbol{\gamma}}, \boldsymbol{\Sigma}_{\boldsymbol{\gamma}})}\mathbbm{1}_{(0, \mathcal{L}^*(\tilde{\mathbf{x}}, \boldsymbol{\mu}_{\boldsymbol{\gamma}}, \boldsymbol{\Sigma}_{\boldsymbol{\gamma}}))}(y) \text{d}y\mathcal{E}_{\mathbf{x},\boldsymbol{\gamma}, \theta}(\text{d}\tilde{\mathbf{x}})\text{d}\theta.
    \end{align}

    Noticing that $\int_0^{\mathcal{L}^*(\mathbf{x}, \boldsymbol{\mu}_{\boldsymbol{\gamma}}, \boldsymbol{\Sigma}_{\boldsymbol{\gamma}})}\mathbbm{1}_{(0, \mathcal{L}^*(\tilde{\mathbf{x}}, \boldsymbol{\mu}_{\boldsymbol{\gamma}}, \boldsymbol{\Sigma}_{\boldsymbol{\gamma}}))}(y) \text{d}y = \min\left(\mathcal{L}^*(\tilde{\mathbf{x}}, \boldsymbol{\mu}_{\boldsymbol{\gamma}}, \boldsymbol{\Sigma}_{\boldsymbol{\gamma}}), \mathcal{L}^*(\mathbf{x}, \boldsymbol{\mu}_{\boldsymbol{\gamma}}, \boldsymbol{\Sigma}_{\boldsymbol{\gamma}}) \right)$, and rearranging terms, we have
    $$H_{\boldsymbol{\gamma}}(\mathbf{x}, A) \ge \frac{1}{2\pi}\int_{0}^{2\pi}\int_{A}\min\left(\frac{\mathcal{L}^*(\tilde{\mathbf{x}}, \boldsymbol{\mu}_{\boldsymbol{\gamma}}, \boldsymbol{\Sigma}_{\boldsymbol{\gamma}})}{\mathcal{L}^*(\mathbf{x}, \boldsymbol{\mu}_{\boldsymbol{\gamma}}, \boldsymbol{\Sigma}_{\boldsymbol{\gamma}})}, 1 \right)\mathcal{E}_{\mathbf{x},\boldsymbol{\gamma}, \theta}(\text{d}\tilde{\mathbf{x}})\text{d}\theta.$$
    Since $$\min\left(\frac{\mathcal{L}^*(\tilde{\mathbf{x}}, \boldsymbol{\mu}_{\boldsymbol{\gamma}}, \boldsymbol{\Sigma}_{\boldsymbol{\gamma}})}{\mathcal{L}^*(\mathbf{x}, \boldsymbol{\mu}_{\boldsymbol{\gamma}}, \boldsymbol{\Sigma}_{\boldsymbol{\gamma}})}, 1 \right) \ge \frac{\inf_{\mathbf{t} \in C}\mathcal{L}^*(\mathbf{t}, \boldsymbol{\mu}_{\boldsymbol{\gamma}}, \boldsymbol{\Sigma}_{\boldsymbol{\gamma}})}{\sup_{\mathbf{t} \in C}\mathcal{L}^*(\mathbf{t}, \boldsymbol{\mu}_{\boldsymbol{\gamma}}, \boldsymbol{\Sigma}_{\boldsymbol{\gamma}})}:= \beta > 0 \quad \forall \tilde{\mathbf{x}} \in A$$
    where $\beta > 0$ by Assumption \ref{assumption2} in the main manuscript, we have that  
    $$H_{\boldsymbol{\gamma}}(\mathbf{x}, A) \ge \frac{\beta}{2\pi}\int_{0}^{2\pi}\int_{A}\mathcal{E}_{\mathbf{x},\boldsymbol{\gamma}, \theta}(\text{d}\tilde{\mathbf{x}})\text{d}\theta.$$
    Recall that $\int_{A}\mathcal{E}_{\mathbf{x}, \theta, \boldsymbol{\gamma}}(\text{d}\tilde{\mathbf{x}}) = \int_A k g_{\mathbf{x}, \boldsymbol{\gamma}}\left( (\tilde{\mathbf{x}} - \boldsymbol{\mu}_{\mathbf{x},\boldsymbol{\gamma}, \theta})^{\top} \boldsymbol{\Sigma}_{\mathbf{x},\boldsymbol{\gamma}, \theta}^{-1}(\tilde{\mathbf{x}} - \boldsymbol{\mu}_{\mathbf{x},\boldsymbol{\gamma}, \theta}) \right) \lambda (\text{d}\tilde{\mathbf{x}})$ for some $k > 0$, where $k$ depends on the choice of elliptical distribution, adaptive parameters, and $\mathbf{x}$. We will omit the dependence of $k$ on these values for readability. Let $\theta \in [\frac{\pi}{4}, \frac{\pi}{2}]$ (i.e. $0.5 \le \sin^2\theta \le 1$).
    
     Consider the case where $\mathcal{E}$ is in the family of multivariate Gaussian distributions. Thus, we know that $g_{\mathbf{x}, \boldsymbol{\gamma}}(t) = \exp(-0.5 t)$. Thus, due to the fact that $\mathbf{x}$ lies in a bounded space, and $\boldsymbol{\mu}_{\boldsymbol{\gamma}}$ and $\boldsymbol{\Sigma}_{\boldsymbol{\gamma}}$ lie in compact spaces (Assumption \ref{assumption1}), there exists a $\kappa < \infty$ such that $(\tilde{\mathbf{x}} - \boldsymbol{\mu}_{\mathbf{x}, \theta, \boldsymbol{\gamma}})^{\top} \boldsymbol{\Sigma}_{\theta, \boldsymbol{\gamma}}^{-1}(\tilde{\mathbf{x}} - \boldsymbol{\mu}_{\mathbf{x}, \theta, \boldsymbol{\gamma}}) \le \kappa$ for all $\tilde{\mathbf{x}} \in C$ and $\boldsymbol{\gamma} \in \mathcal{Y}$. Thus, there exists a $g^* = k\exp(-0.5\kappa) > 0$ such that $g^* \le k g_{\mathbf{x}, \boldsymbol{\gamma}}\left( (\tilde{\mathbf{x}} - \boldsymbol{\mu}_{\mathbf{x},\boldsymbol{\gamma}, \theta})^{\top} \boldsymbol{\Sigma}_{\mathbf{x},\boldsymbol{\gamma}, \theta}^{-1}(\tilde{\mathbf{x}} - \boldsymbol{\mu}_{\mathbf{x},\boldsymbol{\gamma}, \theta}) \right)$ for all $\tilde{\mathbf{x}} \in C$ and $\boldsymbol{\gamma} \in \mathcal{Y}$. 
    
    Consider the case where $\mathcal{E}$ is in the family of symmetric multivariate Pearson type VII distributions. From Property \ref{prop: Conditional_Marginal}, we have that $g_{\mathbf{x}, \boldsymbol{\gamma}}(t) = g\left(t + (\mathbf{x} - \boldsymbol{\mu}_{\boldsymbol{\gamma}})^{\top}\boldsymbol{\Sigma}_{\boldsymbol{\gamma}}^{-1}(\mathbf{x} - \boldsymbol{\mu}_{\boldsymbol{\gamma}})\right) = \left( 1 + \frac{t + (\mathbf{x} - \boldsymbol{\mu}_{\boldsymbol{\gamma}})^{\top}\boldsymbol{\Sigma}_{\boldsymbol{\gamma}}^{-1}(\mathbf{x} - \boldsymbol{\mu}_{\boldsymbol{\gamma}})}{m}\right)^{-M}$. Similarly to the Gaussian case, we have that there exists a $\kappa < \infty$ such that $(\tilde{\mathbf{x}} - \boldsymbol{\mu}_{\mathbf{x}, \theta, \boldsymbol{\gamma}})^{\top} \boldsymbol{\Sigma}_{\theta, \boldsymbol{\gamma}}^{-1}(\tilde{\mathbf{x}} - \boldsymbol{\mu}_{\mathbf{x}, \theta, \boldsymbol{\gamma}}) \le \kappa < \infty$ for all $\tilde{\mathbf{x}} \in C$. Using the same reasoning, namely that  $\mathbf{x}$ lies in a bounded space ($\mathbf{x} \in C$), and $\boldsymbol{\mu}_{\boldsymbol{\gamma}}$ and $\boldsymbol{\Sigma}_{\boldsymbol{\gamma}}$ lie in compact spaces (Assumption \ref{assumption1}), there exists $\kappa_{\mathbf{x}}$ such that $(\mathbf{x} - \boldsymbol{\mu}_{\boldsymbol{\gamma}})^{\top}\boldsymbol{\Sigma}_{\boldsymbol{\gamma}}^{-1}(\mathbf{x} - \boldsymbol{\mu}_{\boldsymbol{\gamma}}) < \kappa_{\mathbf{x}}< \infty$ for all $\mathbf{x} \in C$ and $\boldsymbol{\gamma} \in \mathcal{Y}$. Thus, there exists $g^* = k\left( 1 + \frac{\kappa + \kappa_{\mathbf{x}}}{m}\right)^{-M} > 0$ such that  $g^* \le k g_{\mathbf{x}, \boldsymbol{\gamma}}\left( (\tilde{\mathbf{x}} - \boldsymbol{\mu}_{\mathbf{x},\boldsymbol{\gamma}, \theta})^{\top} \boldsymbol{\Sigma}_{\mathbf{x},\boldsymbol{\gamma}, \theta}^{-1}(\tilde{\mathbf{x}} - \boldsymbol{\mu}_{\mathbf{x},\boldsymbol{\gamma}, \theta}) \right)$ for all $\tilde{\mathbf{x}}, \mathbf{x} \in C$ and $\boldsymbol{\gamma} \in \mathcal{Y}$.
    
    Using this, we have
    $$\int_A k g_{\mathbf{x}, \boldsymbol{\gamma}}\left( (\tilde{\mathbf{x}} - \boldsymbol{\mu}_{\mathbf{x},\boldsymbol{\gamma}, \theta})^{\top} \boldsymbol{\Sigma}_{\mathbf{x},\boldsymbol{\gamma}, \theta}^{-1}(\tilde{\mathbf{x}} - \boldsymbol{\mu}_{\mathbf{x},\boldsymbol{\gamma}, \theta}) \right) \lambda (\text{d}\tilde{\mathbf{x}})\ge g^* \lambda(A) \quad \text{for } \theta \in\left[\frac{\pi}{4}, \frac{\pi}{2}\right].$$

    Therefore, we have 
    \begin{align}
        \nonumber H_{\boldsymbol{\gamma}}(\mathbf{x}, A) &\ge \frac{\beta}{2\pi}\int_{0}^{2\pi}\int_{A}\mathcal{E}_{\mathbf{x},\boldsymbol{\gamma}, \theta}(\text{d}\tilde{\mathbf{x}})\text{d}\theta \\
        \nonumber &\ge \frac{\beta}{2\pi}\int_{\pi / 4}^{\pi / 2}\int_{A}\mathcal{E}_{\mathbf{x},\boldsymbol{\gamma}, \theta}(\text{d}\tilde{\mathbf{x}})\text{d}\theta\\
        \label{eq: pos_measure} & \ge \frac{\beta g^*\lambda(C)}{8} \times \frac{\lambda(A)}{\lambda(C)} := \delta \nu_{\boldsymbol{\gamma}}(A),
    \end{align}
    where $\delta := \frac{\beta g^*\lambda(C)}{8} > 0$ and $\nu_{\boldsymbol{\gamma}}(A) :=  \frac{\lambda(A)}{\lambda(C)}$ is a probability measure on $C$.
\end{proof}

\subsection{Proposition 4}
\begin{proposition}
    \label{prop: drift}
    Suppose Assumptions \labelcref{assumption1,assumption2,assumption3,assumption4,assumption6} hold. Define $V(\mathbf{x}):= 1 + \left[\left(\mathbf{x} - \boldsymbol{\mu}_0 \right)^{\top} \mathbf{A}^{-1}\left(\mathbf{x} - \boldsymbol{\mu}_0 \right)\right]^{1/2}$ where $\mathbf{A}$ is defined as in Assumption \ref{assumption6} if $\mathcal{X}$ is not bounded, otherwise let $\mathbf{A} = \mathbf{I}$.  Then, there exist $\phi < 1$, $b < \infty$, and a set $C= B_{\tilde{R}}(\boldsymbol{\mu}_0, \mathbf{A}) := \left\{\mathbf{x}\in \mathcal{X} \mid q_{\mathbf{x}}(\boldsymbol{\mu}_0, \mathbf{A}) < \tilde{R} \right\}$ ($\tilde{R} > 0$), such that $H_{\boldsymbol{\gamma}}V(\mathbf{x}) \le \phi V(\mathbf{x}) + b \mathbbm{1}_{C}(\mathbf{x})$ for all $\mathbf{x} \in \mathcal{X}$.
\end{proposition}

\begin{proof}
    By Assumption \ref{assumption6}, if $\mathcal{X}$ is not bounded, there exists $R \in (0, \infty)$, $\alpha \in (0, 1)$, $\xi \in (0 , \sqrt{\alpha})$, $\psi > 0$, and a positive definite matrix $\mathbf{A}$ such that when $\mathbf{x} \in B_R^C(\boldsymbol{\mu}_0, \mathbf{A}) :=\left\{\mathbf{x} \in \mathcal{X} \mid q_{\mathbf{x}}(\boldsymbol{\mu}_0, \mathbf{A}) \ge R \right\}$, where $q_{\mathbf{x}}(\boldsymbol{\mu}_0, \mathbf{A}) := (\mathbf{x} - \boldsymbol{\mu}_{0})^{\top}\mathbf{A}^{-1}(\mathbf{x} - \boldsymbol{\mu}_{0})$, the following holds: 
    \begin{align}
        \nonumber \textbf{Elliptical Subcover:} & \quad \left\{\mathbf{y}\in \mathcal{X} \mid q_{\mathbf{y}}(\boldsymbol{\mu}_0, \mathbf{A}) < \alpha q_{\mathbf{x}}(\boldsymbol{\mu}_0, \mathbf{A}) \right\} \subseteq\mathcal{X}_{\boldsymbol{\gamma}}(\mathcal{L}^*(\mathbf{x}, \boldsymbol{\mu}_{\boldsymbol{\gamma}}, \boldsymbol{\Sigma}_{\boldsymbol{\gamma}})),\\
        \nonumber \textbf{Diminishing Tails: } & \;\; \frac{\max_{\mathbf{x}\in \{ \mathbf{x} \in \mathcal{X} \mid q_{\mathbf{x}}(\boldsymbol{\mu}_0, \mathbf{A}) = R_2\}}\mathcal{L}^*(\mathbf{x}, \boldsymbol{\mu}_{\boldsymbol{\gamma}}, \boldsymbol{\Sigma}_{\boldsymbol{\gamma}})}{\max_{\mathbf{y}\in \{ \mathbf{y} \in \mathcal{X}\mid q_{\mathbf{y}}(\boldsymbol{\mu}_0, \mathbf{A}) = R_1\}}\mathcal{L}^*(\mathbf{y}, \boldsymbol{\mu}_{\boldsymbol{\gamma}}, \boldsymbol{\Sigma}_{\boldsymbol{\gamma}})} \le \left(1 + \left[R_2 - R_1\right]  \right)^{-1},
    \end{align}
    for all $R_2 \ge R_1 \ge R$ and $\boldsymbol{\gamma} \in \mathcal{Y}$. Throughout this proof, we will use the following useful properties. For $\mathbf{x}, \mathbf{z}, \boldsymbol{\mu}_{0}, \boldsymbol{\mu}_{\boldsymbol{\gamma}} \in \mathbb{R}^{P}$ and some positive definite $P \times P$ matrices $\mathbf{A}, \mathbf{B}$, the following inequalities hold:
    \begin{enumerate}[label=(I\arabic*), ref=(I\arabic*)]
        \item \label{ineq: triangle} $\left[(\mathbf{x} + \mathbf{z})^{\top}\mathbf{A}^{-1} (\mathbf{x} + \mathbf{z})\right]^{1/2} \le \left[\mathbf{x}^{\top}\mathbf{A}^{-1}\mathbf{x}\right]^{1/2} +  \left[\mathbf{z}^{\top}\mathbf{A}^{-1}\mathbf{z}\right]^{1/2},$
        \item \label{ineq: recenter} $\displaystyle\begin{aligned}[t]
        \nonumber \left[(\mathbf{x} - \boldsymbol{\mu}_{0})^{\top} \mathbf{A}^{-1} (\mathbf{x} - \boldsymbol{\mu}_{0})\right]^{1/2} & = \left[(\mathbf{x} - \boldsymbol{\mu}_{0} + \boldsymbol{\mu}_{\boldsymbol{\gamma}} - \boldsymbol{\mu}_{\boldsymbol{\gamma}})^{\top} \mathbf{A}^{-1} (\mathbf{x} - \boldsymbol{\mu}_{0}+ \boldsymbol{\mu}_{\boldsymbol{\gamma}} - \boldsymbol{\mu}_{\boldsymbol{\gamma}})\right]^{1/2}\\
        \nonumber & \le \left[(\mathbf{x} -\boldsymbol{\mu}_{\boldsymbol{\gamma}})^{\top} \mathbf{A}^{-1}(\mathbf{x} -\boldsymbol{\mu}_{\boldsymbol{\gamma}})\right]^{1/2} + \frac{\lVert\boldsymbol{\mu}_{\boldsymbol{\gamma}} - \boldsymbol{\mu}_{0}\rVert_2}{\sqrt{\lambda_{\min}(\mathbf{A})}}.\end{aligned}$
    \end{enumerate}
    Note that the first property can be seen by noticing that $\left[(\mathbf{x} + \mathbf{z})^{\top}\mathbf{A}^{-1} (\mathbf{x} + \mathbf{z})\right]^{1/2} = \lVert\mathbf{A}^{-1/2}(\mathbf{x} + \mathbf{z}) \rVert_2$ and applying the triangle inequality. Define the set $C:= B_{\tilde{R}}(\boldsymbol{\mu}_0, \mathbf{A})$ for some fixed $\tilde{R} > R$ if $\mathcal{X}$ is not bounded, and $C:= \mathcal{X}$ (i.e., $\tilde{R}$ large enough such that $B_{\tilde{R}}(\boldsymbol{\mu}_0, \mathbf{A}) = \mathcal{X}$) if $\mathcal{X}$ is bounded. We will show that the geometric drift condition is satisfied by first considering $\mathbf{x} \in C$ and then $\mathbf{x} \not\in C$, later specifying possible values of $\tilde{R}$.

    \vspace{1em}
    Suppose $\mathbf{x} \in C$. By definition, we have that 
    $$H_{\boldsymbol{\gamma}}V(\mathbf{x}) = \int_{\mathbb{R}^P}(1 + q_{\mathbf{y}}(\boldsymbol{\mu}_0, \mathbf{A})^{1/2})H_{\boldsymbol{\gamma}}(\mathbf{x}, \text{d}\mathbf{y}).$$

    Let $\mathbf{Z}_{\mathbf{x}}:= \mathbf{Z} \mid \mathbf{X} = \mathbf{x} \sim \mathcal{E}_{P}(\boldsymbol{\mu}_{\boldsymbol{\gamma}}, \boldsymbol{\Sigma}_{\boldsymbol{\gamma},\mathbf{x}}, g_{\mathbf{x}, \boldsymbol{\gamma}})$ and $\mathbf{Y} := p_{\mathbf{x}, \mathbf{Z}_{\mathbf{x}}, \boldsymbol{\gamma}} (\theta)$. Thus, using \labelcref{ineq: triangle,ineq: recenter}, we have
    \begin{align}
        \nonumber \left[q_{\mathbf{Y}}(\boldsymbol{\mu}_0, \mathbf{A})\right]^{1/2} & =  \left[q_{\cos(\theta)\mathbf{x} + \sin(\theta) (\mathbf{Z}_{\mathbf{x}} -\boldsymbol{\mu}_{\boldsymbol{\gamma}})  + (1 - \cos(\theta))\boldsymbol{\mu}_{\boldsymbol{\gamma}}}(\boldsymbol{\mu}_0, \mathbf{A})\right]^{1/2}\\ 
        \nonumber & \le \lvert\cos(\theta)\rvert\left[(\mathbf{x} - \boldsymbol{\mu}_0)^{\top} \mathbf{A}^{-1} (\mathbf{x} - \boldsymbol{\mu}_0)\right]^{1/2} + \lvert\sin(\theta)\rvert \left[(\mathbf{Z}_{\mathbf{x}} - \boldsymbol{\mu}_{\boldsymbol{\gamma}})^{\top}\mathbf{A}^{-1}(\mathbf{Z}_{\mathbf{x}} - \boldsymbol{\mu}_{\boldsymbol{\gamma}})\right]^{1/2}  \\
        \nonumber & + \lvert1 - \cos(\theta)\rvert \left[\boldsymbol{\mu}_{\boldsymbol{\gamma}}^{\top} \mathbf{A}^{-1} \boldsymbol{\mu}_{\boldsymbol{\gamma}}\right]^{1/2} + 2 \left[\boldsymbol{\mu}_{0}^{\top} \mathbf{A}^{-1}\boldsymbol{\mu}_{0}\right]^{1/2}\\
        \nonumber & \le \left[q_{\mathbf{x}}(\boldsymbol{\mu}_0, \mathbf{A})\right]^{1/2} +  \left[q_{\mathbf{Z}_{\mathbf{x}}}(\boldsymbol{\mu}_{\boldsymbol{\gamma}}, \mathbf{A})\right]^{1/2} + \frac{2}{\sqrt{\lambda_{\min}(\mathbf{A})}}\lVert \boldsymbol{\mu}_{\boldsymbol{\gamma}}\rVert_2 + \frac{2}{\sqrt{\lambda_{\min}(\mathbf{A})}}\lVert  \boldsymbol{\mu}_0 \rVert_2  \quad \forall \theta \in [0, 2\pi).
    \end{align}
    From Assumption \ref{assumption1}, we have that  $\lVert  \boldsymbol{\mu}_0 \rVert_2 < R_{\mu}$ and $\lVert \boldsymbol{\mu}_{\boldsymbol{\gamma}}\rVert_2 < R_{\mu}$. Thus we have that 
    $$\left[q_{\mathbf{Y}}(\boldsymbol{\mu}_0, \mathbf{A})\right]^{1/2} \le \left[q_{\mathbf{x}}(\boldsymbol{\mu}_0, \mathbf{A})\right]^{1/2} +  \left[q_{\mathbf{Z}_{\mathbf{x}}}(\boldsymbol{\mu}_{\boldsymbol{\gamma}}, \mathbf{A})\right]^{1/2} + \frac{4}{\sqrt{\lambda_{\min}(\mathbf{A})}}R_{\mu} \quad \text{a.s.}$$
    Using this, we have that 
    \begin{align}
        \nonumber H_{\boldsymbol{\gamma}}V(\mathbf{x}) & = \int_{\mathbb{R}^P}(1 + \left[q_{\mathbf{y}}(\boldsymbol{\mu}_0, \mathbf{A})\right]^{1/2})H_{\boldsymbol{\gamma}}(\mathbf{x}, \text{d}\mathbf{y})\\
        \label{eq: bound_x_in_C} & \le 1 + \left[q_{\mathbf{x}}(\boldsymbol{\mu}_0, \mathbf{A})\right]^{1/2}  + \frac{3}{\sqrt{\lambda_{\min}(\mathbf{A})}}R_{\mu} + \int_{\mathbb{R}^P} \left[q_{\mathbf{z}}(\boldsymbol{\mu}_{\boldsymbol{\gamma}}, \mathbf{A})\right]^{1/2} \mathcal{E}_{\mathbf{Z}\mid\mathbf{x}}(\text{d}\mathbf{z}).
    \end{align}

    \begin{align}
        \nonumber E_{\mathcal{E}_{\mathbf{Z}\mid\mathbf{x}}}\left[\left[q_{\mathbf{Z}_{\mathbf{x}}}(\boldsymbol{\mu}_{\boldsymbol{\gamma}}, \mathbf{A}) \right]^{1/2}\right]  & \le E_{\mathcal{E}_{\mathbf{Z}\mid\mathbf{x}}}\left[\frac{1}{\sqrt{\lambda_{\min}(\mathbf{A})}} \lVert \mathbf{Z}_{\mathbf{x}} - \boldsymbol{\mu}_{\boldsymbol{\gamma}} \rVert_2 \right] \\ 
        \nonumber & = \frac{1}{\sqrt{\lambda_{\min}(\mathbf{A})}} E_{\mathcal{E}_{\mathbf{Z}\mid\mathbf{x}}}\left[\text{tr} \left\{(\mathbf{Z}_{\mathbf{x}} - \boldsymbol{\mu}_{\boldsymbol{\gamma}}) (\mathbf{Z}_{\mathbf{x}} - \boldsymbol{\mu}_{\boldsymbol{\gamma}})^{\top}\right\}^{1/2} \right]\\
        \nonumber &\le \frac{1}{\sqrt{\lambda_{\min}(\mathbf{A})} }\left[\text{tr} \left\{E_{\mathcal{E}_{\mathbf{Z}\mid\mathbf{x}}}\left[(\mathbf{Z}_{\mathbf{x}} - \boldsymbol{\mu}_{\boldsymbol{\gamma}}) (\mathbf{Z}_{\mathbf{x}} - \boldsymbol{\mu}_{\boldsymbol{\gamma}})^{\top} \right]\right\}\right]^{1/2}\\
        \label{eq: expected_q_zx} & = \frac{1}{\sqrt{\lambda_{\min}(\mathbf{A})}} \left[\text{tr}\left\{\text{cov}_{\mathcal{E}_{\mathbf{Z}\mid\mathbf{x}}}\left[ \mathbf{Z}_{\mathbf{x}}\right]\right\}\right]^{1/2}
    \end{align}
    From Equations \labelcref{eq: var_gaussian,eq: var_pearson}, we have that 
    \begin{align}
        \nonumber \text{\text{cov}}(\mathbf{Z} \mid \mathbf{X} = \mathbf{x}) & = \boldsymbol{\Sigma}_{\boldsymbol{\gamma}}
    \end{align}
    if $\mathcal{E}$ is in the family of multivariate Gaussian distributions and 
    \begin{align}
        \nonumber \text{cov}(\mathbf{Z} \mid \mathbf{X} = \mathbf{x}) = \left(\frac{m + (\mathbf{x} - \boldsymbol{\mu}_{\boldsymbol{\gamma}})^{\top}\boldsymbol{\Sigma}_{\boldsymbol{\gamma}}^{-1}(\mathbf{x} - \boldsymbol{\mu}_{\boldsymbol{\gamma}})}{2M  - 2}\right)\boldsymbol{\Sigma}_{\boldsymbol{\gamma}}
    \end{align}
    if $\mathcal{E}$ is in the family of symmetric multivariate Pearson type VII distributions.
    Thus, from Assumption \ref{assumption1}, if $\mathcal{E}$ is in the family of multivariate Gaussian distributions, we have 
    \begin{equation}
    \label{eq: K_Sigma_gaussian}
    \text{tr}\left\{\text{cov}_{\mathcal{E}_{\mathbf{Z}\mid\mathbf{x}}}\left[ \mathbf{Z}_{\mathbf{x}}\right]\right\} \le P k_{\max},
    \end{equation}
    and if $\mathcal{E}$ is in the family of symmetric multivariate Pearson type VII distributions, we have
    \begin{align}
        \nonumber \text{tr}\left\{\text{cov}_{\mathcal{E}_{\mathbf{Z}\mid\mathbf{x}}}\left[ \mathbf{Z}_{\mathbf{x}}\right]\right\} & = \text{tr}\left\{ \left(\frac{m + (\mathbf{x} - \boldsymbol{\mu}_{\boldsymbol{\gamma}})^{\top}\boldsymbol{\Sigma}_{\boldsymbol{\gamma}}^{-1}(\mathbf{x} - \boldsymbol{\mu}_{\boldsymbol{\gamma}})}{2M -  2}\right)\boldsymbol{\Sigma}_{\boldsymbol{\gamma}}\right\}\\
        \nonumber & \le \left(\frac{m + \lVert\mathbf{x} - \boldsymbol{\mu}_{\boldsymbol{\gamma}}\rVert_2^2\frac{1}{k_{\min}}}{2M  - 2}\right)Pk_{\max} \\
        \nonumber & \le \left(\frac{m + \left(2\lVert\mathbf{x} - \boldsymbol{\mu}_0\rVert_2^2  + 8 R_{\mu}^2\right)\frac{1}{k_{\min}}}{2M  - 2}\right)Pk_{\max} \\
        \label{eq: K_Sigma_pearson} & \le \left(\frac{m + \left(2\tilde{R}\lambda_{\max}(\mathbf{A})  + 8 R_{\mu}^2\right)\frac{1}{k_{\min}}}{2M  - 2}\right)Pk_{\max}
    \end{align}
    
    Thus we have that $E_{\mathcal{E}_{\mathbf{Z}\mid\mathbf{x}}}\left[\left[q_{\mathbf{Z}_{\mathbf{x}}}(\boldsymbol{\mu}_{\boldsymbol{\gamma}}, \mathbf{A})\right]^{1/2} \right] \le \kappa_{\Sigma} < \infty$, where $\kappa_{\Sigma}$ can be defined by Equations \labelcref{eq: expected_q_zx,eq: K_Sigma_gaussian,eq: K_Sigma_pearson}, depending on the choice of elliptical distribution. Thus, from Equation \ref{eq: bound_x_in_C} and recalling that $\mathbf{x} \in C$, we have that
    \begin{align}
        \nonumber H_{\boldsymbol{\gamma}}V(\mathbf{x}) &\le 1 + \left[q_{\mathbf{x}}(\boldsymbol{\mu}_0, \mathbf{A})\right]^{1/2}  + \frac{4}{\sqrt{\lambda_{\min}(\mathbf{A})}}R_{\mu} + \int_{\mathbb{R}^P} \left[q_{\mathbf{z}}(\boldsymbol{\mu}_{\boldsymbol{\gamma}}, \mathbf{A})\right]^{1/2} \mathcal{E}_{\mathbf{Z}\mid\mathbf{x}}(\text{d}\mathbf{z}) \\
        \nonumber & \le 1 + q_{\mathbf{x}}(\boldsymbol{\mu}_0, \mathbf{A})^{1/2}  + \frac{4}{\sqrt{\lambda_{\min}(\mathbf{A})}}R_{\mu} +\kappa_{\Sigma}\\
        \nonumber & \le \phi V(\mathbf{x})  + (1 - \phi) \left(1 + \tilde{R} \right) + \frac{4}{\sqrt{\lambda_{\min}(\mathbf{A})}}R_{\mu} +\kappa_{\Sigma},
    \end{align}
    for some $\phi \in (0, 1)$ defined later in the proof. Letting
    \begin{equation}
    \label{eq: def_b}
        b:= (1 - \phi) \left(1 + \tilde{R} \right) + \frac{4}{\sqrt{\lambda_{\min}(\mathbf{A})}}R_{\mu} +\kappa_{\Sigma} < \infty,
    \end{equation} gives us the desired result of 
    \begin{equation}
        \label{eq: geometric_drift_C}
        H_{\boldsymbol{\gamma}}V(\mathbf{x}) \le \phi V(\mathbf{x}) + b \mathbbm{1}_{C}(\mathbf{x}) \text{ for all } \mathbf{x} \in C.
    \end{equation}

\begin{remark}[Bounded $\mathcal{X}$]
    If $\mathcal{X}$ is bounded, then $C=\mathcal{X}$ (by construction), and we have shown the desired result of $H_{\boldsymbol{\gamma}}V(\mathbf{x}) \le \phi V(\mathbf{x}) + b \mathbbm{1}_{C}(\mathbf{x})$ for all $\mathbf{x} \in \mathcal{X}$. The remaining part of the proof is for the case when $\mathcal{X}$ is not bounded. 
\end{remark}

\begin{remark}
    Equation \ref{eq: geometric_drift_C} shows that for any $\tilde{R} < \infty$, we can obtain the upper bound on $H_{\boldsymbol{\gamma}}V(\mathbf{x})$ for $\mathbf{x} \in B_{\tilde{R}}(\boldsymbol{\mu}_0, \mathbf{A})$. The goal of the remaining part of the proof is to show that if $\mathbf{x}$ is sufficiently far (i.e., $\mathbf{x} \not\in B_{\tilde{R}}(\boldsymbol{\mu}_0, \mathbf{A})$ for some $\tilde{R}$), the Markov chain (under fixed $\boldsymbol{\gamma} \in \mathcal{Y}$) will \textit{drift} towards values of $\mathbf{x}$ that have lower values of $V(\mathbf{x})$ (i.e., $H_{\boldsymbol{\gamma}}V(\mathbf{x}) \le \phi V(\mathbf{x})$). However, achieving tight bounds on $H_{\boldsymbol{\gamma}}V(\mathbf{x})$ is challenging due to the intricate nature of the transition kernel. Thus we will rely on bounding $H_{\boldsymbol{\gamma}}V(\mathbf{x})$ by  (1) lower bounding the probability of transitioning within some ellipse ($B_{\tilde{\alpha}q_{\mathbf{x}}(\boldsymbol{\mu}_0, \mathbf{A})}(\boldsymbol{\mu}_0, \mathbf{A})$ for $\tilde{\alpha} \le \alpha$) on the first iteration of the while-loop of the shrinkage algorithm (Algorithm 2 in the main manuscript) and (2) upper bounding the probability of transitioning outside the covering set ($B^C_{\alpha^*q_{\mathbf{x}}(\boldsymbol{\mu}_0, \mathbf{A})}(\boldsymbol{\mu}_0, \mathbf{A})$ for $\alpha^* \ge \frac{1}{\alpha}$) through the random variable $y$ (the threshold). Although these bounds are not tight, they are sufficient to show that the adaptive algorithm is ergodic.
\end{remark}    
    
    \vspace{1em}
 Suppose $\mathbf{x} \not\in C$. Letting $Z := \left[\left( \mathbf{Y} - \boldsymbol{\mu}_0\right)^{\top} \mathbf{A}^{-1}\left( \mathbf{Y} - \boldsymbol{\mu}_0\right)\right]^{1/2}$ and noticing that $Z$ is a non-negative real-valued random variable, we have $\int_{\mathbb{R}^P}\left[\left( \mathbf{Y} - \boldsymbol{\mu}_0\right)^{\top} \mathbf{A}^{-1}\left( \mathbf{Y} - \boldsymbol{\mu}_0\right)\right]^{1/2}H_{\boldsymbol{\gamma}}(\mathbf{x}, \text{d}\mathbf{y}) = \int_{0}^\infty \text{pr}\{Z \ge t\} \text{d}t$ where $\text{pr}\{Z \ge t\} = 1 - H_{\boldsymbol{\gamma}}(\mathbf{x}, B_{t^2}(\boldsymbol{\mu}_0, \mathbf{A}))$, where $B_{t^2}(\boldsymbol{\mu}_0, \mathbf{A}) := \{\mathbf{x} \mid \left( \mathbf{x} - \boldsymbol{\mu}_0\right)^{\top} \mathbf{A}^{-1}\left( \mathbf{x} - \boldsymbol{\mu}_0\right) < t^2\}$. Thus, we have 
\begin{equation}
    \label{eq: reform_geom_drift}
H_{\boldsymbol{\gamma}}V(\mathbf{x}) = 1 + \int_{0}^{\infty} H_{\boldsymbol{\gamma}}(\mathbf{x}, B_{t^2}^C(\boldsymbol{\mu}_0, \mathbf{A})) \text{d}t.
\end{equation}
Let $q_{\mathbf{x}}(\boldsymbol{\mu}_0, \mathbf{A}):=\left( \mathbf{x} - \boldsymbol{\mu}_0\right)^{\top} \mathbf{A}^{-1}\left( \mathbf{x} - \boldsymbol{\mu}_0\right)$.  Consider $H_{\boldsymbol{\gamma}}(\mathbf{x}, B_{\alpha q_{\mathbf{x}}(\boldsymbol{\mu}_0, \mathbf{A})}(\boldsymbol{\mu}_0, \mathbf{A}))$ where $B_{\alpha q_{\mathbf{x}}(\boldsymbol{\mu}_0, \mathbf{A})}(\boldsymbol{\mu}_0, \mathbf{A}) := \left\{\mathbf{y}\in \mathcal{X} \mid q_{\mathbf{y}}(\boldsymbol{\mu}_0, \mathbf{A})< \alpha q_{\mathbf{x}}(\boldsymbol{\mu}_0, \mathbf{A}) \right\}$. As argued in the proof of Proposition \ref{prop: minorization}, the probability of transitioning from $\mathbf{x}$ to $B_{\alpha q_{\mathbf{x}}(\boldsymbol{\mu}_0, \mathbf{A})}(\boldsymbol{\mu}_0, \mathbf{A})$ is greater than or equal to the probability of transitioning from $\mathbf{x}$ to $B_{\alpha q_{\mathbf{x}}(\boldsymbol{\mu}_0, \mathbf{A})}(\boldsymbol{\mu}_0, \mathbf{A})$ in the first iteration of the while loop in Algorithm 1 of the main manuscript. Since $B_{\alpha q_{\mathbf{x}}(\boldsymbol{\mu}_0, \mathbf{A})}(\boldsymbol{\mu}_0, \mathbf{A}) \subseteq \mathcal{X}_{\boldsymbol{\gamma}}(\mathcal{L}^*(\mathbf{x}, \boldsymbol{\mu}_{\boldsymbol{\gamma}}, \boldsymbol{\Sigma}_{\boldsymbol{\gamma}}))$ by Assumption \ref{assumption6}, we have that we will accept any proposed move into $B_{\alpha q_{\mathbf{x}}(\boldsymbol{\mu}_0, \mathbf{A})}(\boldsymbol{\mu}_0, \mathbf{A})$ with probability 1. Thus, letting $p_{\mathbf{x}, \mathbf{Z}_\mathbf{x}, \boldsymbol{\gamma}}(\tilde{\theta}):= (\mathbf{x} - \boldsymbol{\mu}_{\boldsymbol{\gamma}}) \cos(\tilde{\theta}) + (\mathbf{Z}_{\mathbf{x}} - \boldsymbol{\mu}_{\boldsymbol{\gamma}}) \sin(\tilde{\theta}) + \boldsymbol{\mu}_{\boldsymbol{\gamma}}$ be the proposed first move, where $\tilde{\theta} \sim \mathcal{U}_{[0, 2\pi)}$ and $\mathbf{Z}_{\mathbf{x}} := \mathbf{Z}\mid \mathbf{X} = \mathbf{x} \sim \mathcal{E}_{P}(\boldsymbol{\mu}_{\boldsymbol{\gamma}}, \boldsymbol{\Sigma}_{\boldsymbol{\gamma},\mathbf{x}}, g_{\mathbf{x}, \boldsymbol{\gamma}})$, and using \ref{ineq: recenter} twice, we have 
\begin{align}
\label{eq: first_prop}
\nonumber H_{\boldsymbol{\gamma}}&(\mathbf{x}, B_{\alpha q_{\mathbf{x}}(\boldsymbol{\mu}_0, \mathbf{A})}(\boldsymbol{\mu}_0, \mathbf{A})) \\
&\nonumber \ge  \text{pr}\left(\left[\left( p_{\mathbf{x}, \mathbf{Z}_\mathbf{x}, \boldsymbol{\gamma}}(\tilde{\theta}) - \boldsymbol{\mu}_{0} \right)^{\top}\mathbf{A}^{-1} \left( p_{\mathbf{x}, \mathbf{Z}_\mathbf{x}, \boldsymbol{\gamma}}(\tilde{\theta}) - \boldsymbol{\mu}_{0} \right)\right]^{1/2} \le \left[\alpha q_{\mathbf{x}}(\boldsymbol{\mu}_0, \mathbf{A})\right]^{1/2}\right)\\
\nonumber & \ge \text{pr}\left(\left[\left( p_{\mathbf{x}, \mathbf{Z}_\mathbf{x}, \boldsymbol{\gamma}}(\tilde{\theta}) - \boldsymbol{\mu}_{\boldsymbol{\gamma}} \right)^{\top}\mathbf{A}^{-1} \left( p_{\mathbf{x}, \mathbf{Z}_\mathbf{x}, \boldsymbol{\gamma}}(\tilde{\theta}) - \boldsymbol{\mu}_{\boldsymbol{\gamma}} \right)\right]^{1/2} \le \left[\alpha q_{\mathbf{x}}(\boldsymbol{\mu}_0, \mathbf{A})\right]^{1/2} - \frac{2R_{\mu}}{\sqrt{\lambda_{\min}(\mathbf{A})}}\right)\\
& \ge \text{pr}\left(\left[\left( p_{\mathbf{x}, \mathbf{Z}_\mathbf{x}, \boldsymbol{\gamma}}(\tilde{\theta}) - \boldsymbol{\mu}_{\boldsymbol{\gamma}} \right)^{\top}\mathbf{A}^{-1} \left( p_{\mathbf{x}, \mathbf{Z}_\mathbf{x}, \boldsymbol{\gamma}}(\tilde{\theta}) - \boldsymbol{\mu}_{\boldsymbol{\gamma}} \right)\right]^{1/2} \le \left[\alpha q_{\mathbf{x}}(\boldsymbol{\mu}_{\boldsymbol{\gamma}}, \mathbf{A})\right]^{1/2} - \frac{4R_{\mu}}{\sqrt{\lambda_{\min}(\mathbf{A})}}\right)
\end{align}
where the term on the right is the probability of proposing a move in $B_{\alpha q_{\mathbf{x}}(\boldsymbol{\mu}_0, \mathbf{A})}(\boldsymbol{\mu}_0, \mathbf{A})$ on the first iteration of the while loop. 
Consider the stochastic representation of $p_{\mathbf{x}, \mathbf{Z}_\mathbf{x}, \boldsymbol{\gamma}}(\tilde{\theta})$ for fixed $\tilde{\theta} \in [0, 2\pi)$. Using Equations \ref{eq: stochastic_gaussian} and \ref{eq: stochastic_pearson}, we have

$$p_{\mathbf{x}, \mathbf{Z}_\mathbf{x}, \boldsymbol{\gamma}}(\tilde{\theta}) \overset{d}{=} (\mathbf{x} - \boldsymbol{\mu}_{\boldsymbol{\gamma}}) \cos(\tilde{\theta}) + (\mathcal{R}\boldsymbol{\Lambda} \mathcal{U}^{(P)}) \sin(\tilde{\theta}) + \boldsymbol{\mu}_{\boldsymbol{\gamma}}.$$
Thus, using \ref{ineq: triangle}, we have
{\small \begin{align}
    \nonumber \left[\left( p_{\mathbf{x}, \mathbf{Z}_\mathbf{x}, \boldsymbol{\gamma}}(\tilde{\theta}) - \boldsymbol{\mu}_{\boldsymbol{\gamma}} \right)^{\top} \mathbf{A}^{-1} \left( p_{\mathbf{x}, \mathbf{Z}_\mathbf{x}, \boldsymbol{\gamma}}(\tilde{\theta}) - \boldsymbol{\mu}_{\boldsymbol{\gamma}} \right)\right]^{1/2}  \le & \mathcal{R}  \lvert\sin(\tilde{\theta})\rvert\left[\left(\mathcal{U}^{(P)}\right)^{\top}\boldsymbol{\Lambda}^{\top} \mathbf{A}^{-1}\boldsymbol{\Lambda} \mathcal{U}^{(P)}\right]^{1/2}\\
    \nonumber & + \left[\left((\mathbf{x} - \boldsymbol{\mu}_{\boldsymbol{\gamma}}) \cos(\tilde{\theta}) +  \boldsymbol{\mu}_{\boldsymbol{\gamma}}\right)^{\top} \mathbf{A}^{-1}\left((\mathbf{x} - \boldsymbol{\mu}_{\boldsymbol{\gamma}}) \cos(\tilde{\theta}) +  \boldsymbol{\mu}_{\boldsymbol{\gamma}}\right)\right]^{1/2}\\
    \nonumber \le  & \mathcal{R} \lvert\sin(\tilde{\theta})\rvert\left[\lambda_{\max}\left(\boldsymbol{\Lambda}^{\top} \mathbf{A}^{-1}\boldsymbol{\Lambda}\right)\right]^{1/2}\lVert \mathcal{U}^{(P)}\rVert_2\\
    \nonumber & + \lvert\cos(\tilde{\theta})\rvert \left[(\mathbf{x} - \boldsymbol{\mu}_{\boldsymbol{\gamma}})^{\top}\mathbf{A}^{-1}(\mathbf{x} - \boldsymbol{\mu}_{\boldsymbol{\gamma}})\right]^{1/2} + \left[\boldsymbol{\mu}_{\boldsymbol{\gamma}}^{\top}\mathbf{A}^{-1}\boldsymbol{\mu}_{\boldsymbol{\gamma}}\right]^{1/2}\\
    \nonumber \le & \mathcal{R}\lvert\sin(\tilde{\theta})\rvert\, \left(\frac{k_{\max}}{\lambda_{\min}(\mathbf{A})}\right)^{1/2} +\lvert\cos(\tilde{\theta})\rvert \left[(\mathbf{x} - \boldsymbol{\mu}_{\boldsymbol{\gamma}})^{\top}\mathbf{A}^{-1}(\mathbf{x} - \boldsymbol{\mu}_{\boldsymbol{\gamma}})\right]^{1/2} \\ 
    \nonumber & + \left[\boldsymbol{\mu}_{\boldsymbol{\gamma}}^{\top}\mathbf{A}^{-1}\boldsymbol{\mu}_{\boldsymbol{\gamma}}\right]^{1/2},
\end{align}}
where we use the fact that $\lVert \mathcal{U}^{(P)}\rVert_2 = 1$ and that the eigenvalues of $\boldsymbol{\Lambda}^{\top} \mathbf{A}^{-1}\boldsymbol{\Lambda}$ coincide with those of $\mathbf{A}^{-1}\boldsymbol{\Sigma}_{\boldsymbol{\gamma}}$, so that $\lambda_{\max}\left(\boldsymbol{\Lambda}^{\top} \mathbf{A}^{-1}\boldsymbol{\Lambda}\right) \le \lambda_{\max}(\boldsymbol{\Sigma}_{\boldsymbol{\gamma}})\lambda_{\max}(\mathbf{A}^{-1}) \le \frac{k_{\max}}{\lambda_{\min}(\mathbf{A})}$ by Assumption \ref{assumption1}.

Thus we have 
\begin{align}
    \nonumber \text{pr}\left(\left[q_{p_{\mathbf{x}, \mathbf{Z}_\mathbf{x}, \boldsymbol{\gamma}}(\tilde{\theta})}(\boldsymbol{\mu}_{\boldsymbol{\gamma}}, \mathbf{A})\right]^{1/2} >  t\right)  & \le \text{pr}\left(\mathcal{R} >  \frac{t - \lvert\cos(\tilde{\theta})\rvert \left[(\mathbf{x} - \boldsymbol{\mu}_{\boldsymbol{\gamma}})^{\top}\mathbf{A}^{-1}(\mathbf{x} - \boldsymbol{\mu}_{\boldsymbol{\gamma}})\right]^{1/2} - \left[\boldsymbol{\mu}_{\boldsymbol{\gamma}}^{\top}\mathbf{A}^{-1}\boldsymbol{\mu}_{\boldsymbol{\gamma}}\right]^{1/2}}{\lvert\sin(\tilde{\theta})\rvert\, \sqrt{\frac{k_{\max}}{\lambda_{\min}(\mathbf{A})}}}\right)\\
    \nonumber & =  \text{pr}\left(\mathcal{R}^2 > \frac{\left(t - \lvert\cos(\tilde{\theta})\rvert \left[(\mathbf{x} - \boldsymbol{\mu}_{\boldsymbol{\gamma}})^{\top}\mathbf{A}^{-1}(\mathbf{x} - \boldsymbol{\mu}_{\boldsymbol{\gamma}})\right]^{1/2} - \left[\boldsymbol{\mu}_{\boldsymbol{\gamma}}^{\top}\mathbf{A}^{-1}\boldsymbol{\mu}_{\boldsymbol{\gamma}}\right]^{1/2}\right)^2}{\sin^2(\tilde{\theta})\, \left(\frac{k_{\max}}{\lambda_{\min}(\mathbf{A})}\right)} \right),
\end{align}

for $t > \lvert\cos(\tilde{\theta})\rvert \left[(\mathbf{x} - \boldsymbol{\mu}_{\boldsymbol{\gamma}})^{\top}\mathbf{A}^{-1}(\mathbf{x} - \boldsymbol{\mu}_{\boldsymbol{\gamma}})\right]^{1/2} + \left[\boldsymbol{\mu}_{\boldsymbol{\gamma}}^{\top}\mathbf{A}^{-1}\boldsymbol{\mu}_{\boldsymbol{\gamma}}\right]^{1/2}$.
Letting $\tilde{\alpha} > 0$, we have
\begin{align}
    \nonumber &\text{pr}\left(\left[q_{p_{\mathbf{x}, \mathbf{Z}_\mathbf{x}, \boldsymbol{\gamma}}(\tilde{\theta})}(\boldsymbol{\mu}_{\boldsymbol{\gamma}}, \mathbf{A})\right]^{1/2} > \left[\tilde{\alpha}q_{\mathbf{x}}(\boldsymbol{\mu}_{\boldsymbol{\gamma}}, \mathbf{A})\right]^{1/2} - \frac{4R_{\mu}}{\sqrt{\lambda_{\min}(\mathbf{A})}} \right) \\
    \label{eq: radius_prob} & \le \text{pr}\left(\mathcal{R}^2 > \frac{\left( \left( \sqrt{\tilde{\alpha}} -\lvert\cos(\tilde{\theta})\rvert\right) \left[(\mathbf{x} - \boldsymbol{\mu}_{\boldsymbol{\gamma}})^{\top}\mathbf{A}^{-1}(\mathbf{x} - \boldsymbol{\mu}_{\boldsymbol{\gamma}})\right]^{1/2} - \frac{5R_{\mu}}{\sqrt{\lambda_{\min}(\mathbf{A})}}\right)^2}{\sin^2(\tilde{\theta}) \left(\frac{k_{\max}}{\lambda_{\min}(\mathbf{A})}\right)} \right),
\end{align}
where we use the fact that $[\boldsymbol{\mu}_{\boldsymbol{\gamma}}^{\top}\mathbf{A}^{-1} \boldsymbol{\mu}_{\boldsymbol{\gamma}}]^{1/2} \le \frac{R_\mu}{\sqrt{\lambda_{\min}(\mathbf{A})}}$.

\vspace{1em}
\textbf{Symmetric Multivariate Pearson Type VII}

\vspace{1em}
Consider the case where the  elliptical distribution, $\mathcal{E}$, used in the adaptive scheme is a symmetric multivariate Pearson type VII distribution. Thus, from Equation \ref{eq: radius_prob}, we have that
\begin{align}
    \nonumber \text{pr}\left(\frac{\mathcal{R}^2}{m + q_{\mathbf{x}}(\boldsymbol{\mu}_{\boldsymbol{\gamma}}, \boldsymbol{\Sigma}_{\boldsymbol{\gamma}})} > \frac{\left( \left( \sqrt{\tilde{\alpha}} -\lvert\cos(\tilde{\theta})\rvert\right) \left[(\mathbf{x} - \boldsymbol{\mu}_{\boldsymbol{\gamma}})^{\top}\mathbf{A}^{-1}(\mathbf{x} - \boldsymbol{\mu}_{\boldsymbol{\gamma}})\right]^{1/2} - \frac{5R_{\mu}}{\sqrt{\lambda_{\min}(\mathbf{A})}}\right)^2}{\sin^2(\tilde{\theta}) \left(\frac{k_{\max}}{\lambda_{\min}(\mathbf{A})}\right) \left(m + q_{\mathbf{x}}(\boldsymbol{\mu}_{\boldsymbol{\gamma}}, \boldsymbol{\Sigma}_{\boldsymbol{\gamma}})\right)} \right).
\end{align}
Notice that
\begin{align}
    \nonumber \left(\frac{k_{\max}}{\lambda_{\min}(\mathbf{A})}\right) q_{\mathbf{x}}(\boldsymbol{\mu}_{\boldsymbol{\gamma}}, \boldsymbol{\Sigma}_{\boldsymbol{\gamma}})
    & \le \left(\frac{k_{\max}}{\lambda_{\min}(\mathbf{A})}\right)\frac{\lVert \mathbf{x} - \boldsymbol{\mu}_{\boldsymbol{\gamma}}\rVert_2^2}{k_{\min}}\\
    \nonumber & \le \left(\frac{k_{\max}\lambda_{\max}(\mathbf{A})}{k_{\min}\lambda_{\min}(\mathbf{A})}\right) (\mathbf{x} -  \boldsymbol{\mu}_{\boldsymbol{\gamma}})^{\top}\mathbf{A}^{-1}(\mathbf{x} -  \boldsymbol{\mu}_{\boldsymbol{\gamma}}).
\end{align}
We want to ensure that
$$(\mathbf{x} -  \boldsymbol{\mu}_{\boldsymbol{\gamma}})^{\top}\mathbf{A}^{-1}(\mathbf{x} -  \boldsymbol{\mu}_{\boldsymbol{\gamma}}) > R^* \ge \max\left\{\frac{25R_\mu^2}{\lambda_{\min}(\mathbf{A})\xi^2}, \frac{mk_{\min}}{\psi \lambda_{\max}(\mathbf{A})}\right\},$$
so let 
\begin{align}
\label{eq: def_R_tilde_Pearson}
\nonumber (\mathbf{x} -  \boldsymbol{\mu}_{0})^{\top}\mathbf{A}^{-1}(\mathbf{x} -  \boldsymbol{\mu}_{0}) > \tilde{R} & \ge 2R^* +  \frac{8R_{\mu}^2}{\lambda_{\min}(\mathbf{A})} \\
&\ge \max\left\{\frac{50R_\mu^2}{\lambda_{\min}(\mathbf{A})\xi^2} +  \frac{8R_{\mu}^2}{\lambda_{\min}(\mathbf{A})} , \frac{2mk_{\min}}{\psi \lambda_{\max}(\mathbf{A})} +  \frac{8R_{\mu}^2}{\lambda_{\min}(\mathbf{A})} , R\right\},
\end{align}
we have 
\begin{align}
    \nonumber & \text{pr}\left(\frac{\mathcal{R}^2}{m + q_{\mathbf{x}}(\boldsymbol{\mu}_{\boldsymbol{\gamma}}, \boldsymbol{\Sigma}_{\boldsymbol{\gamma}})} > \frac{\left( \left( \sqrt{\tilde{\alpha}} -\lvert\cos(\tilde{\theta})\rvert\right) \left[(\mathbf{x} - \boldsymbol{\mu}_{\boldsymbol{\gamma}})^{\top}\mathbf{A}^{-1}(\mathbf{x} - \boldsymbol{\mu}_{\boldsymbol{\gamma}})\right]^{1/2} - \frac{5R_{\mu}}{\sqrt{\lambda_{\min}(\mathbf{A})}}\right)^2}{\sin^2(\tilde{\theta}) \left(\frac{k_{\max}}{\lambda_{\min}(\mathbf{A})}\right) \left(m + q_{\mathbf{x}}(\boldsymbol{\mu}_{\boldsymbol{\gamma}}, \boldsymbol{\Sigma}_{\boldsymbol{\gamma}})\right)} \right)\\
   \label{eq: beta_bound}  & \le \text{pr}\left(\frac{\mathcal{R}^2}{m + q_{\mathbf{x}}(\boldsymbol{\mu}_{\boldsymbol{\gamma}}, \boldsymbol{\Sigma}_{\boldsymbol{\gamma}})} > \frac{ \left(\left( \sqrt{\tilde{\alpha}} - \lvert \cos(\tilde{\theta})\rvert\right) - \xi\right)^2}{( 1 + \psi)\left(\frac{k_{\max}\lambda_{\max}(\mathbf{A})}{k_{\min}\lambda_{\min}(\mathbf{A})}\right)\sin^2(\tilde{\theta})} \right).
\end{align}
Since $\mathcal{R}^2 / (m + q_{\mathbf{x}}(\boldsymbol{\mu}_{\boldsymbol{\gamma}}, \boldsymbol{\Sigma}_{\boldsymbol{\gamma}})) \sim BeII(P/2, M - P/2)$, we have that $\text{pr}\left(\mathcal{R}^2 / (m + q_{\mathbf{x}}(\boldsymbol{\mu}_{\boldsymbol{\gamma}}, \boldsymbol{\Sigma}_{\boldsymbol{\gamma}}))  > r \right) = 1 - I_{\frac{r}{1+r}}\left(P/2, M - P/2\right)$, where $I_x(\alpha, \beta)$ is the regularized incomplete beta function. Define the following quantities as follows:
$$g(\tilde{\alpha}, \tilde{\theta}, \xi, \psi) := \frac{ \left(\left( \sqrt{\tilde{\alpha}} - \lvert \cos(\tilde{\theta})\rvert\right) - \xi\right)^2}{( 1 + \psi)\left(\frac{k_{\max}\lambda_{\max}(\mathbf{A})}{k_{\min}\lambda_{\min}(\mathbf{A})}\right)\sin^2(\tilde{\theta})},$$
$$\Theta_{\alpha}^{\xi} := \left\{\theta \bigm\vert \lvert\cos(\theta)\rvert < \sqrt{\alpha} - \xi\right\},$$
$$F_1(\alpha, M, \xi, \psi):= \int_{\xi^2}^{\alpha} \frac{1}{4\pi\sqrt{\tilde{\alpha}}} \int_{\Theta_{\tilde{\alpha}}^{\xi}} I_{\frac{g(\tilde{\alpha}, \tilde{\theta}, \xi, \psi)}{1 + g(\tilde{\alpha}, \tilde{\theta}, \xi, \psi)}}(P/2, M - P/2) \text{d}\tilde{\theta} \text{d}\tilde{\alpha},$$
$$F_{\alpha}(M,\xi, \psi) := \frac{1}{2\pi} \int_{\Theta_{\alpha}^{\xi}}I_{\frac{g(\alpha, \tilde{\theta}, \xi, \psi)}{1 + g(\alpha, \tilde{\theta}, \xi, \psi)}}(P/2, M - P/2) \text{d}\tilde{\theta}.$$
Notice that $F_{\alpha}(M, \xi, \psi)$ is equal to $\text{pr}\left(\frac{\mathcal{R}^2}{m + q_{\mathbf{x}}(\boldsymbol{\mu}_{\boldsymbol{\gamma}}, \boldsymbol{\Sigma}_{\boldsymbol{\gamma}})} < \frac{\left(\left( \sqrt{\alpha} - \lvert \cos(\tilde{\theta})\rvert\right) - \xi\right)^2}{(1 + \psi)\left(\frac{k_{\max}\lambda_{\max}(\mathbf{A})}{k_{\min}\lambda_{\min}(\mathbf{A})}\right)\sin^2(\tilde{\theta})} \right)$ when marginalizing out $\tilde{\theta}$. Using these quantities along with Equations \labelcref{eq: first_prop,eq: radius_prob,eq: beta_bound}, we have
\begin{equation}
    \label{eq: pearson_second_bound}
    \int_{\xi q_{\mathbf{x}}(\boldsymbol{\mu}_0, \mathbf{A})^{1/2}}^{\sqrt{\alpha} q_{\mathbf{x}}(\boldsymbol{\mu}_0, \mathbf{A})^{1/2}} H_{\boldsymbol{\gamma}}(\mathbf{x}, B_{t^2}^C(\boldsymbol{\mu}_0, \mathbf{A})) \text{d}t \le (\sqrt{\alpha} - \xi) q_{\mathbf{x}}(\boldsymbol{\mu}_0, \mathbf{A})^{1/2} - F_1(\alpha, M, \xi, \psi) q_{\mathbf{x}}(\boldsymbol{\mu}_0, \mathbf{A})^{1/2}
\end{equation}
and 
\begin{align}
    \nonumber \int_{\sqrt{\alpha q_{\mathbf{x}}(\boldsymbol{\mu}_0, \mathbf{A})}}^{\sqrt{\frac{1}{\alpha} q_{\mathbf{x}}(\boldsymbol{\mu}_0, \mathbf{A})}} H_{\boldsymbol{\gamma}}(\mathbf{x}, B_{t^2}^C(\boldsymbol{\mu}_0, \mathbf{A})) \text{d}t &\le \left(\frac{1}{\sqrt{\alpha}} - \sqrt{\alpha}\right)q_{\mathbf{x}}(\boldsymbol{\mu}_0, \mathbf{A})^{1/2} \left(1 - F_{\alpha}(M,\xi, \psi)\right)\\
    \label{eq: pearson_third_bound} & < \frac{F_1(\alpha, M, \xi, \psi)\left[q_{\mathbf{x}}(\boldsymbol{\mu}_0, \mathbf{A})\right]^{1/2}}{2},
\end{align}
where the second inequality of Equation \ref{eq: pearson_third_bound} comes from Assumption \ref{assumption6}. Notice that for all $\mathbf{y}$ with $q_{\mathbf{y}}(\boldsymbol{\mu}_0, \mathbf{A}) > (1 / \alpha) q_{\mathbf{x}}(\boldsymbol{\mu}_0, \mathbf{A})$, we have $\frac{\mathcal{L}^*(\mathbf{y}, \boldsymbol{\mu}_{\boldsymbol{\gamma}}, \boldsymbol{\Sigma}_{\boldsymbol{\gamma}})}{\mathcal{L}^*(\mathbf{x}, \boldsymbol{\mu}_{\boldsymbol{\gamma}}, \boldsymbol{\Sigma}_{\boldsymbol{\gamma}})} \le \left(1 + q_{\mathbf{y}}(\boldsymbol{\mu}_0, \mathbf{A}) - (1/\alpha)q_{\mathbf{x}}(\boldsymbol{\mu}_0, \mathbf{A}) \right)^{-1}$. To see this, let $\mathbf{y} \in \mathcal{X}$ and define $R_2 = q_{\mathbf{y}}(\boldsymbol{\mu}_0, \mathbf{A})$. For any $R_1 \in ((1/\alpha)q_{\mathbf{x}}(\boldsymbol{\mu}_0, \mathbf{A}), R_2)$ and $\mathbf{w} \in \{\mathbf{x} \in \mathcal{X} \mid  q_{\mathbf{x}}(\boldsymbol{\mu}_0, \mathbf{A}) = R_1 \}$, we have that $\mathcal{L}^*(\mathbf{w}, \boldsymbol{\mu}_{\boldsymbol{\gamma}}, \boldsymbol{\Sigma}_{\boldsymbol{\gamma}}) < \mathcal{L}^*(\mathbf{x}, \boldsymbol{\mu}_{\boldsymbol{\gamma}}, \boldsymbol{\Sigma}_{\boldsymbol{\gamma}})$ from the elliptical subcover property of Assumption \ref{assumption6}. Using the diminishing tails condition of Assumption \ref{assumption6}, we also have that $$\mathcal{L}^*(\mathbf{y}, \boldsymbol{\mu}_{\boldsymbol{\gamma}}, \boldsymbol{\Sigma}_{\boldsymbol{\gamma}}) \le \max_{\mathbf{w} \in \{\mathbf{w} \in \mathcal{X} \mid q_{\mathbf{w}}(\boldsymbol{\mu}_0, \mathbf{A}) = R_2\}}\mathcal{L}^*(\mathbf{w}, \boldsymbol{\mu}_{\boldsymbol{\gamma}}, \boldsymbol{\Sigma}_{\boldsymbol{\gamma}})\le (1 + R_2 -R_1)^{-1}\mathcal{L}^*(\mathbf{x}, \boldsymbol{\mu}_{\boldsymbol{\gamma}}, \boldsymbol{\Sigma}_{\boldsymbol{\gamma}}).$$
Letting $R_1$ approach $(1 / \alpha) q_{\mathbf{x}}(\boldsymbol{\mu}_0, \mathbf{A})$ gives us the desired result. Let us consider a transition from $\mathbf{x}$ to $B_{t}^C(\boldsymbol{\mu}_0, \mathbf{A})$ for $t^2 \ge \frac{1}{\alpha} q_{\mathbf{x}}(\boldsymbol{\mu}_0, \mathbf{A})$. We can bound this probability, as for this transition to occur we need (1) for $y$, (the threshold) to be such that $y \in (0, \mathcal{L}^*(\mathbf{x}, \boldsymbol{\mu}_{\boldsymbol{\gamma}}, \boldsymbol{\Sigma}_{\boldsymbol{\gamma}}))$ and (2) for the sampling scheme to not accept a move into $B_{\alpha q_{\mathbf{x}}(\boldsymbol{\mu}_0, \mathbf{A})}(\boldsymbol{\mu}_0, \mathbf{A})$ on the first iteration of the while loop. Using this and the fact that $y \sim \mathcal{U}_{(0, \mathcal{L}^*(\mathbf{x}, \boldsymbol{\mu}_{\boldsymbol{\gamma}}, \boldsymbol{\Sigma}_{\boldsymbol{\gamma}}))}$, we have
\begin{align}
   \nonumber \int_{\sqrt{\frac{1}{\alpha} q_{\mathbf{x}}(\boldsymbol{\mu}_0, \mathbf{A})}}^{\infty} H_{\boldsymbol{\gamma}}(\mathbf{x}, B_{t^2}^C(\boldsymbol{\mu}_0, \mathbf{A})) \text{d}t & \le \left( 1 - F_{\alpha}(M,\xi, \psi)\right) \int_{\sqrt{\frac{1}{\alpha} q_{\mathbf{x}}(\boldsymbol{\mu}_0, \mathbf{A})}}^{\infty}\left(1 + t^2 - \frac{1}{\alpha} q_{\mathbf{x}}(\boldsymbol{\mu}_0, \mathbf{A})\right)^{-1} \text{d}t \\
    \label{eq: pearson_fourth_bound} & \le \left( 1 - F_{\alpha}(M,\xi, \psi)\right)\frac{\pi}{2}.
\end{align}
Notice that regardless of the class of elliptical distribution chosen (multivariate Gaussian or symmetric multivariate Pearson type VII), $H_{\boldsymbol{\gamma}}(\mathbf{x}, B_{t^2}^C(\boldsymbol{\mu}_0, \mathbf{A})) \le 1$ for any $t \ge 0$, leading to 
\begin{equation}
    \label{eq: first_bound}
    \int_{0}^{\xi q_{\mathbf{x}}(\boldsymbol{\mu}_0, \mathbf{A})^{1/2}} H_{\boldsymbol{\gamma}}(\mathbf{x}, B_{t^2}^C(\boldsymbol{\mu}_0, \mathbf{A})) \text{d}t \le \xi q_{\mathbf{x}}(\boldsymbol{\mu}_0, \mathbf{A})^{1/2}.
\end{equation}
Thus using Equations \labelcref{eq: first_bound,eq: pearson_second_bound,eq: pearson_third_bound,eq: pearson_fourth_bound}, we have
\begin{align}
\nonumber H_{\boldsymbol{\gamma}}V(\mathbf{x}) & = 1 + \int_{0}^{\infty} H_{\boldsymbol{\gamma}}(\mathbf{x}, B_{t^2}^C(\boldsymbol{\mu}_0, \mathbf{A})) \text{d}t \\
\nonumber & \le 1 + q_{\mathbf{x}}(\boldsymbol{\mu}_0, \mathbf{A})^{1/2}\left( 1 - \left(F_1(\alpha, M, \xi, \psi)\left(1 - \frac{1}{2} -\frac{\left(1 - F_{\alpha}(M,\xi, \psi)\right) \pi}{2F_1(\alpha, M, \xi, \psi)q_{\mathbf{x}}(\boldsymbol{\mu}_0, \mathbf{A})^{1/2}}\right)\right)\right).
\end{align}
Notice that $0<F_1(\alpha, M, \xi, \psi) \le 1$ and $0 < F_{\alpha}(M,\xi, \psi) \le 1$. Letting 
\begin{equation}
\label{eq: def_R_tilde_Pearson2}
 \tilde{R} > \max\left\{\frac{50R_\mu^2}{\lambda_{\min}(\mathbf{A})\xi^2} +  \frac{8R_{\mu}^2}{\lambda_{\min}(\mathbf{A})} , \frac{2mk_{\min}}{\psi \lambda_{\max}(\mathbf{A})} +  \frac{8R_{\mu}^2}{\lambda_{\min}(\mathbf{A})} , R, \frac{4\pi^2}{\left[F_1(\alpha, M, \xi, \psi)\right]^2}\right\},
\end{equation}
we have 
{\small$$0 < \left( 1 - \left(F_1(\alpha, M, \xi, \psi)\left(1 - \frac{1}{2} -\frac{\left(1 - F_{\alpha}(M,\xi, \psi)\right) \pi}{2F_1(\alpha, M, \xi, \psi)q_{\mathbf{x}}(\boldsymbol{\mu}_0, \mathbf{A})^{1/2}}\right)\right)\right)< 1 -F_1(\alpha, M, \xi, \psi)\left(\frac{1}{2} - \frac{1 - F_{\alpha}(M,\xi, \psi)}{4} \right) < 1.$$}
Thus, defining
$$\phi := \frac{1 + \sqrt{\tilde{R}}\left[1 -F_1(\alpha, M, \xi, \psi)\left(\frac{1}{2} - \frac{1 - F_{\alpha}(M,\xi, \psi)}{4} \right)\right]}{1 + \sqrt{\tilde{R}}} < 1,$$
gives us the desired result of
\begin{equation}
    \label{eq: pearson_drift2} H_{\boldsymbol{\gamma}}V(\mathbf{x}) \le \phi V(\mathbf{x}) \text{ for all } \mathbf{x} \not\in C.
\end{equation}

\vspace{1em}
\textbf{Multivariate Gaussian}

\vspace{1em}
Consider the case where  the elliptical distribution, $\mathcal{E}$, used in the adaptive schemes is a multivariate gaussian distribution. From Equation \ref{eq: radius_prob}, we are interested in the quantity:
$$\text{pr}\left(\mathcal{R}^2 > \frac{\left( \left( \sqrt{\tilde{\alpha}} -\lvert\cos(\tilde{\theta})\rvert\right) \left[(\mathbf{x} - \boldsymbol{\mu}_{\boldsymbol{\gamma}})^{\top}\mathbf{A}^{-1}(\mathbf{x} - \boldsymbol{\mu}_{\boldsymbol{\gamma}})\right]^{1/2} - \frac{5R_{\mu}}{\sqrt{\lambda_{\min}(\mathbf{A})}}\right)^2}{\sin^2(\tilde{\theta}) \left(\frac{k_{\max}}{\lambda_{\min}(\mathbf{A})}\right)} \right).$$
From Section \ref{sec: Mult_Gaussian}, we have that 
\begin{equation}
\label{eq: CDF_R_gaussian}
\text{pr}(\mathcal{R}^2 \le t) = \int_0^{t}f(u) \text{d}u := F(t) \quad t > 0,
\end{equation}
where $F(t)$ is the CDF of a Gamma distribution with shape equal to $P/2$ and rate equal to $1/2$.
Thus we have
\begin{align}
    \nonumber & \text{pr}\left(\left[q_{p_{\mathbf{x}, \mathbf{Z}_\mathbf{x}, \boldsymbol{\gamma}}(\tilde{\theta})}(\boldsymbol{\mu}_{\boldsymbol{\gamma}}, \mathbf{A})\right]^{1/2} > \left[\tilde{\alpha}q_{\mathbf{x}}(\boldsymbol{\mu}_{\boldsymbol{\gamma}}, \mathbf{A})\right]^{1/2} - \frac{4R_{\mu}}{\sqrt{\lambda_{\min}(\mathbf{A})}} \right) \\
\nonumber \le &  1 - \frac{1}{2\pi}\int_{ \Theta_{\tilde{\alpha}}^G}\text{pr}\left(\mathcal{R}^2 \le \frac{\left( \left( \sqrt{\tilde{\alpha}} -\lvert\cos(\tilde{\theta})\rvert\right) \left[(\mathbf{x} - \boldsymbol{\mu}_{\boldsymbol{\gamma}})^{\top}\mathbf{A}^{-1}(\mathbf{x} - \boldsymbol{\mu}_{\boldsymbol{\gamma}})\right]^{1/2} - \frac{5R_{\mu}}{\sqrt{\lambda_{\min}(\mathbf{A})}}\right)^2}{\sin^2(\tilde{\theta}) \left(\frac{k_{\max}}{\lambda_{\min}(\mathbf{A})}\right)} \right)\text{d}\tilde{\theta}\\
\nonumber \le &  1 - \frac{1}{2\pi}\int_{\Theta_{\tilde{\alpha}}^G}\text{pr}\left(\mathcal{R}^2 \le \frac{ \left( \left( \sqrt{\tilde{\alpha}} -\lvert\cos(\tilde{\theta})\rvert\right) \sqrt{R^*} - \frac{5R_{\mu}}{\sqrt{\lambda_{\min}(\mathbf{A})}}\right)^2}{\left(\frac{k_{\max}}{\lambda_{\min}(\mathbf{A})}\right)} \right)\text{d}\tilde{\theta} \\
\label{eq: bound_radius_gaussian} \le &  1 -\frac{1}{2\pi}\int_{\Theta_{\tilde{\alpha}}^G}\text{pr}\left(\mathcal{R}^2 \le \frac{ \left( \sqrt{\tilde{\alpha}} - \lvert\cos(\tilde{\theta})\rvert  - C_1(R^*)\right)^2 \lambda_{\min}(\mathbf{A})}{k_{\max}} R^*\right)\text{d}\tilde{\theta},
\end{align}
where $C_1(R^*) := \frac{5R_{\mu}}{\sqrt{\lambda_{\min}(\mathbf{A})R^*}}$, and $\Theta_{\alpha}^G$ is defined below. Define the following quantities, where superscript $G$ denotes the quantities associated with the multivariate Gaussian distribution:
$$g^G(\tilde{\alpha}, \tilde{\theta}, R^*) := \frac{ \left( \sqrt{\tilde{\alpha}} - \lvert\cos(\tilde{\theta})\rvert  - C_1(R^*)\right)^2\lambda_{\min}(\mathbf{A})}{k_{\max}} R^*,$$
$$\Theta_{\alpha}^G := \left\{\theta \bigm\vert \lvert\cos(\theta) \rvert < \sqrt{\alpha} - C_1(R^*)\right\},$$
$$F_1^G(\alpha, R^*):= \int_{C_1(R^*)^2}^{\alpha} \frac{1}{4\pi\sqrt{\tilde{\alpha}}} \int_{\Theta_{\tilde{\alpha}}^G}  F(g^G(\tilde{\alpha}, \tilde{\theta}, R^*))\text{d}\tilde{\theta} \text{d}\tilde{\alpha},$$
$$F_{\alpha}^G(R^*) := \frac{1}{2\pi} \int_{\Theta_{\alpha}^G}  F(g^G(\alpha, \tilde{\theta}, R^*))\text{d}\tilde{\theta},$$
where $F$ is the CDF defined in Equation \ref{eq: CDF_R_gaussian}. Notice that $F_{\alpha}^G(R^*)$ is equal to $\text{pr}\left(\mathcal{R}^2 \le \frac{ \left( \sqrt{\alpha} - \lvert\cos(\tilde{\theta})\rvert  - C_1(R^*)\right)^2\lambda_{\min}(\mathbf{A})}{k_{\max}} R^*\right)$ when marginalizing out $\tilde{\theta}$.  Let $R^* > \left(\frac{5R_{\mu}}{\alpha\sqrt{\lambda_{\min}(\mathbf{A})}}\right)^2$ be such that
\begin{equation}
\label{eq: def_R_tilde_gaussian}
F_1^{G}(\alpha, R^*) \ge \max\left\{2\left( \frac{1}{\sqrt{\alpha}} - \sqrt{\alpha}\right) \left( 1 - F_{\alpha}^G(R^*)\right), 0.1 \right\}
\end{equation}
and
\begin{equation}
\label{eq: def_R_tilde_gaussian2}
0.5 > \frac{\pi}{F_1^G(\alpha, R^*)\sqrt{\tilde{R}}},
\end{equation}
where $\tilde{R} \ge 2R^* + \frac{8R_{\mu}^2}{\lambda_{\min}(\mathbf{A})}$.
Notice that such a $R^*$ (and therefore $\tilde{R}$) exists, as we can pick a large $R^*$ such that $\text{pr}\left(\mathcal{R}^2 \le g^G(\tilde{\alpha}, \tilde{\theta}, R^*) \right)$ is arbitrarily close to 1 for $\alpha > 0.75$. Since $C_1(R^*) \rightarrow 0$ as $R^* \rightarrow \infty$, we can similarly make $R^*$ large enough such that $F_1^{G}(\alpha, R^*)$ is arbitrarily close to $\int_{0}^{\alpha} \frac{\lambda\left(\left\{\theta \bigm\vert \cos^{2}(\theta) <\tilde{\alpha} \right\}\right)}{4\pi\sqrt{\tilde{\alpha}}}\text{d}\tilde{\alpha} = \frac{2}{\pi} \int_{0}^{\sqrt{\alpha}}\sin^{-1}(u) \text{d}u > \frac{\sqrt{3}}{3} - \frac{1}{\pi}$ and $F_{\alpha}^G(R^*)$ is arbitrarily close to $\frac{\lambda\left(\left\{\theta \bigm\vert \cos^{2}(\theta) < \alpha \right\}\right)}{2\pi}  > \frac{2}{3}$ (since $\alpha > 0.75$, and $F_1^{G}(\alpha, R^*)$ and $F_{\alpha}^G(R^*)$ are monotone increasing in $\alpha$). Since $\left(\frac{1}{\sqrt{\alpha}} - \sqrt{\alpha}\right) < \frac{\sqrt{3}}{6}$, we can see that such a $R^*$ exists. Using these quantities along with Equations \labelcref{eq: first_prop,eq: radius_prob,eq: bound_radius_gaussian}, we have
{\small \begin{equation}
    \label{eq: gaussian_second_bound}
    \int_{C_1(R^*) \sqrt{q_{\mathbf{x}}(\boldsymbol{\mu}_0, \mathbf{A})}}^{\sqrt{\alpha q_{\mathbf{x}}(\boldsymbol{\mu}_0, \mathbf{A})}} H_{\boldsymbol{\gamma}}(\mathbf{x}, B_{t^2}^C(\boldsymbol{\mu}_0, \mathbf{A})) \text{d}t \le (\sqrt{\alpha} -C_1(R^*)) q_{\mathbf{x}}(\boldsymbol{\mu}_0, \mathbf{A})^{1/2} - F_1^G(\alpha, R^*) q_{\mathbf{x}}(\boldsymbol{\mu}_0, \mathbf{A})^{1/2}
\end{equation}}
and 
{\small \begin{equation}
    \label{eq: gaussian_third_bound}
    \int_{\sqrt{\alpha q_{\mathbf{x}}(\boldsymbol{\mu}_0, \mathbf{A})}}^{\sqrt{\frac{1}{\alpha} q_{\mathbf{x}}(\boldsymbol{\mu}_0, \mathbf{A})}} H_{\boldsymbol{\gamma}}(\mathbf{x}, B_{t^2}^C(\boldsymbol{\mu}_0, \mathbf{A})) \text{d}t \le \left(\frac{1}{\sqrt{\alpha}} - \sqrt{\alpha}\right)q_{\mathbf{x}}(\boldsymbol{\mu}_0, \mathbf{A})^{1/2} \left(1 - F_{\alpha}^G(R^*)\right) < \frac{F_1^G(\alpha, R^*)q_{\mathbf{x}}(\boldsymbol{\mu}_0, \mathbf{A})^{1/2}}{2}.
\end{equation}}

 Using the same argument as in the multivariate Pearson type VII setting, we have $\frac{\mathcal{L}^*(\mathbf{y}, \boldsymbol{\mu}_{\boldsymbol{\gamma}}, \boldsymbol{\Sigma}_{\boldsymbol{\gamma}})}{\mathcal{L}^*(\mathbf{x}, \boldsymbol{\mu}_{\boldsymbol{\gamma}}, \boldsymbol{\Sigma}_{\boldsymbol{\gamma}})} \le \left(1 + q_{\mathbf{y}}(\boldsymbol{\mu}_0, \mathbf{A}) - (1/\alpha)q_{\mathbf{x}}(\boldsymbol{\mu}_0, \mathbf{A}) \right)^{-1}$ for all $\mathbf{y}$ such that $q_{\mathbf{y}}(\boldsymbol{\mu}_0, \mathbf{A}) > (1 / \alpha) q_{\mathbf{x}}(\boldsymbol{\mu}_0, \mathbf{A})$. Thus, we can bound the probability of transitioning from $\mathbf{x}$ to $B_{t^2}^C(\boldsymbol{\mu}_0, \mathbf{A})$ by
\begin{align}
    \nonumber \int_{\sqrt{\frac{1}{\alpha} q_{\mathbf{x}}(\boldsymbol{\mu}_0, \mathbf{A})}}^{\infty} H_{\boldsymbol{\gamma}}(\mathbf{x}, B_{t^2}^C(\boldsymbol{\mu}_0, \mathbf{A})) \text{d}t &\le \left( 1 - F_{\alpha}^G(R^*)\right) \int_{\sqrt{\frac{1}{\alpha} q_{\mathbf{x}}(\boldsymbol{\mu}_0, \mathbf{A})}}^{\infty}\left(1 + t^2 - \frac{1}{\alpha} q_{\mathbf{x}}(\boldsymbol{\mu}_0, \mathbf{A})\right)^{-1} \text{d}t \\
    \label{eq: gaussian_fourth_bound} & \le \left( 1 - F_{\alpha}^G(R^*)\right)\frac{\pi}{2}.
\end{align}
 Lastly, we have that 
\begin{equation}
    \label{eq: first_bound_gaussian}
    \int_{0}^{C_1(R^*)\sqrt{q_{\mathbf{x}}(\boldsymbol{\mu}_0, \mathbf{A})}} H_{\boldsymbol{\gamma}}(\mathbf{x}, B_{t^2}^C(\boldsymbol{\mu}_0, \mathbf{A})) \text{d}t \le C_1(R^*) q_{\mathbf{x}}(\boldsymbol{\mu}_0, \mathbf{A})^{1/2}.
\end{equation}

Thus using Equations \labelcref{eq: first_bound_gaussian,eq: gaussian_second_bound,eq: gaussian_third_bound,eq: gaussian_fourth_bound}, we have
\begin{align}
\nonumber H_{\boldsymbol{\gamma}}V(\mathbf{x}) & = 1 + \int_{0}^{\infty} H_{\boldsymbol{\gamma}}(\mathbf{x}, B_{t^2}^C(\boldsymbol{\mu}_0, \mathbf{A})) \text{d}t \\
\nonumber & \le 1 + q_{\mathbf{x}}(\boldsymbol{\mu}_0, \mathbf{A})^{1/2}\left( 1 - \left(F_1^G(\alpha, R^*)\left(1 - \frac{1}{2} -\frac{(1 - F_{\alpha}^G(R^*)) \pi}{2F_1^G(\alpha, R^*)q_{\mathbf{x}}(\boldsymbol{\mu}_0, \mathbf{A})^{1/2}}\right)\right)\right).\\
\end{align}
Notice that $0.1\le F_1^G(\alpha, R^*) \le 1$, $0 < F_{\alpha}^G(R^*)\le1$, and $\frac{\pi}{F_1^G(\alpha, R^*)q_{\mathbf{x}}(\boldsymbol{\mu}_0, \mathbf{A})^{1/2}} < 0.5$ (Equation \ref{eq: def_R_tilde_gaussian2}), so $0 < \left( 1 - \left(F_1^G(\alpha, R^*)\left(1 - \frac{1}{2} -\frac{(1 - F_{\alpha}^G(R^*)) \pi}{2F_1^G(\alpha, R^*)q_{\mathbf{x}}(\boldsymbol{\mu}_0, \mathbf{A})^{1/2}}\right)\right)\right) \le 1- \frac{1}{40}$. Thus, letting $\phi:= \frac{1 + \sqrt{\tilde{R}}\left(0.975 \right)}{1 + \sqrt{\tilde{R}}} < 1$ gives us the desired result of
\begin{equation}
    \label{eq: gaussian_drift2} H_{\boldsymbol{\gamma}}V(\mathbf{x}) \le \phi V(\mathbf{x}) \text{ for all } \mathbf{x} \not\in C.
\end{equation}

\vspace{1em}
Thus, using Equations \labelcref{eq: geometric_drift_C,eq: pearson_drift2,eq: gaussian_drift2}, we have defined a $b < \infty$ (Equation \ref{eq: def_b}), $\phi < 1$, and a set $C:= \left\{\mathbf{y} \in \mathcal{X} \mid \left(\mathbf{y} - \boldsymbol{\mu}_0\right)^{\top}\mathbf{A}^{-1}\left(\mathbf{y} - \boldsymbol{\mu}_0\right) \le \tilde{R}\right\}$ ($\tilde{R}$ is defined in Equations \labelcref{eq: def_R_tilde_Pearson2,eq: def_R_tilde_gaussian,eq: def_R_tilde_gaussian2} if $\mathcal{X}$ is not bounded, and $\tilde{R}$ such that $C= \mathcal{X}$ if $\mathcal{X}$ is bounded) such that $H_{\boldsymbol{\gamma}}V(\mathbf{x}) \le \phi V(\mathbf{x}) + b \mathbbm{1}_{C}(\mathbf{x})$ for all $\mathbf{x} \in \mathcal{X}$.
\end{proof}

\subsection{Theorem 2}
\begin{theorem}[Ergodicity]
    \label{thm: ergodicitiy}
    Suppose Assumptions \labelcref{assumption1,assumption2,assumption3,assumption4,assumption6} hold. Then the adaptive scheme proposed in Algorithm 1 in the main manuscript is ergodic.
\end{theorem} 
\begin{proof}
    From Theorem \ref{thm: reversibility}, we have that for every $\boldsymbol{\gamma} \in \mathcal{Y}$, $\mu$ is stationary for the transition kernel $H_{\boldsymbol{\gamma}}$. From Propositions \labelcref{prop: minorization,prop: drift}, we see that the family of Markov chain transition kernels is simultaneously strongly aperiodically geometrically ergodic. Lastly, since we adapt increasingly rarely through the AirMCMC scheme \citep{chimisov2018air}, we have that diminishing adaptation holds. Thus, directly applying Theorem 3 in \citet{roberts2007coupling}, the proposed adaptive algorithm is ergodic.
\end{proof}

\section{Sampling Scheme used for Case Studies}
\label{sec: algorithm}
In this section, we will give an outline of the sampling algorithm used in the case studies. We will denote one iteration of the AGESS algorithm, as specified in Algorithm 2 in the main manuscript, as $\texttt{AGESS\_TRANSITION}(\mathbf{x}_i, \boldsymbol{\mu}_{\boldsymbol{\gamma}}, \boldsymbol{\Sigma}_{\boldsymbol{\gamma}}, t_6)$, where $\mathbf{x}_i$ is the current state of the Markov Chain and $t_6$ denotes using a multivariate $t$-distribution with 6 degrees of freedom as the elliptical distribution of choice. In our simulation studies, we set $\epsilon_A = 0.05$ and $\epsilon_B = 0.05$ ($P \ge 10$; $\epsilon_A = 0.1$ for $P < 10$), controlling the probability of performing 1-dimensional transitions ($\epsilon_B$) and the probability of using a non-adaptive transition kernel ($\epsilon_A$).  

Although we adapt increasingly rarely using the AirMCMC framework \citep{chimisov2018air}, we keep track of a set of \textit{background adaptive parameters}, which update after every iteration of the Markov chain. Note that we only update the actual adaptive parameters on iterations in the set $\{N_j\}_{j=1}^{\infty}$, where $N_j:= \sum_{i=1}^j \lfloor i^\beta \rfloor$. In our simulation studies, we set $\beta = 0.5$, to allow for high levels of adaptation, especially at the beginning of the Markov chain. The set of background adaptive parameters ($\tilde{\boldsymbol{\mu}}_{\boldsymbol{\gamma}}$ and $\tilde{\boldsymbol{\Sigma}}_{\boldsymbol{\gamma}}$), are updated on every iteration according to the weights $w_i$ which diminish as $i \rightarrow \infty$. In general, the weights that diminished slowly tended to lead to better sampling performance, at the cost of longer burn-in times. Thus, we choose the adaptive weights according to the dimension of the target distribution. Specifically, in our simulation studies we let $w(i) = i^{-\max\left\{2/3, (\sqrt[3]{P} - 1)/\sqrt[3]{P} \right\}}$; allowing for a balanced tradeoff between longer burn-in times and more efficient sampling across various dimensions of target distributions.

\begin{algorithm}
\caption{AGESS in Practice}
\label{alg: AGESS}
\hspace*{\algorithmicindent} \textbf{Input}: initial state $\mathbf{x}_1$, initial mean vector $\boldsymbol{\mu}_0$, initial scale matrix $\boldsymbol{\Sigma}_0$, likelihood function $\mathcal{L}(\cdot)$, $N$, family of elliptical distributions $\mathcal{E}$, $\epsilon_A$, $\epsilon_B$, $N_{burn}$, $\beta$ \\
\hspace*{\algorithmicindent} \textbf{Output}:  Markov chain $\{\mathbf{x}_t|1 \le t \le N\}$
\begin{algorithmic}
\State $\boldsymbol{\mu}_{\boldsymbol{\gamma}} \gets \boldsymbol{\mu}_0$ \Comment{Set the current adaptive parameters}
\State $\boldsymbol{\Sigma}_{\boldsymbol{\gamma}} \gets \boldsymbol{\Sigma}_0$
\State $\tilde{\boldsymbol{\mu}}_{\boldsymbol{\gamma}} \gets \boldsymbol{\mu}_0$ \Comment{Set background adaptive parameters for next update}
\State $\tilde{\boldsymbol{\Sigma}}_{\boldsymbol{\gamma}} \gets \boldsymbol{\Sigma}_0$

\State $d_w = \max\left\{\frac{2}{3}, \left(\frac{\sqrt[3]{P} - 1}{\sqrt[3]{P}}\right) \right\}$ \Comment{Set rate for background adaptation}
\State $i \gets 2$
\While{$i \le N$}
    \If{$P \ge 10$} \Comment{Moderate-dimensional scenario}
        \If{$i \le 0.1 \times N_{burn}$} \Comment{1-D updates for fast convergence in burn-in}
            \For{$p$ in $1$ to $P$}
                \State $\mathbf{x}_{i}[p] \gets \texttt{AGESS\_TRANSITION}(\mathbf{x}_{i-1}[p], \boldsymbol{\mu}_{\boldsymbol{\gamma}}[p], \boldsymbol{\Sigma}_{\boldsymbol{\gamma}}[p,p], t_6)$
            \EndFor
        \Else \Comment{Iterations outside of 10\% burn-in}
            \State $p_{\varepsilon} \gets \mathcal{U}_{[0,1]}$
            \If{$p_{\varepsilon} \le \epsilon_A + \epsilon_B$}
                \State $p_B \gets Bernoulli(\epsilon_B / (\epsilon_A + \epsilon_B))$
                \If{$p_B ==1$} \Comment{1-D updates w.p. $\epsilon_B$}
                    \For{$p$ in $1$ to $P$}
                        \State $\mathbf{x}_{i}[p] \gets \texttt{AGESS\_TRANSITION}(\mathbf{x}_{i-1}[p], \boldsymbol{\mu}_{\boldsymbol{\gamma}}[p], \boldsymbol{\Sigma}_{\boldsymbol{\gamma}}[p,p], t_6)$
                    \EndFor
                \Else \Comment{Non-adaptive updates w.p. $\epsilon_A$}
                    \State $\mathbf{x}_{i} \gets \texttt{AGESS\_TRANSITION}(\mathbf{x}_{i-1}, \boldsymbol{\mu}_{0}, \boldsymbol{\Sigma}_{0}, t_6)$
                \EndIf
            \Else \Comment{Adaptive updates w.p. $1 - \epsilon_A - \epsilon_B$}
                \State $\mathbf{x}_{i} \gets \texttt{AGESS\_TRANSITION}(\mathbf{x}_{i-1}, \boldsymbol{\mu}_{\boldsymbol{\gamma}}, \boldsymbol{\Sigma}_{\boldsymbol{\gamma}}, t_6)$
            \EndIf
        \EndIf
    \Else \Comment{Low-dimensional scenario}
        \State $p_A \gets \mathcal{U}_{[0,1]}$
        \If{$p_A \le \epsilon_A$}\Comment{Non-adaptive updates w.p. $\epsilon_A$}
            \State $\mathbf{x}_{i} \gets \texttt{AGESS\_TRANSITION}(\mathbf{x}_{i-1}, \boldsymbol{\mu}_{0}, \boldsymbol{\Sigma}_{0}, t_6)$
        \Else \Comment{Adaptive updates w.p. $1 - \epsilon_A$}
            \State $\mathbf{x}_{i} \gets \texttt{AGESS\_TRANSITION}(\mathbf{x}_{i-1}, \boldsymbol{\mu}_{\boldsymbol{\gamma}}, \boldsymbol{\Sigma}_{\boldsymbol{\gamma}}, t_6)$
        \EndIf
    \EndIf
    \State $w_i = i^{-d_w}$
    \State $\tilde{\boldsymbol{\mu}}_{\boldsymbol{\gamma}} \gets [1 - w_i]\times\tilde{\boldsymbol{\mu}}_{\boldsymbol{\gamma}} + w_i \times \mathbf{x}_{i}$  \Comment{Update background adaptive mean}
    \State $\tilde{\boldsymbol{\Sigma}}_{\boldsymbol{\gamma}} \gets [1 - w_i] \times \tilde{\boldsymbol{\Sigma}}_{\boldsymbol{\gamma}} + w_i \times (\mathbf{x}_{i} - \tilde{\boldsymbol{\mu}}_{\boldsymbol{\gamma}})(\mathbf{x}_{i} - \tilde{\boldsymbol{\mu}}_{\boldsymbol{\gamma}})^{\top}$ \Comment{Update background adaptive covariance}
    \If{$i \in \{N_j\}_{j=1}^{\infty}$ ($N_j := \sum_{l=1}^j n_l$ where $n_l := \lfloor l^\beta \rfloor$)} \Comment{AirMCMC \citep{chimisov2018air}}
        \State $\boldsymbol{\mu}_{\boldsymbol{\gamma}} \gets \tilde{\boldsymbol{\mu}}_{\boldsymbol{\gamma}}$  \Comment{Update mean}
        \State $\boldsymbol{\Sigma}_{\boldsymbol{\gamma}} \gets \tilde{\boldsymbol{\Sigma}}_{\boldsymbol{\gamma}}$ \Comment{Update Scale}
    \EndIf
    \State $i \gets i + 1$
\EndWhile
\end{algorithmic}
\end{algorithm}

\section{Practical Considerations}
\label{sec: Practical}
Although we showed in the main manuscript that the proposed adaptive algorithm is ergodic---meaning that the constructed Markov chain will \textit{asymptotically} converge to the stationary distribution---we are generally interested in sampling schemes that (1) converge to the stationary distribution at a reasonably fast rate and (2) generate efficient samples from the posterior distribution once the chain has converged. In this subsection, we discuss some of the practical considerations that must be made when applying the adaptive algorithm and explain how these decisions influence the performance of the adaptive algorithm. 

\subsection{Choice of Elliptical Distribution} 
The choice of elliptical distributions is crucial as it will affect the integrability and behavior of the tails of the \textit{transformed likelihood}. Since the adaptive algorithm requires us to draw $\mathbf{Z}$ from its conditional distribution, we would preferably select an elliptical distribution whose conditional distributions are easy to sample from. Although in this scenario a multivariate Gaussian distribution may seem appealing due to the fact that the conditional distribution is equal to the marginal distribution, multivariate Gaussian distributions have tails that exponentially decay. When the likelihood function has tails that do not decay fast enough, such as tails that decay at polynomial or sub-exponential rates, the choice of a Gaussian elliptical distribution will lead to integrability problems of the \textit{transformed likelihood} and will also violate the regularity conditions used to prove that the adaptive algorithm is ergodic (Assumption \ref{assumption6}). To avoid these problems, we advocate for the use of a multivariate $t$-distribution, as the tails decay at a polynomial rate. In particular, we used a multivariate $t$-distribution with $\nu = 6$ for all case studies.

\subsection{Adaptation Scheme} 
The adaptation scheme is a key variable that affects the rate at which the Markov chain converges, as well as the sampling efficiency once the Markov chain converges. When considering adaptation schemes, the practitioner must specify (1) $\beta$, which controls the rate at which we adapt increasingly rarely using AirMCMC \citep{chimisov2018air}, along with (2) the method used to adapt the parameters. Through experimentation, we found that adapting frequently---particularly at the beginning of the Markov chain---was essential for obtaining an efficient sampling scheme. Consequently, in our case studies we chose $\beta = 0.5$. To adapt the parameters, we used a weighted adaptation scheme that was computed in the background on each iteration with a computational cost of $\mathcal{O}(P^2)$. The weights specified in the adaptation scheme depended on the dimension of the target distribution and diminish based on the number of MCMC iterations. Together, using the AirMCMC framework \citep{chimisov2018air} to adapt increasingly rarely based on the weighted adaptation scheme led to efficient sampling across the wide variety of target distributions considered in the case studies.

\subsection{Blocking Variables} 
To speed up convergence of the Markov chain, particularly when the target distribution is at least of moderate dimension, we suggest grouping the parameters into smaller blocks and performing updates on the smaller blocks. 
For our case studies, we use a general algorithm in moderate- to high-dimensional settings ($P \ge 10$) that does not take advantage of any model-specific blocking structure. Specifically, for the first 10\% of the burn-in period, we perform 1-dimensional updates to efficiently traverse the sample space and learn the geometry of the target distribution. After that period, we randomly update the entire $P$-dimensional state vector with probability $1 - \epsilon_B$ and perform 1-dimensional updates for all of our parameters with probability $\epsilon_B$, where $\epsilon_B = 0.05$ in our case studies. We generally recommend randomly mixing the blocking structure in moderate-dimensional settings, as it allows the sampler to have reasonable performance in cases where we cannot accurately learn the geometry of the target distribution. Although we utilize simple 1-dimensional blocks, further gains in sampling efficiency may be realized by adopting more advanced blocking strategies, such as those described in \citet{andrieu2008tutorial}, or, when possible, by exploiting the model structure to define the blocks.

\subsection{Adaptive and Non-Adaptive Kernels} 
When considering an adaptive MCMC scheme, one valid concern when targeting a multimodal target distribution is that the proposal distribution adapts to the geometry of the local mode, causing the Markov chain to get stuck in the local mode and have poor mixing. To create a more robust sampling scheme, we construct our Markov chain using both adaptive and non-adaptive kernels. Specifically, for each iteration, we choose the adaptive transition kernel with probability $1 - \epsilon_A$ and choose a non-adaptive transition kernel with probability $\epsilon_A$ (i.e., $\boldsymbol{\gamma} = (\boldsymbol{\mu}_0, \boldsymbol{\Sigma}_0)$). In our case studies, we set $\epsilon_A = 0.1$ for low-dimensional target distributions ($P < 10$) and $\epsilon_A = 0.05$ for moderate- to high-dimensional target distributions ($P \ge 10$), giving us the ability to traverse multimodal target distributions while maintaining an efficient algorithm. Although we chose $\epsilon_A$ in the low-dimensional setting so that it is equal to $\epsilon_A + \epsilon_B$ in the moderate- to high-dimensional settings, we found little sensitivity to setting $\epsilon_A = 0.05$ in all settings.

\subsection{Transformations of Variables}
In Bayesian inference, it is common to deal with variables that have support on the positive real line, for example, when working with scale parameters. In such cases, $\mathcal{L}^*$ may not be bounded away from $0$ on any interval of the form $(0, x)$ ($x \in \mathbb{R}_+$), which would violate the regularity conditions used to prove ergodicity (Assumptions \labelcref{assumption2,assumption6}). In these situations, we apply a log transformation to any parameter that has support on the positive real line, which often ensures that the regularity conditions are satisfied for the transformed variable.


\section{Additional Case Studies and Potential Applications}
\label{sec: add_case}
\subsection{Banana and Twin Banana Distributions}
\label{sec: Banana}
The \textit{Banana} distribution \citep{long2013banana, cameron2014recursive} is a common two-dimensional distribution that is non-convex, non-elliptical, and has long tails; often leading to poor sampling performance using common MCMC algorithms. In this simulation study, we will compare the sampling performance of the elliptical slice sampler (ESS), generalized elliptical slice sampler (GESS), adaptive generalized elliptical slice sampler (AGESS), Hamiltonian Monte Carlo (HMC), and adaptive random walk (ARW). Specifically, we will first determine whether the samplers can adequately explore the posterior distribution, especially the tails of the distribution. Among the samplers that can reliably fully explore the posterior distribution, we will compare their sampling performance by calculating the effective sample size per second \citep{vats2019multivariate, vats2021revisiting}.

The Banana distribution on $\boldsymbol{\theta} := (\theta_1, \theta_2)$ can be specified in the following hierarchical way:
$$Y_i \sim \mathcal{N}\left((\theta_1 - \mu_1) + (\theta_2 - \mu_2)^2, 1\right)\quad \theta_1 \sim \mathcal{N}(0, 4)\quad \theta_2 \sim \mathcal{N}(0.5, 4),$$
where $Y_i$ is generated from $\mathcal{N}(1, 1)$. Although there are various ways to construct a Banana distribution, this is the construction that we will use in our simulation study. We conservatively ran each sampling scheme for 200,000 iterations and discarded the first 100,000 iterations as burn-in. The remaining 100,000 iterations were used to calculate the effective sample size per second to compare the performance of the samplers. We generated 100 datasets and used randomly generated $\mu_1$ and $\mu_2$ such that $\mu_1,\mu_2 \sim \mathcal{N}(0,9)$. The randomly generated mean parameters change the centering and curvature of the posterior distributions, allowing us to consider the sampling performance over a diverse set of banana distributions.

\begin{figure}
    \centering
    \includegraphics[width=0.99\linewidth]{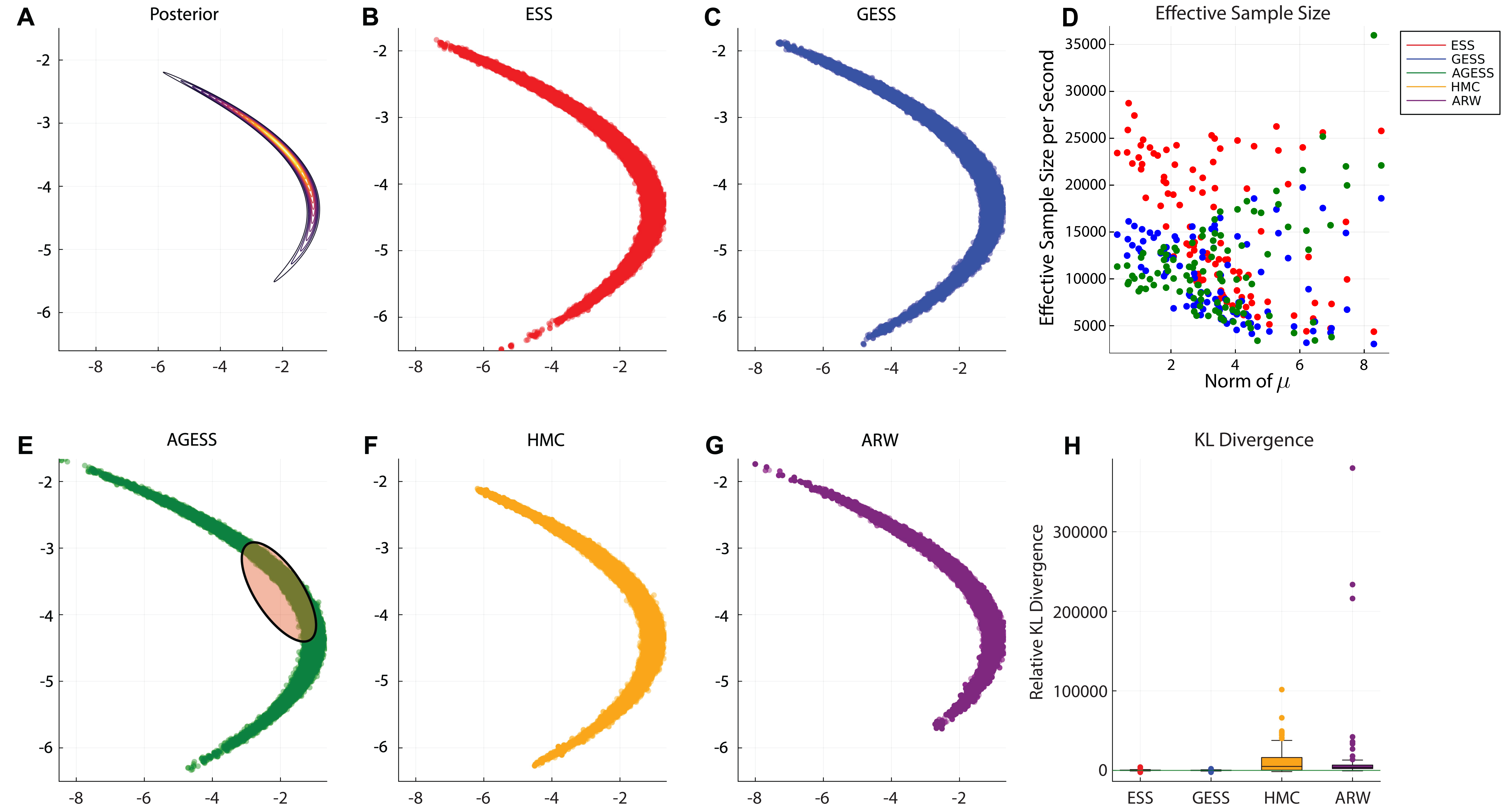}
    \caption{Visualizations of the target distribution (\textbf{Subfigure A}), along with the MCMC samples generated from ESS (\textbf{Subfigure B}), GESS (\textbf{Subfigure C}), AGESS (\textbf{Subfigure E}), HMC (\textbf{Subfigure F}) and ARW (\textbf{Subfigure G}) for one of the 100 datasets. The adaptive covariance structure of AGESS is represented by the orange ellipse in \textbf{Subfigure E}. The effective sample size per second for ESS, GESS, and AGESS can be visualized in \textbf{Subfigure D}, while \textbf{Subfigure H} contains the relative KL divergence.}
    \label{fig: Banana}
\end{figure}

Visualizations of the MCMC samples from one dataset can be found in Figure \ref{fig: Banana}. These visualizations illustrate that ESS, GESS, and AGESS appear to be able to explore the tails of the banana better than HMC and ARW---at least on this run. To quantify how well each sampler can explore the target distribution, we will compare the estimated Kullback–Leibler (KL) divergence between the target distribution and the distribution of the MCMC samples. Specifically, we will approximate the target distribution through numerical integration and will use kernel density estimation to estimate the density of the MCMC samples when estimating the KL divergence. Letting $P$ denote the approximated target distribution and $Q_{\mathcal{S}}$ denote the estimated density of the MCMC samples ($\mathcal{S} \in \{\text{ESS}, \text{ GESS}, \text{ AGESS}, \text{ HMC}, \text{ ARW}\}$), we define the relative KL divergence as $D_{\text{KL}}\infdiv{P}{Q_{\mathcal{S}'}} - D_{\text{KL}}\infdiv{P}{Q_{\text{AGESS}}}$ ($\mathcal{S}' \in \{\text{ESS}, \text{ GESS}, \text{ HMC}, \text{ ARW}\}$). Subfigure H of Figure \ref{fig: Banana} contains a visualization of the relative KL divergences calculated over the 100 datasets. From this, it is apparent that ESS, GESS, and AGESS are consistently able to explore the target distribution, whereas HMC and ARW often have trouble adequately exploring the target distribution, particularly the tail areas. Indeed, tails are areas of high curvature that can often cause mixing problems and cause HMC to not be geometrically ergodic \citep{betancourt2017conceptual}.

Subfigure D of Figure \ref{fig: Banana} provides a visualization of the sampling efficiency among the samplers that can reliably explore the target distribution. When the target distribution is close to the origin, ESS and GESS tend to be more efficient samplers compared to AGESS. In these scenarios, the prior is a reasonable distribution from which to construct the ellipse of possible next states. We can see that AGESS only becomes more efficient as the target distribution becomes further from the prior distribution. In these cases, the curvature of the target distribution tends to become slightly less extreme, and adaptation becomes more important for efficient sampling; outweighing the slight increase in computational complexity of the adaptive scheme.

\begin{figure}
    \centering
    \includegraphics[width=0.99\linewidth]{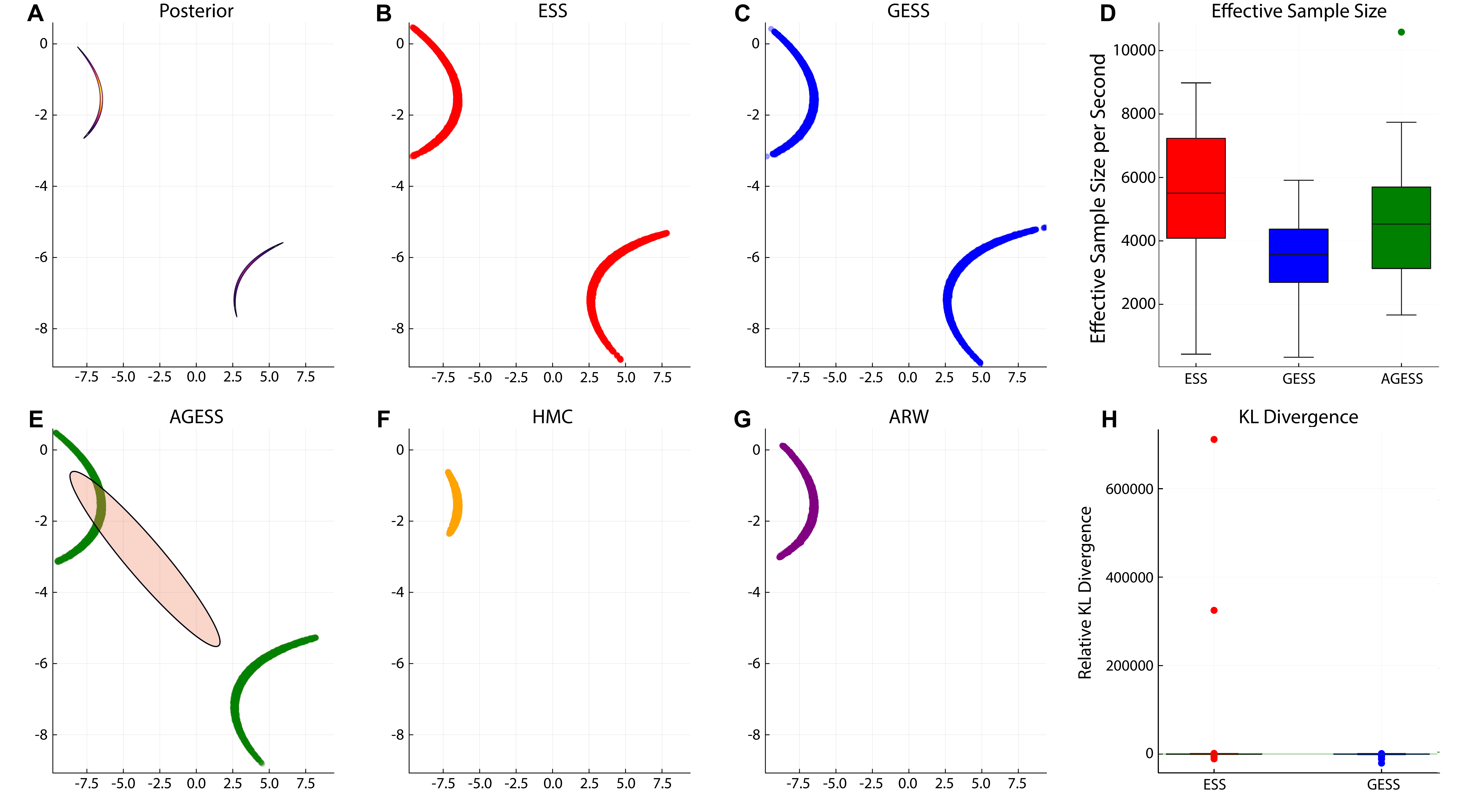}
    \caption{Visualization of one of the twin banana target distributions considered (\textbf{Subfigure A}), along with the MCMC samples generated from ESS (\textbf{Subfigure B}), GESS (\textbf{Subfigure C}), AGESS (\textbf{Subfigure E}), HMC (\textbf{Subfigure F}), and ARW (\textbf{Subfigure G}).  The adaptive covariance structure of AGESS is represented by the orange ellipse in \textbf{Subfigure E}. Visualizations of the effective sample size per second and relative KL divergence can be found in \textbf{Subfigures D} and \textbf{H}, respectively.}
    \label{fig: TwinBanana}
\end{figure}

While the Banana distribution is a good example of non-convex and non-elliptical distribution, it is still a unimodal distribution. To study the performance of the MCMC algorithms over complex multimodal target distributions, we propose a \textit{Twin Banana} distribution. Similarly to the Banana distribution, the Twin Banana distribution on $\boldsymbol{\theta}:= (\theta_1, \theta_2)$ can be specified through the following hierarchical formulation
$$Y_i \sim \mathcal{N}\left(0.1(\theta_1 - \mu_1)^2  - 0.5(\theta_2 - \mu_2)^4 - 10(\theta_1 - \mu_1)(\theta_2 - \mu_2), 100\right),$$
$$\theta_1 \sim \mathcal{N}(0, 4)\quad \theta_2 \sim \mathcal{N}(0, 4),$$
where $Y_i$ is generated from $\mathcal{N}(100, 100)$. While target distributions consisting of two banana distributions have been proposed in the literature \citep{dharamshi2024sampling}, we note that the constructed Twin Banana distribution consists of two \textit{isolated} banana distributions. In other words, jumping from one mode to the other would require the sampler to traverse a large area of almost negligible posterior probability; see Figure \ref{fig: TwinBanana}. The challenging target distribution will test each sampler's ability to make global moves between two non-convex, non-elliptical distributions. In this simulation study, we ran each sampling scheme for 500,000 iterations, discarding the first 250,000 iterations as burn-in. Similarly to the standard Banana simulation study, we simulated 100 datasets and used randomly generated mean parameters such that $\mu_1, \mu_2 \sim \mathcal{N}(0,9)$.

As illustrated in Figure \ref{fig: TwinBanana}, HMC and ARW are unable to explore both modes of the target distribution. Since HMC uses the local geometry to inform the transitions, it is not surprising that HMC has trouble traversing between two isolated modes. Similarly, ARW converges to one of the two modes, and the transition kernel adapts to the local geometry of the single mode; making it unlikely to ever switch between modes. As a result, ARW and HMC are unable to construct Markov chains with transition kernels that converge to the stationary distribution in a reasonable amount of iterations and therefore were not included in effective sample size calculations. The relative KL divergence illustrates that ESS, GESS, and AGESS were able to adequately explore the target distribution most of the time. Two exceptions can be identified by the relative KL divergence, where ESS failed to adequately explore one of the modes; potentially illustrating the benefit of using the heavier-tailed T-distribution for the proposal distribution. The effective sample size per second was found to be relatively similar between ESS, GESS, and AGESS, with perhaps AGESS being slightly more efficient. However, unlike in the standard banana simulation study, we did not find substantial changes in the effective sample size as the mean parameters ($\mu_1, \mu_2$) changed. Overall, this simulation study illustrates that, even with the proposed adaptive scheme, AGESS is relatively robust---it is able to sample from non-elliptical and multimodal target distributions with regions of high curvature.

\begin{table}[t]
\centering
\caption{Performance metrics (mean $\pm$ SD) for the banana and twin banana case studies ($n=100$ replications each).}
\label{tab: banana_summary}
\begin{tabular}{lccc}
\toprule
Sampler & ESS/sec & ESS/iter & ESS/like.\ eval. \\
\addlinespace
\multicolumn{4}{l}{\textit{Banana}} \\
ESS   & 19660 $\pm$ 8579 & 0.03413 $\pm$ 0.01321 & 0.004558 $\pm$ 0.002039 \\
GESS  & 14700 $\pm$ 6019 & 0.03518 $\pm$ 0.01273 & 0.004584 $\pm$ 0.001931 \\
AGESS & 15530 $\pm$ 7093 & 0.03019 $\pm$ 0.00911 & 0.005476 $\pm$ 0.002915 \\
\addlinespace
\multicolumn{4}{l}{\textit{Twin Banana}} \\
ESS   & 5789 $\pm$ 2449 & 0.01421 $\pm$ 0.00561 & 0.001579 $\pm$ 0.0006285 \\
GESS  & 4260 $\pm$ 1703 & 0.01425 $\pm$ 0.00539 & 0.001531 $\pm$ 0.0006039 \\
AGESS & 5359 $\pm$ 2218 & 0.01396 $\pm$ 0.00436 & 0.002039 $\pm$ 0.0008324 \\
\bottomrule
\end{tabular}
\end{table}

\subsection{Multimodal 2-D Distribution}

Here, we consider a two-dimensional multimodal target distribution used in the \texttt{Turing.jl} documentation \citep{ge2018turing,fjelde2025turing}, allowing us to test how different sampling methods can traverse the difficult target distribution. The target distribution can be constructed hierarchically as follows:
\begin{equation}
\begin{gathered}
s^2 \sim \text{Inv-Gamma}(2,3) \qquad m \sim \mathcal{N}(0, s^2);\\
x_i \sim \mathcal{N}(m + 5(\sin(m) + \cos(m)), s^2),
\end{gathered}
\end{equation}
where we observe the vector of data $\mathbf{x} = [1.5, 2.0, 13.0, 2.1, 0.0]$. Using the \texttt{Turing.jl} ecosystem, we compare the performance of AGESS with the following set of sampling methods: (1) particle Gibbs \citep{andrieu2010particle} ($20$ particles), (2) Metropolis-Hastings (using the prior as the proposal distribution), (3) No-U-Turn Sampler \citep{hoffman2014no}, and (4) adaptive random walk \citep{haario2001adaptive}. Each of the sampling methods considered was run for 10,000 MCMC iterations. The code for this case study can be found in the documentation for \texttt{AdaptEllipticalSliceSampler.jl} (the corresponding package associated with this manuscript). 

\begin{figure}[h]
    \centering
    \includegraphics[width=\linewidth]{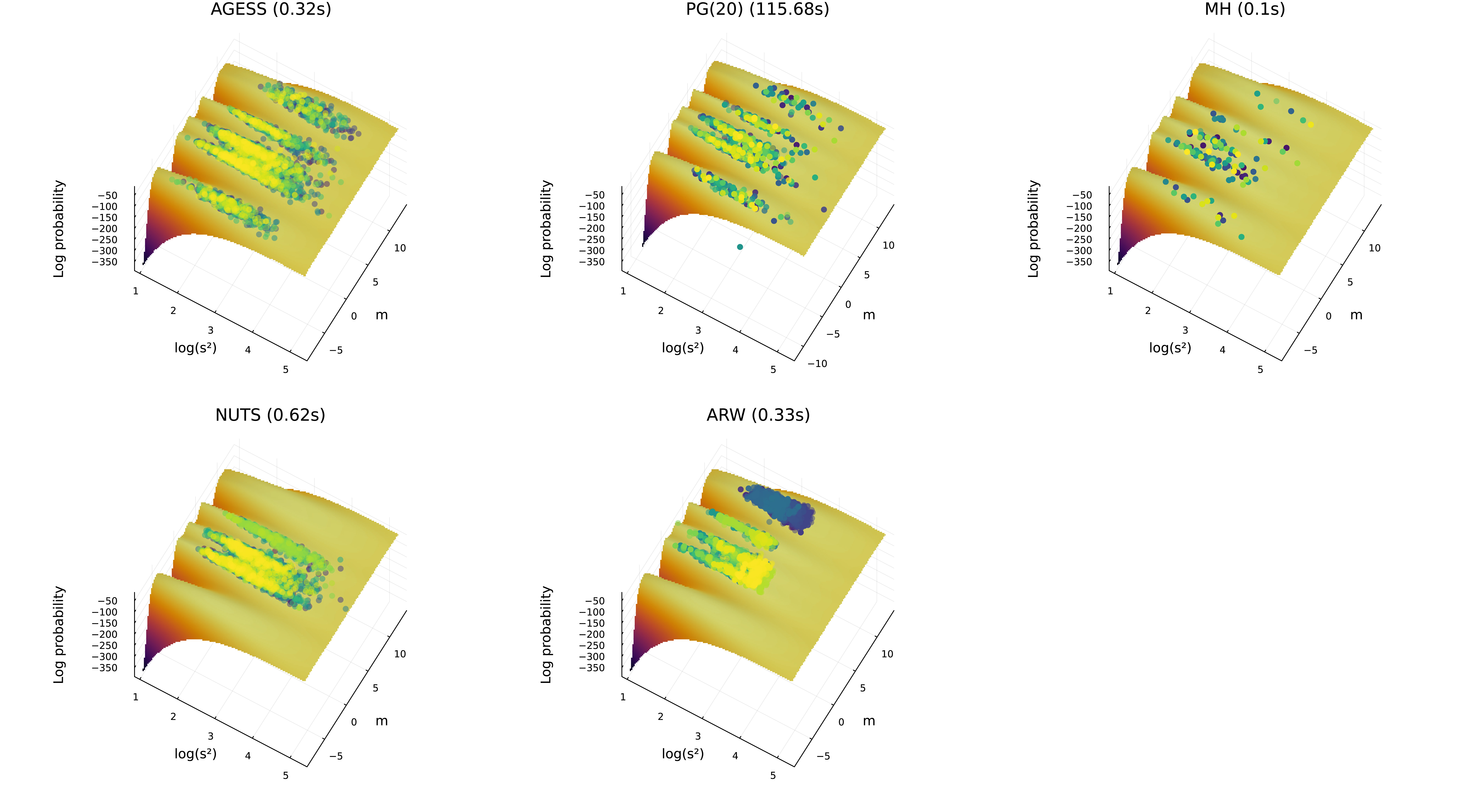}
    \caption{Visualization of the trajectories of one Markov chain generated by each of the various sampling methods. The computation time (in seconds) is included in the title of each subplot.}
    \label{fig: multi_2d}
\end{figure}

As illustrated in Figure \ref{fig: multi_2d}, both the No-U-Turn sampler and the adaptive random walk have difficulty traversing the target distribution. In contrast, AGESS, particle Gibbs, and Metropolis-Hastings were all able to traverse the multimodal distribution, but the efficiency in which they did so greatly differs between the three methods. Metropolis-Hastings is the cheapest to compute, taking only 0.1 seconds to run; however, we can see that it was inefficient in that most of the proposed transitions were rejected (1.8\% acceptance rate). While particle Gibbs increases the efficiency per iteration, we can see a substantial increase in the computation time, as illustrated in Table \ref{tab: multi_efficiency}. We can see that AGESS does a good job traversing the target distribution: only taking roughly three times longer than Metropolis-Hastings, while substantially outperforming the alternative methods in terms of effective sample size per iteration and per second. Overall, this toy example is a good example of how AGESS can traverse modes---at least in low to moderate dimensions. AGESS is not as prone to getting stuck in modes as methods such as HMC and ARW, while scaling well with dimension compared to Metropolis-Hastings and Particle Gibbs (as evident from the other case studies and mixing times derived for an optimally specified elliptical sampler).

\begin{table}
\centering
\caption{Sampling efficiency for the various samplers provided by \texttt{Turing.jl} (10,000 iterations, first 2,500 discarded as burn-in). The No-U-Turn sampler and adaptive random walk are not included as they failed to effectively explore the target distribution.}
\label{tab: multi_efficiency}
\begin{tabular}{lccccc}
\toprule
 Parameter & ESS (bulk) & ESS (tail) & $\hat{R}$ & ESS/sec & Time (s) \\
\addlinespace
\multicolumn{6}{l}{\textit{AGESS}} \\ $s^2$ & 1430  & 1605  & 1.003 & 4469  & 0.32 \\
 $m$   & 2080  & 1451  & 1.000 & 6500 & 0.32 \\
\addlinespace
\multicolumn{6}{l}{\textit{Particle Gibbs (20 particles)}} \\
$s^2$ & 348.6 & 379.7 & 1.001 & 3.01   & 115.68 \\
 $m$   & 370.0 & 265.4 & 1.003 & 3.20   & 115.68 \\
\addlinespace
\multicolumn{6}{l}{\textit{Metropolis--Hastings}} \\
 $s^2$ & 45.1  & 167.2 & 1.043 & 451  & 0.10 \\
 $m$   & 51.9  & 60.4  & 1.055 & 519  & 0.10 \\
\bottomrule
\end{tabular}
\end{table}

\subsection{Potential Utility in Constrained Bayesian Inference}

Although the elliptical slice sampler \citep{murray2010elliptical} and the generalized elliptical slice sampler \citep{nishihara2014parallel} assume that the target distribution is a continuous distribution over $\mathbb{R}^P$, we relax these assumptions and show that our adaptive algorithm is ergodic for fairly general target distributions on $(\mathcal{X}, \mathcal{B}(\mathcal{X}))$, where $\mathcal{X}$ is an open subset of $\mathbb{R}^P$. This allows us to explore the use of the proposed adaptive algorithm to conduct constrained Bayesian inference \citep{gelfand1992bayesian, duan2020bayesian, presman2023distance, zhou2024proximal}, where the parameters of interest are constrained to a set. Currently, most methods focus on relaxing the constraints using distance-to-set penalties \citep{duan2020bayesian, presman2023distance, zhou2024proximal}, thus performing inference on a differentiable approximation of the true posterior distribution. However, as the approximation becomes tighter, the gradient-based sampling methods commonly used in these scenarios can exhibit poor sampling efficiency \citep{duan2020bayesian}. Alternatively, adaptive generalized elliptical slice sampling emerges as a potentially promising alternative: a sampling method that targets the exact posterior distribution for constrained inference problems where the constraint set is an open subset of $\mathbb{R}^P$.

\bibliographystyle{abbrvnat}
\bibliography{AGESS}